\documentclass{jfm}

\usepackage{graphicx}
\usepackage{natbib}
\usepackage{subfig}
\usepackage{color}

\ifCUPmtlplainloaded \else
  \checkfont{eurm10}
  \iffontfound
    \IfFileExists{upmath.sty}
      {\typeout{^^JFound AMS Euler Roman fonts on the system,
                   using the 'upmath' package.^^J}%
       \usepackage{upmath}}
      {\typeout{^^JFound AMS Euler Roman fonts on the system, but you
                   dont seem to have the}%
       \typeout{'upmath' package installed. JFM.cls can take advantage
                 of these fonts,^^Jif you use 'upmath' package.^^J}%
      }
  \else
  \fi
\fi

\ifCUPmtlplainloaded \else
  \checkfont{msam10}
  \iffontfound
    \IfFileExists{amssymb.sty}
      {\typeout{^^JFound AMS Symbol fonts on the system, using the
                'amssymb' package.^^J}%
       \usepackage{amssymb}%
         
       \let\ge=\geqslant  
      }{}
  \fi
\fi

\ifCUPmtlplainloaded \else
  \IfFileExists{amsbsy.sty}
    {\typeout{^^JFound the 'amsbsy' package on the system, using it.^^J}%
     \usepackage{amsbsy}}
    {\providecommand\boldsymbol[1]{\mbox{\boldmath $##1$}}}
\fi

\newcommand\Rey{\mbox{\textit{Re}}}  % Reynolds number
\newcommand{\nom}{Nu_\omega}
\newcommand{\nomh}{\overline{Nu}_\omega}
\newcommand{\rew}{\Rey_w}
\newcommand{\usro}{Ro^{-1}}
\newcommand{\bu}{\boldsymbol{u}}

\newcommand{\mean}[1]{\left \langle #1 \right \rangle}

\newsavebox{\astrutbox}
\sbox{\astrutbox}{\rule[-5pt]{0pt}{20pt}}

\def\mathLarge#1{\mbox{\Large $#1$}}

\title[Optimal Taylor-Couette flow: direct numerical simulations]{Optimal Taylor-Couette flow: \\ direct numerical simulations}

\author[R. Ostilla, R.J.A.M. Stevens, S. Grossmann,  R. Verzicco and D. Lohse]%
{R\ls O\ls D\ls O\ls L\ls F\ls O\ns O\ls S\ls T\ls I\ls L\ls L\ls A$^1$,\ns%
R\ls I\ls C\ls H\ls A\ls R\ls D\ns J.\ns A.\ns M.\ns S\ls T\ls E\ls V\ls E\ls N\ls S$^{1,2}$,\break%
S\ls I\ls E\ls G\ls F\ls R\ls I\ls E\ls D\ns G\ls R\ls O\ls S\ls S\ls M\ls A\ls N\ls N$^3$,\ns
R\ls O\ls B\ls E\ls R\ls T\ls O\ns V\ls E\ls R\ls Z\ls I\ls C\ls C\ls O$^{1,4}$ \break
\and D\ls E\ls T\ls L\ls E\ls F\ns L\ls O\ls H\ls S\ls E$^1$}

\affiliation{$^1$Physics of Fluids, University of Twente, P.O. Box 217, 7500 AE Enschede, The Netherlands\\[\affilskip]
$^2$Johns Hopkins University, Department of Mechanical Engineering, 3400 N. Charles Street, Baltimore, MD 21218\\[\affilskip]
$^3$Department of Physics, University of Marburg, Renthof 6, D-35032 Marburg, Germany\\[\affilskip]
$^4$Dipartimento di Ingegneria Meccanica, University of Rome ``Tor Vergata'', Via del Politecnico 1, Roma 00133, Italy}

\date{\today}
\begin{document}
 
\maketitle

\begin{abstract}
We numerically simulate turbulent Taylor-Couette flow for independently rotating inner and outer cylinders, focusing on the analogy with turbulent Rayleigh-B\'enard flow. Reynolds numbers of $Re_i=8\cdot10^3$ and $Re_o=\pm4\cdot10^3$ of the inner and outer cylinders, respectively, are reached, corresponding to Taylor numbers Ta up to $10^8$. Effective scaling laws for the torque and other system responses are found. Recent experiments with the Twente turbulent Taylor-Couette ($T^3C$) setup and with a similar facility in Maryland at very high Reynolds numbers have revealed an optimum transport at a certain non-zero rotation rate ratio $a = -\omega_o / \omega_i$ of about $a_{opt}=0.33-0.35$.
 For large enough $Ta$ in the numerically accessible range we also find such an optimum transport at non-zero counter-rotation. The position of this maximum is found to shift with the driving, reaching a maximum of $a_{opt}=0.15$ for $Ta=2.5\cdot10^7$. An explanation for this shift is elucidated, consistent with the experimental result that $a_{opt}$ becomes approximately independent of the driving strength for large enough Reynolds numbers. We furthermore numerically calculate
 the angular velocity profiles and visualize the different flow structures  for the various regimes. By writing the equations in a frame co-rotating with the outer cylinder a link is found between the local angular velocity profiles and the global transport quantities.
\end{abstract}

%\begin{keywords}
%Authors should not enter keywords on the manuscript, as these must be chosen by the author during the online submission process and will then be added during the typesetting process (see http://journals.cambridge.org/data/\linebreak[3]relatedlink/jfm-\linebreak[3]keywords.pdf for the full list)
%\end{keywords}

\section{Introduction}

\begin{figure}
 \begin{center}
  \includegraphics[width=0.7\textwidth]{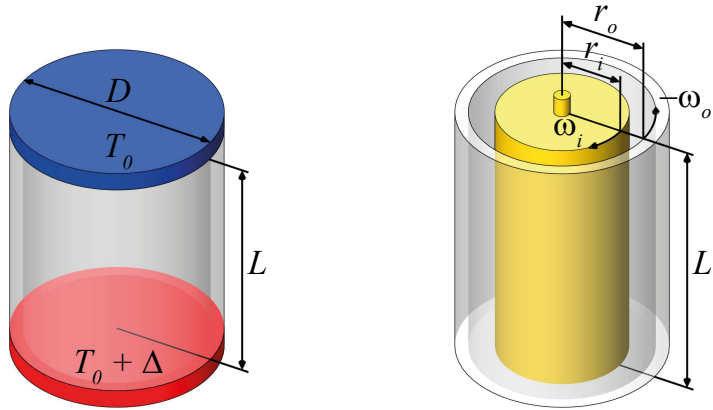}
  \caption{Geometry of the Rayleigh-B\'enard (left) and the Taylor-Couette systems (right). The RB system consists of two cylindrical plates, a hot one at the bottom and a cold one at the top of diameter $D$ separated by a distance $L$. The top plate is at a temperature $T_0$ and the bottom plate is at a temperature $T_0+\Delta$, with $\Delta$ the temperature difference between the plates. The TC system consists of two coaxial cylinders of length $L$. The inner cylinder has the radius $r_i$ and the angular velocity $\omega_i$, while the outer cylinder has the radius $r_o$ and the angular velocity $\omega_o$.}
  \label{fig:RBTCnosplit}
 \end{center}
\end{figure}

Taylor-Couette (TC) flow, i.e.\ the flow in the gap between two independently rotating coaxial cylinders, is 
among the most investigated problems in fluid mechanics, due to its conceptional simplicity and to applications 
in process technology, see e.g.\  \cite{hai94}. Traditionally, the 
driving of this system is expressed by the Reynolds numbers of the inner and outer cylinders, defined by 
$Re_i = r_i \omega_i d /\nu$ and $Re_o = r_o \omega_o d /\nu$, 
where $r_i$ and $r_o$ are the inner and outer cylinder radius, respectively, $d = r_o-r_i$ is the gap width, $\omega_i$ and $\omega_o$ the angular velocities of the inner and outer cylinders, and $\nu$ the kinematic viscosity. In dimensionless numbers the geometry of a TC system is expressed by the radius ratio $\eta = r_i/r_o$ and the aspect 
ratio $\Gamma = L/d$, see figure \ref{fig:RBTCnosplit}. In the limit $\eta\to1$, the flow becomes plane Couette flow. It was shown by  Eckhardt, Grossmann \& Lohse 2007 (from now on referred to as EGL 2007) 
that TC flow has many similarities to Rayleigh-B\'enard (RB) convection, i.e.\ the thermal flow in a fluid layer heated from below and cooled from above, 
which will be discussed in detail below.  

Both RB and TC flows have been popular playgrounds for the development of new concepts in fluid dynamics. Both systems have been used to study instabilities \citep{pfi81,pfi88,cha81,dra81,bus67}, nonlinear dynamics and chaos \citep{lor63,ahl74,beh85,str94}, pattern formation \citep{and86, cro93,bod00}, and turbulence \citep{sig94,gro00,kad01,lat92,ahl09,loh10}. The reasons that RB and TC are so popular include: (i) These systems are mathematically well-defined by the Navier-Stokes equations and the appropriate boundary conditions; (ii) these are closed system and thus exact global balance relations between the driving and the dissipation can be derived; and (iii) they are experimentally and numerically accessible with high precision, thanks to the simple geometries and high symmetries. 

The analogy between TC and RB may be better seen from the exact relations (EGL 2007) between the transport quantities and the energy dissipation rates. For RB flow the conserved quantity that is transported is the thermal flux $J=\langle u_z\theta\rangle_{A, t}-\kappa\partial _z\langle\theta\rangle_{A, t}$ of the temperature field $\theta$, where $\kappa$ is the thermal conductivity of the flow. The first term is then the convective contribution ($u_z$ is the vertical fluid velocity component) and the second term is the diffusive contribution. Here $\langle ... \rangle_{A, t}$ indicates the averaging over time and a horizontal plane. In the state with lowest thermal driving there is not yet convection. Therefore $J \equiv J_0 = \kappa \Delta L^{-1}$ and the corresponding dissipation rate is $\epsilon_{u,0} = 0$ since $\bu = 0$. -- In TC flow, the conserved transport quantity, which is transported from the inner to the outer cylinder (or vice versa) is the flux $J^{\omega}=r^3\left(\langle u_r\omega\rangle_{A, t}-\nu\partial_r\langle\omega\rangle_{A, t}\right)$ of the angular velocity field $\omega$, where the first term is the convective contribution with $u_r$ as the radial fluid velocity component and the second term is the diffusive contribution, cf. EGL 2007. In this case $\langle ... \rangle_{A, t}$ indicates averaging over time and a cylindrical surface with constant radial distance $r$ from the axis. In the state with lowest driving induced by the rotating cylinders and neglecting plate effects from the upper and
lower plates (achieved in the simulations by periodic boundary conditions in axial direction), 
the flow is laminar and purely azimuthal, $u_{\theta}(r) = Ar + B/r$, while $u_r = u_z =0$. This flow provides an angular velocity current $J^{\omega}_0$ (called $J^{\omega}_{lam}$ in EGL2007) and a nonzero dissipation rate, see eqs.(\ref{jzero}) and (\ref{eps-u-0}) in table 1.  

The analogy between RB and TC (EGL 2007) is highlighted when the driving in TC is expressed in terms of the Taylor number $Ta$ and the angular velocity ratio $a=-\omega_o/\omega_i$ of the cylinders, while the response is given by the dimensionless transport current density $J^{\omega}$ divided by the corresponding molecular current density of the angular velocity from the inner to the outer cylinder, called the ``$\omega$-Nusselt number'' $Nu_\omega$. The Taylor number $Ta$ is defined as $Ta = \sigma (r_o - r_i)^2 (r_o + r_i)^2 (\omega_o - \omega_i)^2 / (4 \nu^2)$, or 
\begin{equation}
Ta = (r_a^6 d^2 / r_o^2 r_i^2 \nu^2) (\omega_o - \omega_i)^2. 
\end{equation}
Here \begin{equation}\sigma = r_a^4 / r_g^4\end{equation}
 with $r_a = (r_o + r_i)/2$ the arithmetic and $r_g = \sqrt{r_o r_i}$ the geometric mean radii. $\omega_{o,i}$ are the angular velocities of the outer and inner cylinders, respectively; see also table \ref{table1} for definitions and relations.

\begin{table}
\caption{\small Analogous relations between RB and TC flow, leading to the same effective scaling laws as derived by Eckhardt, Grossmann \& Lohse \citep{eck07b}. 
In RB flow, the dimensionless control parameters 
are the Rayleigh number $Ra = \beta g \Delta L^3 / (\nu \kappa )$, the Prandtl number  $Pr= \nu /\kappa$, 
and the aspect ratio $\Gamma= D/L$, where $\Delta$ is the temperature difference 
between the cold top and hot bottom, $\beta$  the thermal expansion coefficient, 
$g$ the gravitational acceleration, and $\kappa$ the thermal diffusivity, see 
figure \ref{fig:RBTCnosplit}. The response of the system is the heat flux from 
the bottom to the top in terms of the molecular heat flux, known as the Nusselt number $Nu$.
In analogy, for TC flow we define a Nusselt number $Nu_\omega$ as ratio of the total and the purely azimuthal and laminar angular velocity flow. 
$\tilde{\epsilon}_{u,0}$ is the dissipation in the purely diffusive state, equal to zero in RB flow, since the fluid velocity is zero and there is molecular heat transport only, while in TC flow $\tilde{\epsilon}_{u,0}$ is the purely azimuthal and laminar flow dissipation rate. 
\label{table1}}
%\vspace{1mm}
\renewcommand\arraystretch{.92}
\centering
\begin{tabular}{p{.45\textwidth}|p{.45\textwidth}}
\hline\hline
\textbf{~~~~~~~~~~~~~Rayleigh-B\'enard} & \textbf{~~~~~~~~~~~~~~~~~~~~~~~~Taylor-Couette} \\
\hline
 & \\
 %%% %%% %%% %%% %%% %%% %%% %%% %%% %%% %%% %%% %%% %%% %%% %%%
Conserved: temperature flux & Conserved: angular velocity flux \\

\vspace{-2mm} \begin{equation}
J=\langle u_z\Theta\rangle_{A, t}-\kappa\partial _z\langle\Theta\rangle_{A, t}
\end{equation} &

\vspace{-2mm} \begin{equation}
J^{\omega}=r^3\left(\langle u_r\omega\rangle_{A, t}-\nu\partial_r\langle\omega\rangle_{A, t}\right) \label{eq:ch1_J_w}
\label{eq:j-definition}
\end{equation} 
% Add somewhere \thanks{The authors would like to point out that $J^\omega$ appeared first in \cite{bus72} by the name of `torque'}:
\\
 %%% %%% %%% %%% %%% %%% %%% %%% %%% %%% %%% %%% %%% %%% %%% %%%
\vspace{-6mm} Dimensionless transport: & \vspace{-6mm} Dimensionless transport: \\

\vspace{-6mm} \begin{equation}
Nu=\mathLarge{\frac{J}{J_0}}
\end{equation} &

\vspace{-6mm} \begin{equation}
Nu_{\omega}=\mathLarge{\frac{J^{\omega}}{J^{\omega}_0}}\mathrm{~~~}\left(=\mathLarge{\frac{\tau}{2\pi L \rho J^{\omega}_0}}\right)
\end{equation} \\

\vspace{-6mm} \begin{equation}
J_0=\kappa\Delta L^{-1}
\end{equation} &

\vspace{-6mm} \begin{equation} \label{jzero}
J^{\omega}_0=\nu \frac{2 r_i^2 r_o^2}{r_i + r_o} \frac{\omega_i - \omega_o}{d} 
\end{equation} \\
 %%% %%% %%% %%% %%% %%% %%% %%% %%% %%% %%% %%% %%% %%% %%% %%%
\vspace{-4mm} Driven by: & \vspace{-4mm} Driven by: \\

\vspace{-6mm} \begin{equation}
Ra=\mathLarge{\frac{\beta g\Delta L^3}{\kappa \nu}}
\end{equation} &
\vspace{-6mm} \begin{equation}
Ta=\mathLarge{\frac{1}{4}\frac{\sigma(r_o-r_i)^2(r_i+r_o)^2(\omega_i-\omega_o)^2}{\nu^2}}
\end{equation} \\

 %%% %%% %%% %%% %%% %%% %%% %%% %%% %%% %%% %%% %%% %%% %%% %%%
\vspace{-6mm} Exact relation: & \vspace{-6mm} Exact relation:\\
\vspace{-6mm} \begin{equation}
\hspace{-14mm} \tilde{\epsilon}'_u = \tilde{\epsilon}_u - \tilde{\epsilon}_{u,0}
\end{equation} &
\vspace{-6mm} \begin{equation}
\hspace{-14mm} \tilde{\epsilon}'_u = \tilde{\epsilon}_u - \tilde{\epsilon}_{u,0}
\end{equation} \\
\vspace{-6mm} \begin{equation}
\hspace{14mm}=\left(Nu - 1\right)Ra~Pr^{-2} \label{eq:ch1_RB_eps}               
\end{equation} &
\vspace{-6mm} \begin{equation}
\hspace{18mm}=\left(Nu_{\omega} - 1\right)Ta~\sigma^{-2} \label{eq:ch1_TC_eps}
\end{equation} \\
\vspace{-6mm} \begin{equation}
\hspace{-25mm}\tilde{\epsilon}_{u,0} = 0 
\end{equation} &
\vspace{-6mm} \begin{equation}
\hspace{8mm}\tilde{\epsilon}_{u,0} = \frac{d^4}{\nu^3} \cdot \nu \frac{r_i^2 r_o^2}{r_a^2} \left( \frac{\omega_i - \omega_o}{d} \right)^2 \label{eps-u-0}
\end{equation} \\
 %%% %%% %%% %%% %%% %%% %%% %%% %%% %%% %%% %%% %%% %%% %%% %%%
\vspace{-4mm} Prandtl number: & \vspace{-4mm} Pseudo `Prandtl' number:\\
\vspace{-6mm} \begin{equation}
Pr=\nu/\kappa
\end{equation} &
\vspace{-6mm} \begin{equation}
\sigma=\left(1+\mathLarge{\frac{r_i}{r_o}}\right)^4\mathLarge{/}\left(4\mathLarge{\frac{r_i}{r_o}}\right)^2
\end{equation} \\
 %%% %%% %%% %%% %%% %%% %%% %%% %%% %%% %%% %%% %%% %%% %%% %%%
\vspace{-6mm} Scaling: & \vspace{-6mm} Scaling: \\

\vspace{-6mm} \begin{equation}
\mbox{ $Nu\propto Ra^{\gamma}$}
\end{equation} &

\vspace{-6mm} \begin{equation}
\mbox{ $Nu_{\omega}\propto Ta^{\gamma}$}
\end{equation} \\
 %%% %%% %%% %%% %%% %%% %%% %%% %%% %%% %%% %%% %%% %%% %%% %%% 
\hline\hline
\end{tabular}
\end{table}

TC flow has been extensively investigated experimentally \citep{wen33,tay36,smi82,and86,ton90,lat92,lat92a,lew99,gil11,gil11a,pao11,hui12} at low and high $Ta$ numbers for different ratios of the rotation frequencies $a = - \omega_o / \omega_i$, see the phase diagram in figure \ref{fig:PhaseSpaceComparison}. However, up to now most
numerical simulations of TC flow have been restricted to the case of pure inner cylinder rotation \citep{fas84,cou96,don07,don08,pir08}, or eigenvalue study \citep{geb93b}, or counter-rotation at fixed $a$ \citep{don08}. Recent experiments \citep{gil11a,gil11,pao11,hui12} have shown that at fixed $Ta$ an optimal transport is obtained at \emph{non-zero} $a$. \citep{gil11} obtained $a_{opt}=0.33 \pm 0.05$, whereas \cite{pao11} got $a_{opt}\approx0.35$. 

In this paper we use direct numerical simulations (DNS) to study the influence of the rotation ratio $a$ on the flow structures and the corresponding transported angular velocity flux for $Ta$ numbers up to $Ta=10^8$. Our motivation is two-fold: as a first objective, we wish to further investigate the analogy between RB and TC flow by comparing the scaling laws of the global response across the different flow states. Our second objective is to study the optimal transport, which was recently observed in TC experiments \citep{gil11,pao11, gil12}, by using data obtained from DNS.  In DNS we namely have access to the complete velocity field, which is not available in experiments, and this allows us to study this phenomenum in much more detail. Presently, however, in DNS we are restricted to smaller Reynolds numbers as compared to above mentioned recent experiments.

\begin{figure}
 \begin{center}
  \subfloat{\label{fig:PhaseSpaceComplete}\includegraphics[width=0.32\textwidth,trim = 0mm 0mm 0mm 0mm, clip]{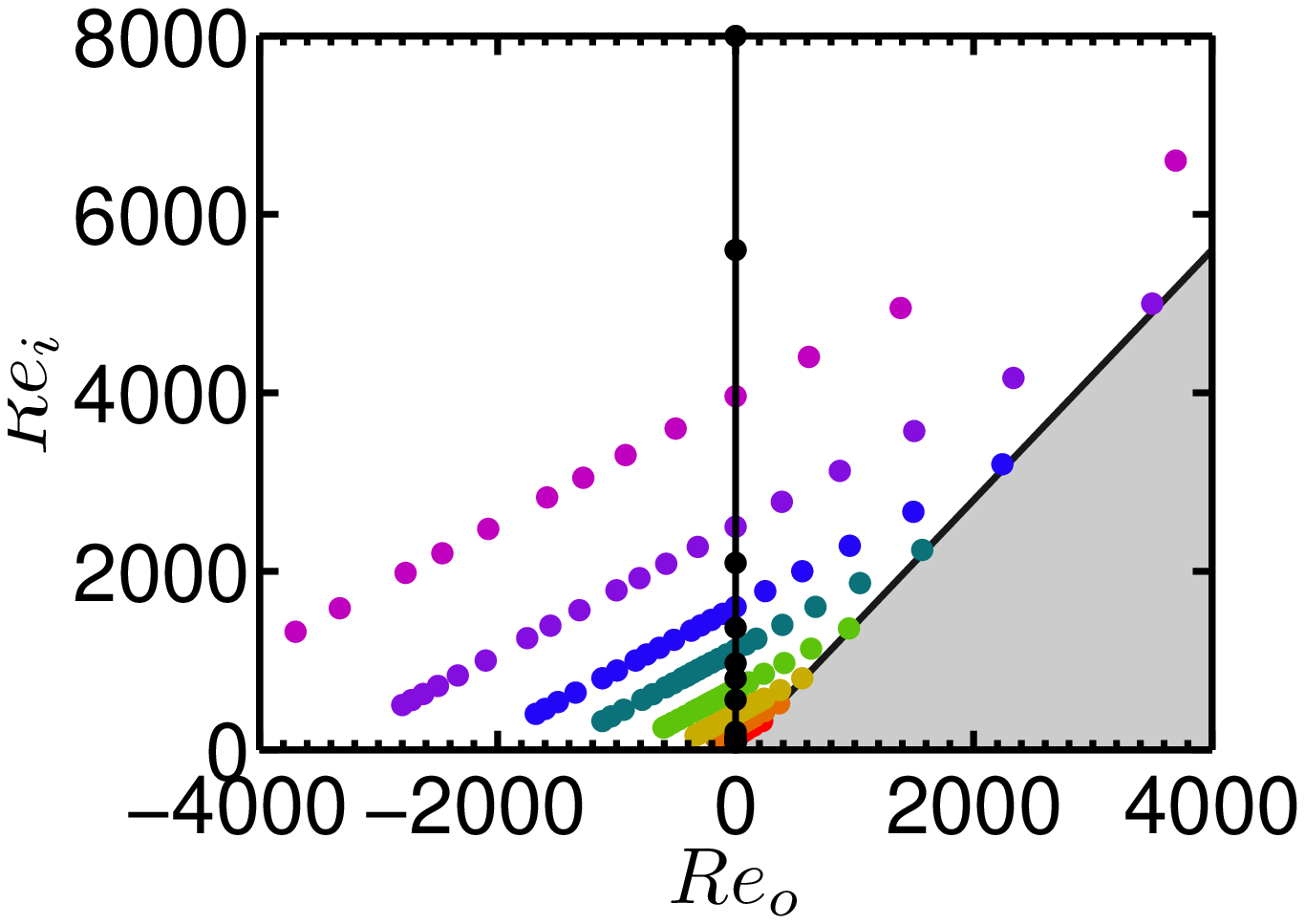}}                
  \subfloat{\label{fig:PhaseSpaceTaa}\includegraphics[width=0.32\textwidth,trim = 0mm 0mm 9mm 0mm, clip]{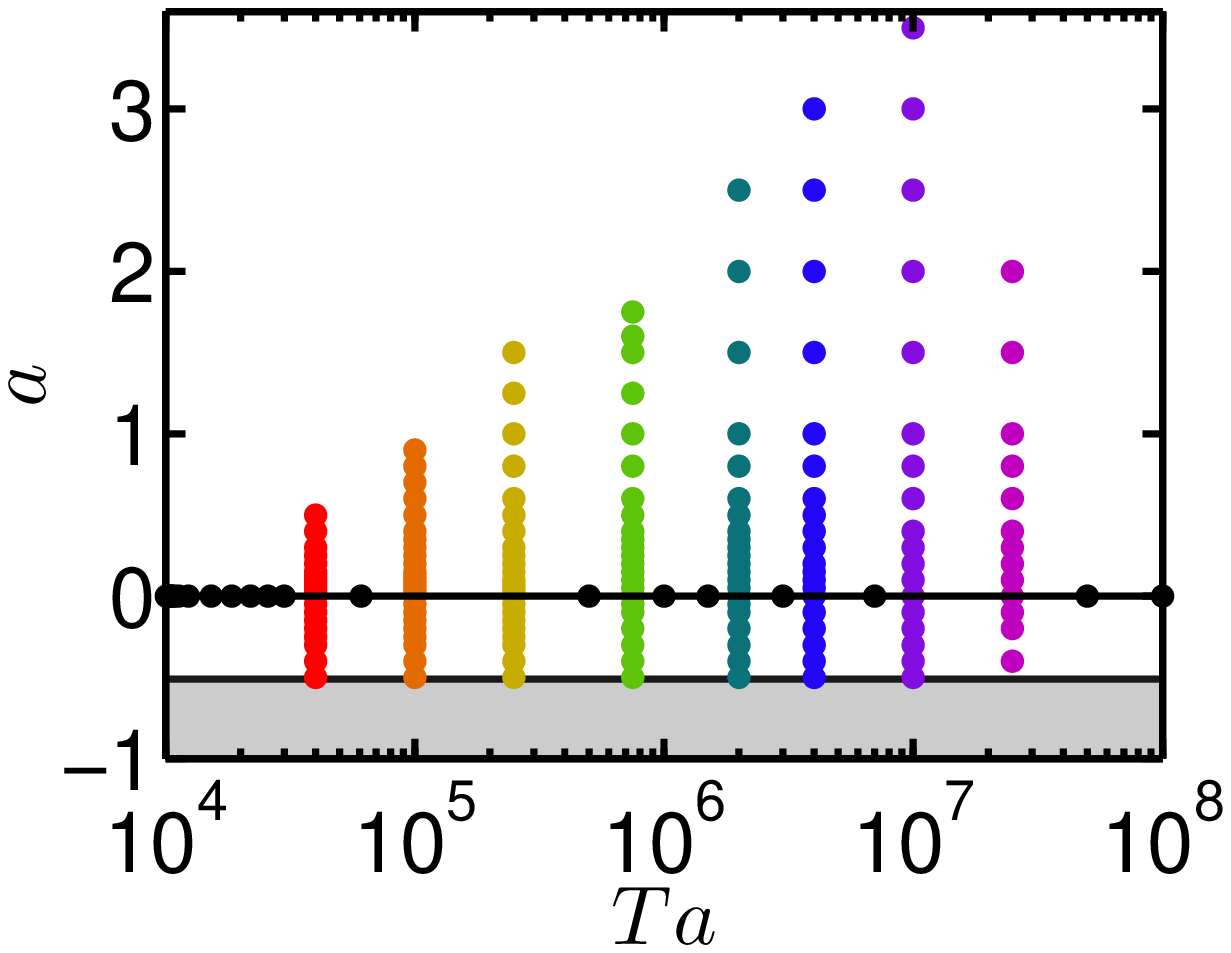}}   
    \subfloat{\label{fig:PhaseSpaceTaRo}\includegraphics[width=0.32\textwidth,trim = 0mm 0mm 9mm 0mm, clip]{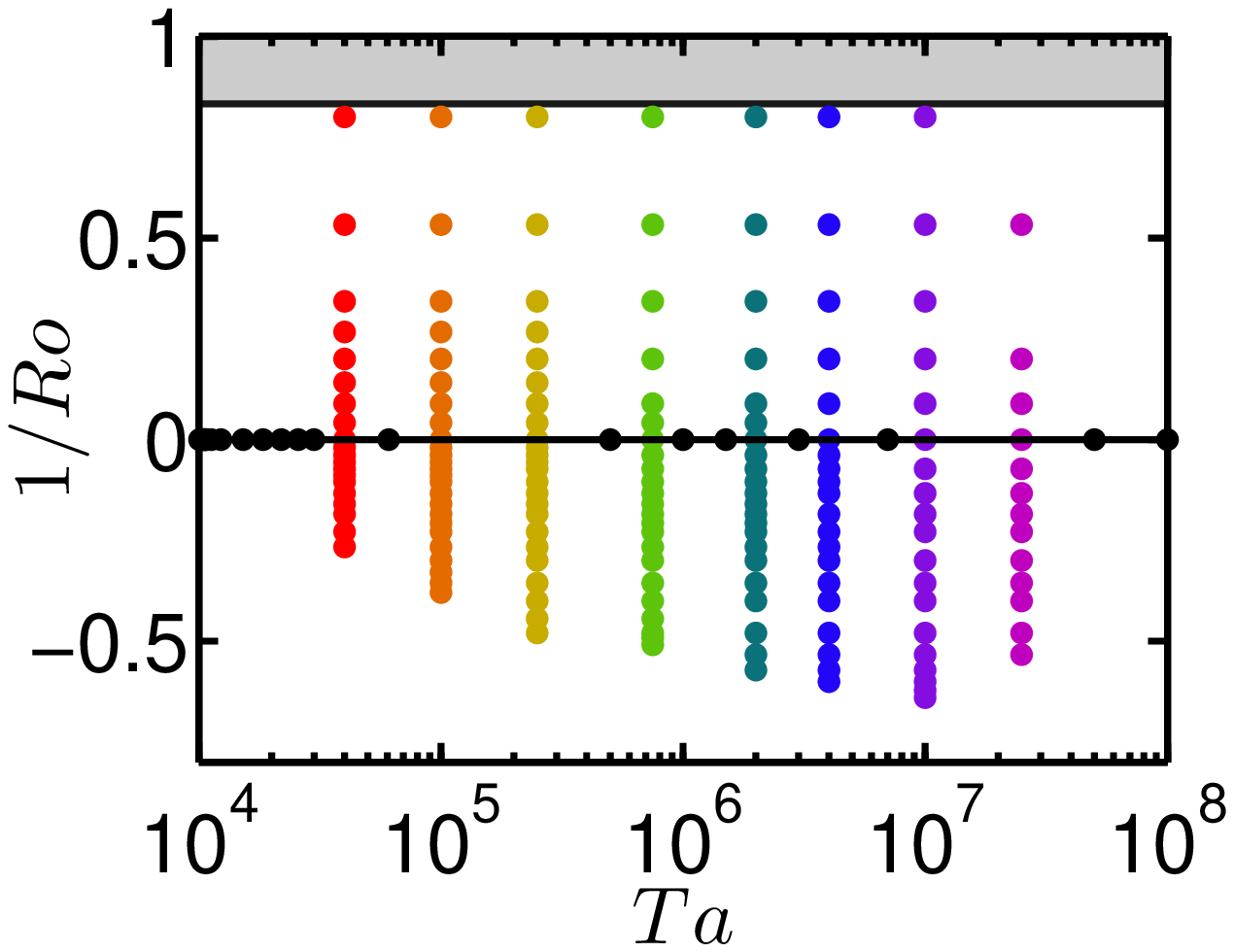}}                 
  \caption{Control parameter phase space which was numerically explored in this paper: (a) classical representation ($Re_o$,$Re_i$) and (b) ($Ta$,$a$) representation
   with $a = - \omega_o / \omega_i$ and (c) ($Ta$, $1/Ro$) representation with $Ro$ defined in eq.\ (\ref{rossby-def}).  
   We fixed $\eta=5/7$, $\Gamma=2\pi$ and employed axial periodicity. The same color code, denoting the Taylor number, is maintained throughout the paper.
   The grey-shaded area, outlines boundary conditions for which the angular momentum of the outer cylinder ($L_o$) is larger than the angular momentum of the inner cylinder ($L_i$). This causes the flow to have an overall transport of angular momentum towards the inner cylinder. In this region, the Rayleigh stability criterium applies, which states that if $dL/dr > 0$ the flow is linearly stable to axisymmetric perturbations.
   }
  \label{fig:PhaseSpace}
 \end{center} 
\end{figure}

\begin{figure}
 \begin{center}
  \includegraphics[width=0.99\textwidth]{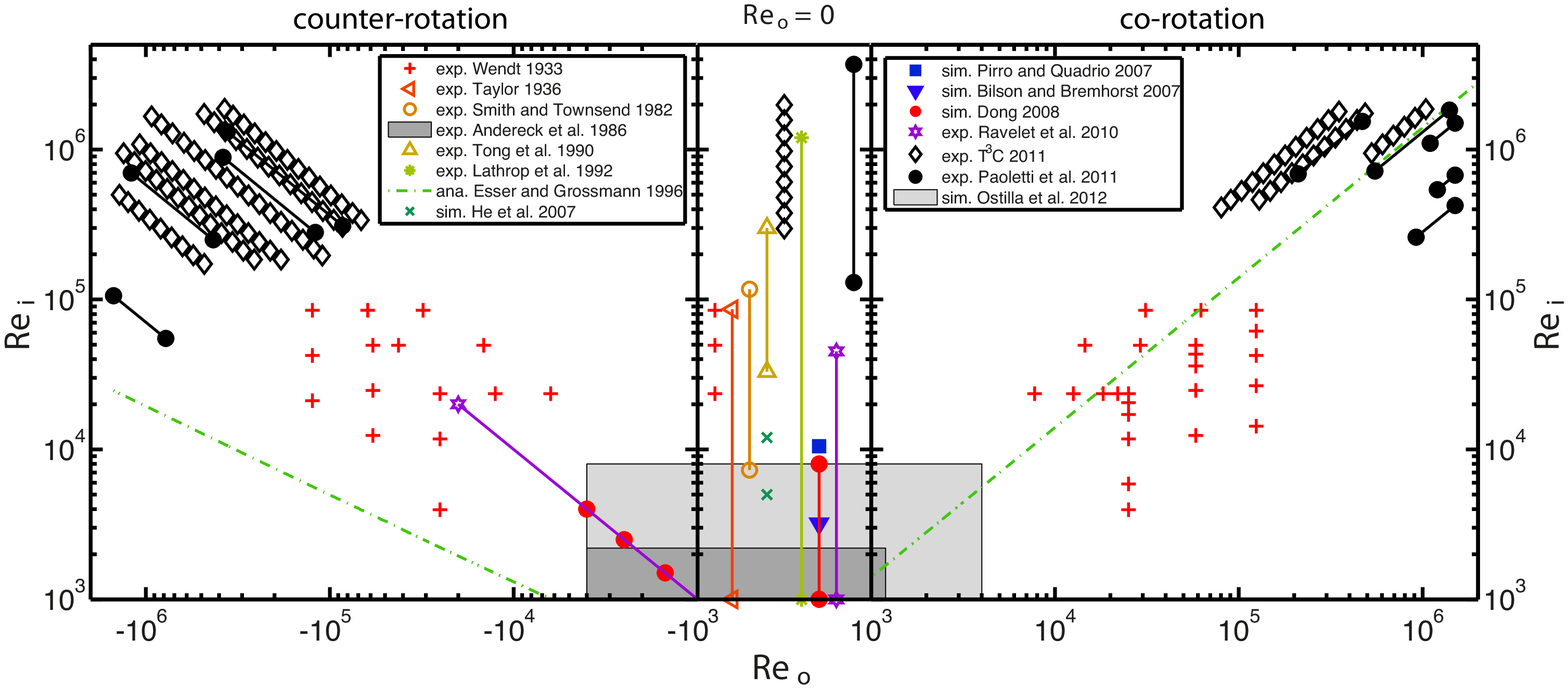}
  \caption{Explored phase space ($Re_o$, $Re_i$) of TC flow with independently rotating 
inner and outer cylinders. Both experimental data \citep{wen33,tay36,smi82,and86,ton90,lat92a,rav10,gil11,pao11} 
and numerical data \citep{pir08,bil07,don07, don08} are shown. Solid lines between markers represent a large density of experiments. The dashed lines are Esser and Grossmann's (1996) 
estimate for the onset of turbulence with $\eta = 0.71$. The dark shaded area indicates the data points in the well studied small Reynolds number regime of pattern formation and spatial temporal chaos (see e.g. \cite{and86,pfi81,cro93}). The light gray area is the region shown in figure \ref{fig:PhaseSpaceComplete}, covered by the present DNS.}
  \label{fig:PhaseSpaceComparison}
 \end{center} 
\end{figure}

 Figure \ref{fig:PhaseSpace} shows the cases which are simulated, in the $(Ta,a)$, the $(Ta,1/Ro)$ and the $(Re_o,Re_i)$ parameter space. Note that a higher density of points has been used in places where the response ($Nu_\omega, Re_w$) shows more variation. All points have been simulated for fixed 
$\Gamma=2\pi$ and $\eta=5/7$ since these are very similar to the parameters of the $T^3C$ setup. There is a significant difference, however, as numerically we take periodic boundary conditions in the axial direction, while the $T^3C$ system is a closed system with solid boundaries at top and bottom which rotate with the outer cylinder. 

In section 2 we start with a description of the numerical method that has been used. In section 3 we will discuss the validation and resolution tests that have been performed. In section 4 the global response, in terms of $Nu_\omega$ and the wind Reynolds number $Re_w$, as functions of the angular velocity ratio $a$ will be discussed. 
In order to understand the global system response we will analyze the coherent structures in section 5 and the boundary layer profiles in section 6, i.e.\ quantities that are difficult to analyze in experiments. This allows us to rationalize the position of the maximum in $Nu_\omega(a)$. 
We conclude with a brief discussion and outlook to future work in section 7.

\cite{bra12} offer a complementary direct numerical simulation of turbulent Taylor-Couette flow: They employ a spectral code with periodic boundary conditions also in azimuthal 
direction and an aspect ratio $\Gamma = 2$ in axial direction rather than $\Gamma = 2\pi$ as we do. Also they find a maximum in the angular velocity transport for 
moderate counter-rotation $a = -\omega_o/\omega_i \approx 0.4$, similar as found in the experiments by \cite{gil11,gil12,pao11} and in the present numerical simulations. 
So the result seems to be very robust and does at least not strongly depend on $Ta$, $\Gamma$ and other details of the flow. \cite{bra12} also offer an analysis of the
PDFs of the local angular velocity fluxes in the different regimes, similarly as has been done in the experiments by \cite{gil12}.

\section{Numerical method}

The Taylor-Couette flow was simulated by solving the Navier-Stokes equations in a rotating frame, which was chosen to rotate with $\Omega=\omega_o \boldsymbol{e}_z$. This way the boundary conditions are simplified: at the inner cylinder the new boundary condition is $u_\theta(r=r_i)=r_i(\omega_i-\omega_o)$, while at the outer cylinder we have a stationary wall
$u_\theta(r=r_o)=0$.
We can choose the characteristic velocity $U \equiv r_i|\omega_i-\omega_o|$ and the characteristic length scale $d$ to non-dimensionalize the equations and boundary conditions. The characteristic velocity $U$ can be written as 
\begin{equation}
U = (\nu / d) \cdot [8 \eta^2 /(1+\eta)^3] \cdot Ta^{1/2}.
\label{characteristic-u}
\end{equation}
\noindent Up to a geometric factor, which is 0.810 for our choice of $\eta$, the characteristic velocity $U$ is thus simply $Ta^{1/2}$, expressed in terms of the molecular velocity $\nu/d$. The non-dimensional variables will be labeled with a hat. In this notation, the non-dimensional inner cylinder velocity boundary condition simplifies to: $\hat{u}_\theta(r=r_i)=(\omega_i-\omega_o)/|\omega_i-\omega_o|$. As $\omega_i-\omega_o$ is positive throughout the range covered in this work, in our coordinate
 system 
 the flow geometry
  is simplified to a pure inner cylinder rotation with the boundary condition  $\hat{u}_{\theta}(r_i) = 1$. The effect of the outer cylinder is felt as a Coriolis force in this rotating frame.

The resulting Navier-Stokes equations in the rotating frame are now
\begin{equation}
 \displaystyle\frac{\partial \hat{\bu}}{\partial \hat{t}} + \hat{\bu}\cdot\hat{\nabla}\hat{\bu} = -\hat{\nabla} \hat{p} +  \left (\displaystyle\frac{f(\eta)}{Ta}\right)^{1/2} \hat{\nabla}^2\hat{\bu} + Ro^{-1} {\boldsymbol{e}_z}\times\hat{\bu}~,
\label{eq:rotatingTC}
\end{equation}
where the Rossby number is defined as
\begin{equation}
 Ro = \displaystyle\frac{|\omega_i-\omega_o|}{2\omega_o}\displaystyle\frac{r_i}{d} = -\displaystyle\frac{|1+a|}{a}\displaystyle\frac{\eta}{2(1-\eta)} 
 \label{rossby-def}
\end{equation}
and $f(\eta)$ as
\begin{equation}
 f(\eta) = \displaystyle\frac{(1+\eta)^3}{8\eta^2}~.
\end{equation}

Equation (\ref{eq:rotatingTC}) is in analogy to the Navier-Stokes equation for a rotating Rayleigh-B\'enard system,
\begin{equation}
 \displaystyle\frac{\partial \hat{\bu}}{\partial \hat{t}} + \hat{\bu}\cdot\hat{\nabla}\hat{\bu}  = -\hat{\nabla}\hat{p} + \left (\displaystyle\frac{Pr}{Ta} \right)^{1/2} \hat{\nabla}^2\hat{\bu} + \hat{\Theta}\boldsymbol{e}_z - Ro^{-1}\boldsymbol{e}_z\times\hat{\bu}~,
\end{equation}
\noindent with the main difference that the Rossby number's sign (carried by $\omega_o$ in eq.\ (\ref{rossby-def}))
is relevant in TC flow. As long as the transport of angular momentum takes place from the inner to the outer cylinder, i.e.\ $\omega_i >\omega_o$, $Ro$ is always negative for counter-rotating cylinders and always positive for co-rotating cylinders. Therefore the sign of $Ro$ affects the flow physics, as it indicates the direction of rotation of the outer cylinder.

These equations were solved using a finite difference solver in cylindrical coordinates. The domain was taken to be periodic in the axial direction. Coordinates were distributed uniformly in the axial and azimuthal direction. In the radial direction, hyperbolic-tangent type clustering was used to cluster points near both walls.  For spatial discretization, a second order scheme was used. Time integration was performed fractionally, using a third order implicit Runge-Kutta method.  More details of the numerical method can be found in \citet{ver96}. Large scale parallelization is obtained with a combination of MPI and OpenMP directives.

In order to quantify the flow, it is useful to continue with the normalized radius $\tilde{r} = (r-r_i)/(r_o-r_i)$ and the normalized height $\tilde{z}=z/(r_o-r_i)$. As an aid to quantification, we define the time- and azimuthally-averaged velocity field as:

\begin{equation}
 \hat{\bar{\bu}}(r,z)=\langle \hat{\bu}(\theta,r,z,t) \rangle_{\theta,t}~.
\end{equation}

% \noindent from which a root mean square fluctuation velocity can be computed

%\begin{equation}
% u_i^{\prime}(r,z) = (\langle (u_i(\theta,r,z,t)- \bar{u}_i(r,z))^2\rangle_{\theta,t})^{1/2}~.
%\end{equation}

\noindent This time- and $\theta$-independent velocity is used to quantify the large time scale circulation through the wind Reynolds number:

\begin{equation}
 Re_w = \displaystyle\frac{U_wd}{\nu} ~\mbox{with}~ U_w = U \langle \hat{\bar{u}}^2_r + \hat{\bar{u}}^2_z\rangle_{r,z}^{1/2}.
\end{equation}
\noindent As mentioned in eq. (\ref{characteristic-u}), $U \propto Re_i - \eta Re_o$ scales as $Ta^{1/2}$; the non-dimensinal transverse velocity fluctuations may or may not lead to corrections of this basic scaling.  

The convective dissipation per unit mass can be calculated either from its definition as a volume average of the local dissipation rate for an incompressible fluid

\begin{equation}
 \epsilon_{u} = \nu \mean{(\nabla \bu)^2}_{V,t} 
 \label{eq:epsilonlocal}
\end{equation}

\noindent or from the global balance (EGL 2007):

\begin{equation}
 \epsilon_{u} - \epsilon_{u,0} = \displaystyle\frac{\nu^3}{d^4}\sigma^{-2}Ta(\nom-1)~,
 \label{eq:epsiloneck}
\end{equation}

\noindent where $\epsilon_{u,0}$ is the volume averaged dissipation rate in the basic, azimuthally symmetric laminar flow, cf. eq. (\ref{eps-u-0}). 

In order to validate the code we have calculated $\epsilon_u$ from both (\ref{eq:epsilonlocal}) and (\ref{eq:epsiloneck}) and checked for sufficient agreement. We define the quantity $\Delta_{\epsilon}$ measuring the relative difference

\begin{equation}
 \Delta_{\epsilon} = \displaystyle\frac{\nu^3d^{-4}\sigma^{-2}Ta(\nom-1)+\epsilon_{u,0} - \nu \mean{(\nabla \bu)^2}_{V,t}}{ \nu \mean{(\nabla \bu)^2}_{V,t} }~.
 \label{eq:epsilondifference}
\end{equation}

\noindent $\Delta_{\epsilon}$ is equal to $0$ analytically, but will deviate when calculated numerically.

The strictest requirement for numerical convergence was that the radial dependence of $\nom(r)$ had to be less than $1\%$. This is a much harder condition to satisfy than torque equality at both inner and outer cylinders, which is satisfied if the $\nom$ at both cylinders is equal. Indeed, in many cases the torques were equal to within $0.01\%$ but $\nom(r)$ was not constant within $1\%$, which meant either a higher resolution had to be chosen or that the simulation had to be run for longer time. The time-average of the energy dissipation calculated locally (equation \ref{eq:epsilonlocal}) was also checked to converge within $1\%$; see section \ref{sec:restests} for more details.

\section{Code validation}

\subsection{Validation against other codes at low Reynolds number}
\label{sec:lowreval}

First of all, the code was validated against numerical results from \citet{fas84} and \citet{pir08}. This comparison was done through $\nom$ measurements at low $Re_i$ numbers, in the range between $60$ and $80$. Only a quarter of the TC system was used, assuming a rotational symmetry of order four. The aspect ratio $\Gamma$ was taken as two, the radius ratio $\eta$ as $0.5$. These geometrical parameters are different than the ones used in the rest of the paper, but they are used here for validation. The resolution of the simulation ($N_\theta$ x $N_r$ x $N_z$) was taken as 32x64x64, the same as in \citet{pir08}. The results can be seen in Table \ref{tab:loreval}. The values show a  match up to three significant figures, or sometimes even higher.

\begin{table}
  \begin{center}
  \def~{\hphantom{0}}
  \begin{tabular}{cccc}
%    $\Rey_i$ & Torque (present study) & Torque \citep{fas84} & Torque \citep{pir08}\\
%    60 & 16.7639 & 16.7551 & 16.7551 \\
%    68 & 16.7657 & 16.7551 & 16.7551 \\
%    70 & 17.1489 & 17.1538 & 17.1542 \\
%    75 & 18.1660 & 18.1628 & 18.1634 \\
%    80 & 19.0594 & 19.0527 & 19.0536 \\
   $\Rey_i$ & $\nom$ (present study) & $\nom$ \citep{fas84} & $\nom$ \citep{pir08}\\
   60 & 1.0005 & 1.0000 & 1.0000 \\
   68 & 1.0006 & 1.0000 & 1.0000 \\
   70 & 1.0235 & 1.0237 & 1.0238 \\
   75 & 1.0835 & 1.0833 & 1.0834 \\
   80 & 1.1375 & 1.1371 & 1.1372 \\
  \end{tabular}
  \caption{$\nom$ for low $Re_i$ number and $Re_o=0$, $\Gamma=2$, $\eta=\frac{1}{2}$ and rotational symmetry of order 4.}
  \label{tab:loreval}
  \end{center}
\end{table}

For the two smallest Reynolds numbers, both references obtain the same result, while we obtain a slightly different result. This probably comes from the fact that they measure the torque directly at the inner cylinder, which we then convert to $\nom$ for comparison, while we measure $\nom$ by taking an average value of $J(r)$ and converting this to a value for the torque and thus $\nom$. The difference between these approaches is probably the origin of the discrepancy. However, as it is very small (below $0.1\%$) it is not worrying.

\subsection{Comparison with experiment}

The code was also validated by comparing responses obtained at higher Taylor numbers versus data from \cite{lew99}. This was done through the Nusselt number for pure inner cylinder rotation ($a=0$) at fixed $\eta=5/7$ and $\Gamma=2\pi$. The overlap between the simulations and experimental data can clearly be seen in the higher Taylor number range, which we have achieved with the numerics. The shift of about 5\% might be attributed to the difference in both aspect ratio and boundary conditions at the top and bottom because of the vertical confinement in the experiment. As we also have an overlap at the lower Taylor range with other numerical simulations as shown in Section \ref{sec:lowreval}, we feel sufficiently confident to proceed with our code.

\begin{figure}
  \subfloat{\label{fig:fig:Ta_Nu_Eta0714_Lews_Norm}\includegraphics[width=0.49\textwidth]{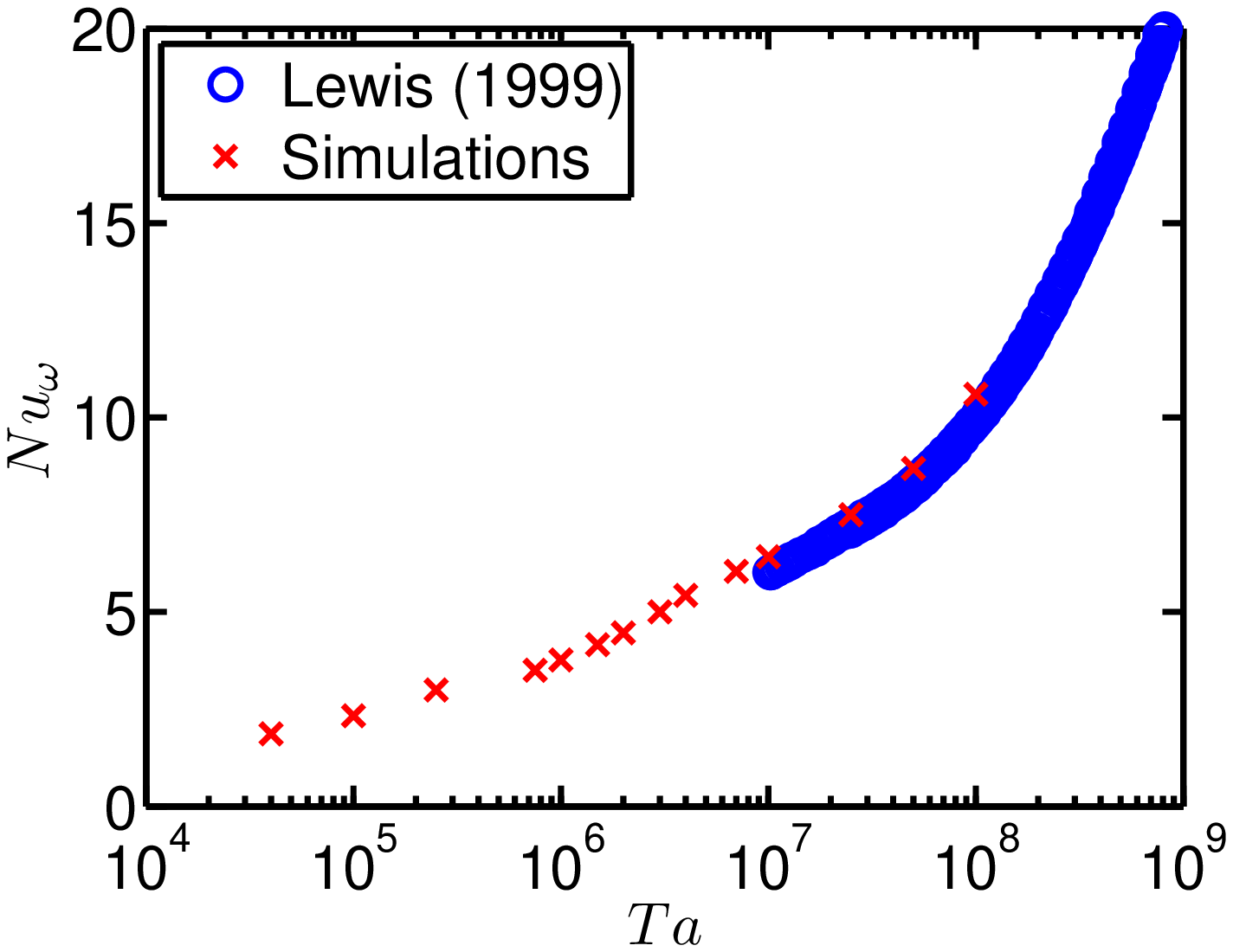}}
  \subfloat{\label{fig:fig:Ta_Nu_Eta0714_Lews_Comp}\includegraphics[width=0.49\textwidth]{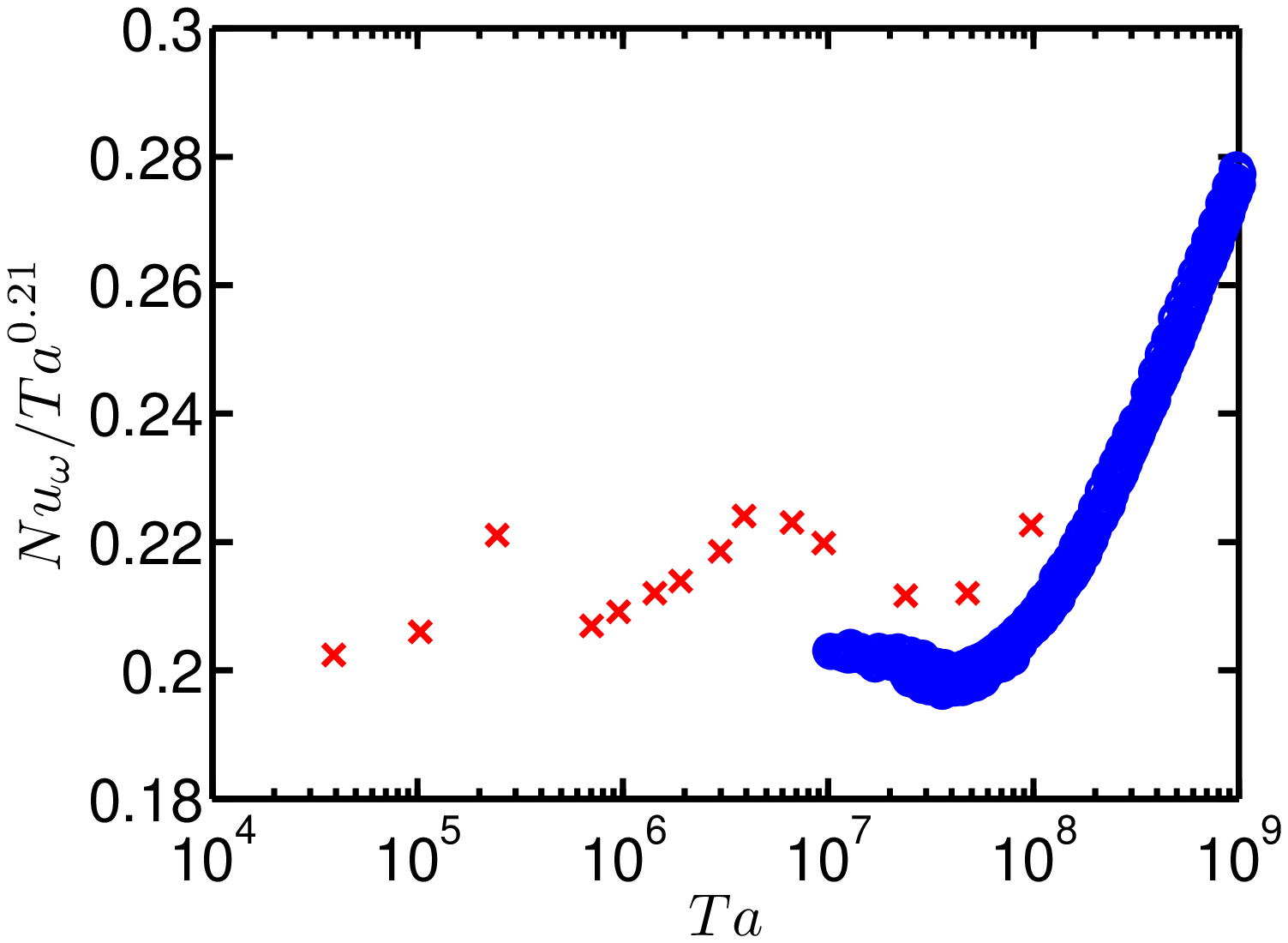}}
  \caption{$\nom$ versus $Ta$ for $\eta=5/7$. Experiments (circles) and numerics agree in shape, but there is a slight shift between the data, which we attribute to the different boundary conditions in lateral direction.}
  \label{fig:Ta_Nu_Eta0714_Lews}
\end{figure}

\subsection{Resolution tests}
\label{sec:restests}

To achieve reliable numerical results, the grid's temporal and spatial resolution have to be adequate. Sufficient temporal resolution is achieved by using an adaptive time step based on a CFL criterium. If the CFL number is too large, the code destabilizes and the velocities grow beyond all limits. As long as the CFL number is low enough to guarantee numerical stability, the results do not change with further lowering of this number.

The requirements for spatial resolution have been studied in \cite{ste10} for RB flow. There it was shown that the effect of underresolved DNS is mainly visible in the convergence of the thermal dissipation rate $\epsilon_{\theta} = \kappa \langle (\partial_i \Theta)^2 \rangle$, which in essence is the thermal Nusselt number $Nu$. The kinetic dissipation rate $\epsilon_u = \nu \langle (\partial_i u_j )^2 \rangle$ turned out to be less sensitive to underresolution. We note that even when if kinetic dissipation rate has converged within $1\%$ the simulation can still be underresolved. It is important to have grid lengths {\em in each direction} of the order of the local Kolmogorov or Batchelor lengths. 

In TC the corresponding fields to $\Theta$ and $u$ are the azimuthal velocity $u_{\theta}$ and the perpendicular components $u_r$ and $u_z$, respectively. But these are more closely related by the Navier-Stokes equations than the $u$,$\Theta$ fields in RB. Therefore we tested the grid spatial resolution at $a = 0$ by calculating $\nom$ beyond onset of Taylor vortices ($\nom > 1$) and $\Delta_{\epsilon}$ from eq.(\ref{eq:epsilondifference}), which analytically is equal to $0$, checking the (relative) difference between the transport ($Nu_{\omega}$) and the dissipation rate ($\epsilon_u$). Both were done for different grid resolutions with increasing Taylor number. For all these simulations we took $\Gamma=2\pi$, $\eta=5/7$ and $a=0$. The results are shown in Table \ref{tbl:resEta0714}.

\begin{table}
  \begin{center} 
  \def~{\hphantom{0}}
  \begin{tabular}{ccccccc}
  $\Rey_i$ & $Ta$ & $N_\theta$ x $N_r$ x $N_z$ & $\nom$ & $\Delta_{\epsilon}$ & Case \\
  $160$  & $3.90\cdot10^4$   & 128x64x64    & 1.86927 & 0.0159 & R \\
  $160$  & $3.90\cdot10^4$   & 256x128x128  & 1.85562 & 0.0074 & E \\
  $260$  & $1.03\cdot10^5$   & 160x80x80    & 2.40536 & 0.0215 & R \\
  $260$  & $1.03\cdot10^5$   & 256x128x128  & 2.40216 & 0.0322 & E \\
  $400$  & $2.44\cdot10^5$   & 100x50x50    & 2.70845 & 0.0392 & U \\
  $400$  & $2.44\cdot10^5$   & 200x100x100  & 2.76208 & 0.0102 & R \\
  $400$  & $2.44\cdot10^5$   & 300x150x150  & 2.77855 & 0.0062 & E \\
  $680$  & $7.04\cdot10^5$   & 256x128x128  & 3.49816 & 0.0147 & R \\
  $680$  & $7.04\cdot10^5$   & 384x192x192  & 3.51268 & 0.0056 & E \\
  $1120$ & $1.91\cdot10^6$   & 192x96x96    & 4.83540 & 0.0949 & U \\
  $1120$ & $1.91\cdot10^6$   & 256x128x128  & 4.46000 & 0.0174 & R \\
  $1120$ & $1.91\cdot10^6$   & 384x192x192  & 4.47765 & 0.0065 & E \\
  $1600$ & $3.90\cdot10^6$   & 300x144x144  & 5.42553 & 0.0216 & R \\
  $1600$ & $3.90\cdot10^6$   & 432x216x216  & 5.37264 & 0.0063 & E \\
  $2500$ & $9.52\cdot10^6$   & 384x192x192  & 6.42160 & 0.0168 & R \\
  $2500$ & $9.52\cdot10^6$   & 641x321x321  & 6.34068 & 0.0078 & E \\
  $3960$ & $2.39\cdot10^7$   & 641x321x321  & 7.46617 & 0.0161 & R \\
  $5600$ & $4.77\cdot10^7$   & 800x400x400  & 8.76601 & 0.0166 & R \\
  $8000$ & $9.75\cdot10^7$   & 1024x500x512 & 10.4360 & 0.0170 & R \\
 \end{tabular}
 \caption{Resolution tests for $\eta=0.714$ and $\Gamma=2\pi$. The first column displays the inner Reynolds number, the second column displays the Taylor number, the third column displays the resolution employed, the fourth column the calculated $\nom$, the fifth column the relative discrepancy $\Delta_{\epsilon}$ between the two different ways of calculating the energy dissipation, and the last column the 'case': (U)nderresolved, (R)esolved and (E)rror reference. The resolution is sufficient for all cases, as the variations are small. $\Delta_{\epsilon}$ turns out to be positive; thus the code gives for the dissipation rate a smaller value for the derivatives-squared-based definition than for the $\nom$-based balance expression.}
 \label{tbl:resEta0714}
 \end{center}
\end{table} 

\begin{figure}
 \begin{center}
  \subfloat{\label{fig:Nurspatialresabs}\includegraphics[width=0.49\textwidth,trim = 2mm 5mm 9mm 13mm, clip]{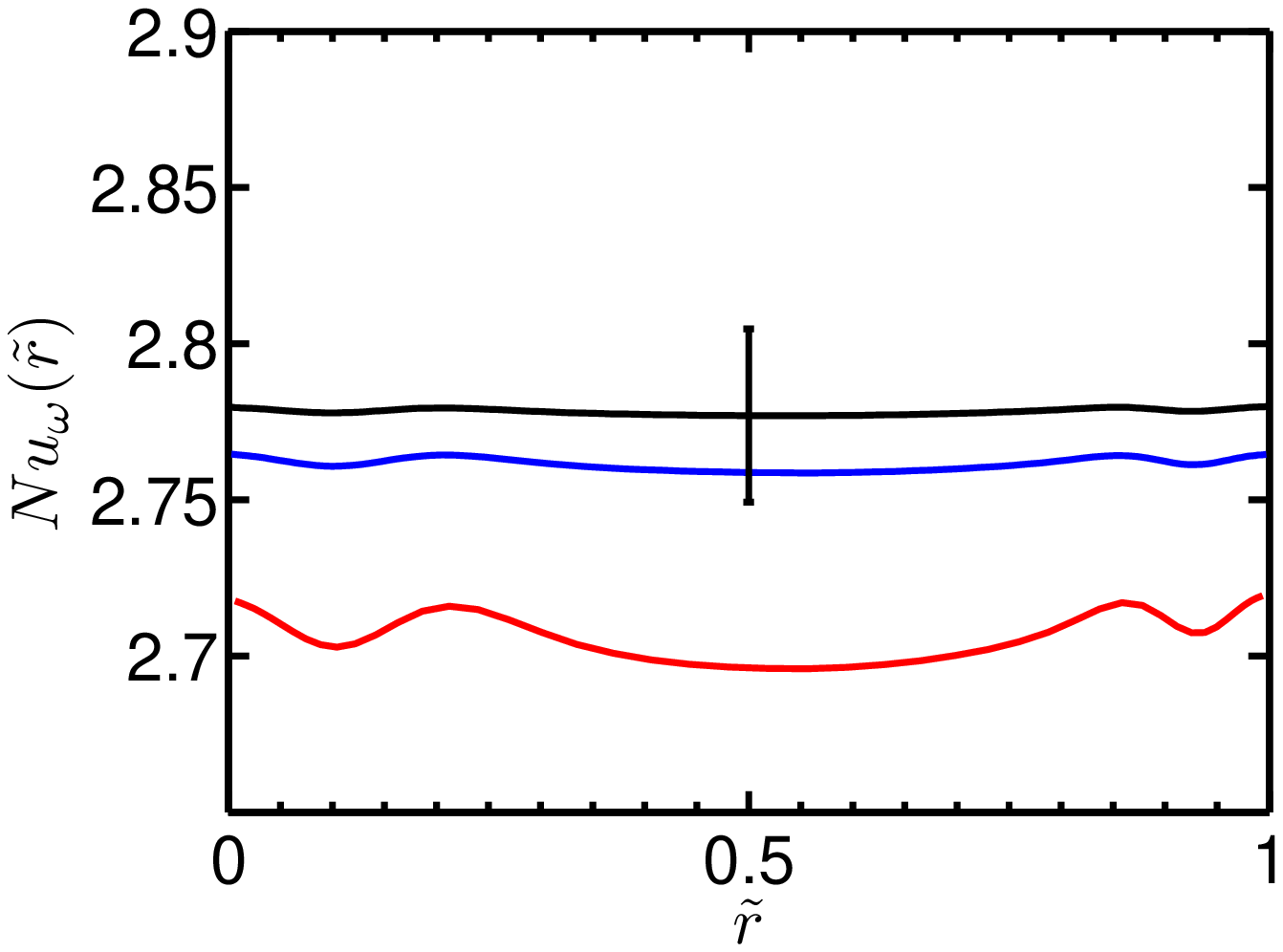}}
  \subfloat{\label{fig:Nurspatialresavg}\includegraphics[width=0.49\textwidth,trim = 0mm 5mm 9mm 13mm, clip]{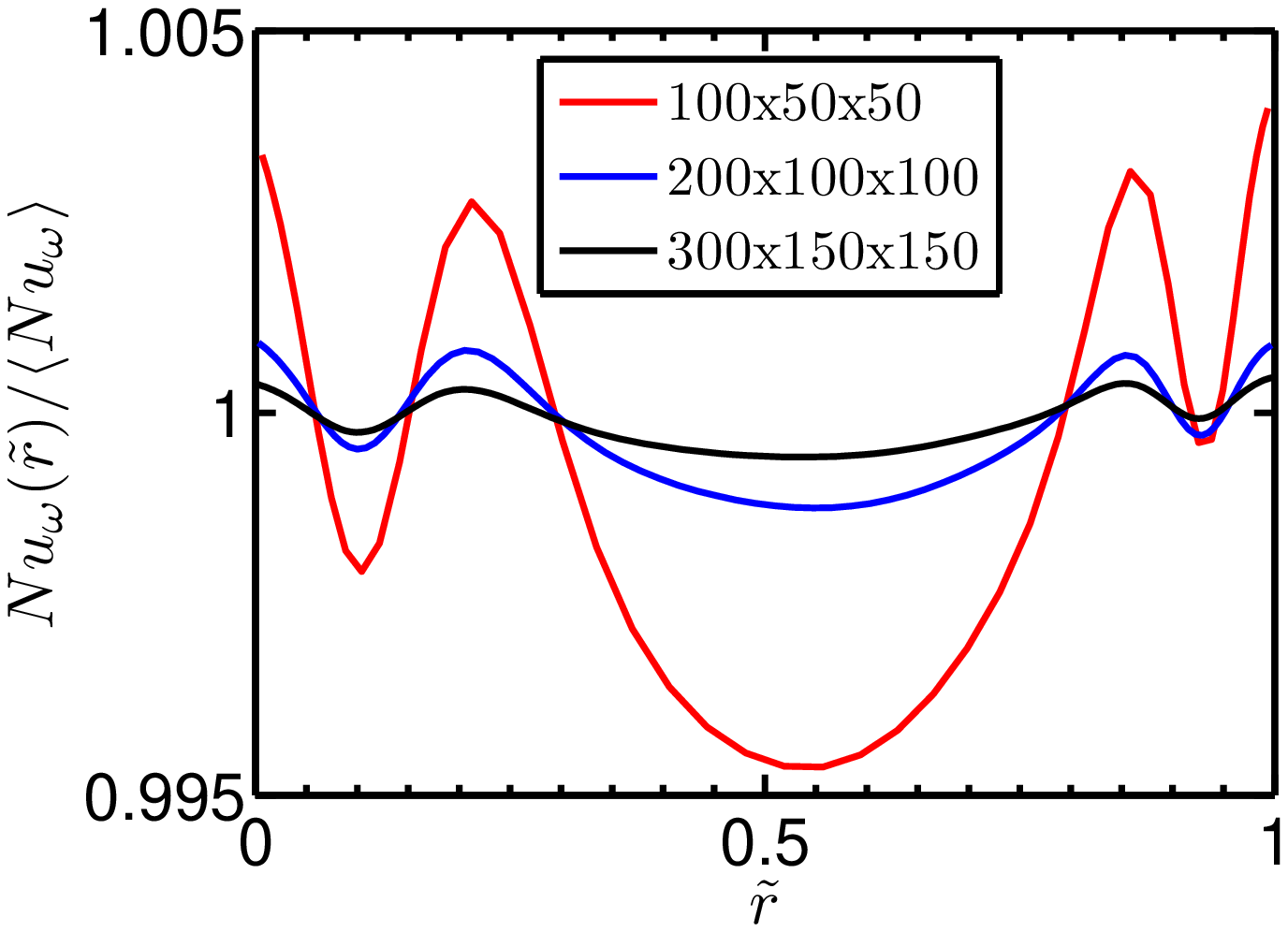}}\\
  \subfloat{\label{fig:Nurspatialresabs2M}\includegraphics[width=0.49\textwidth,trim = 2mm 5mm 9mm 13mm, clip]{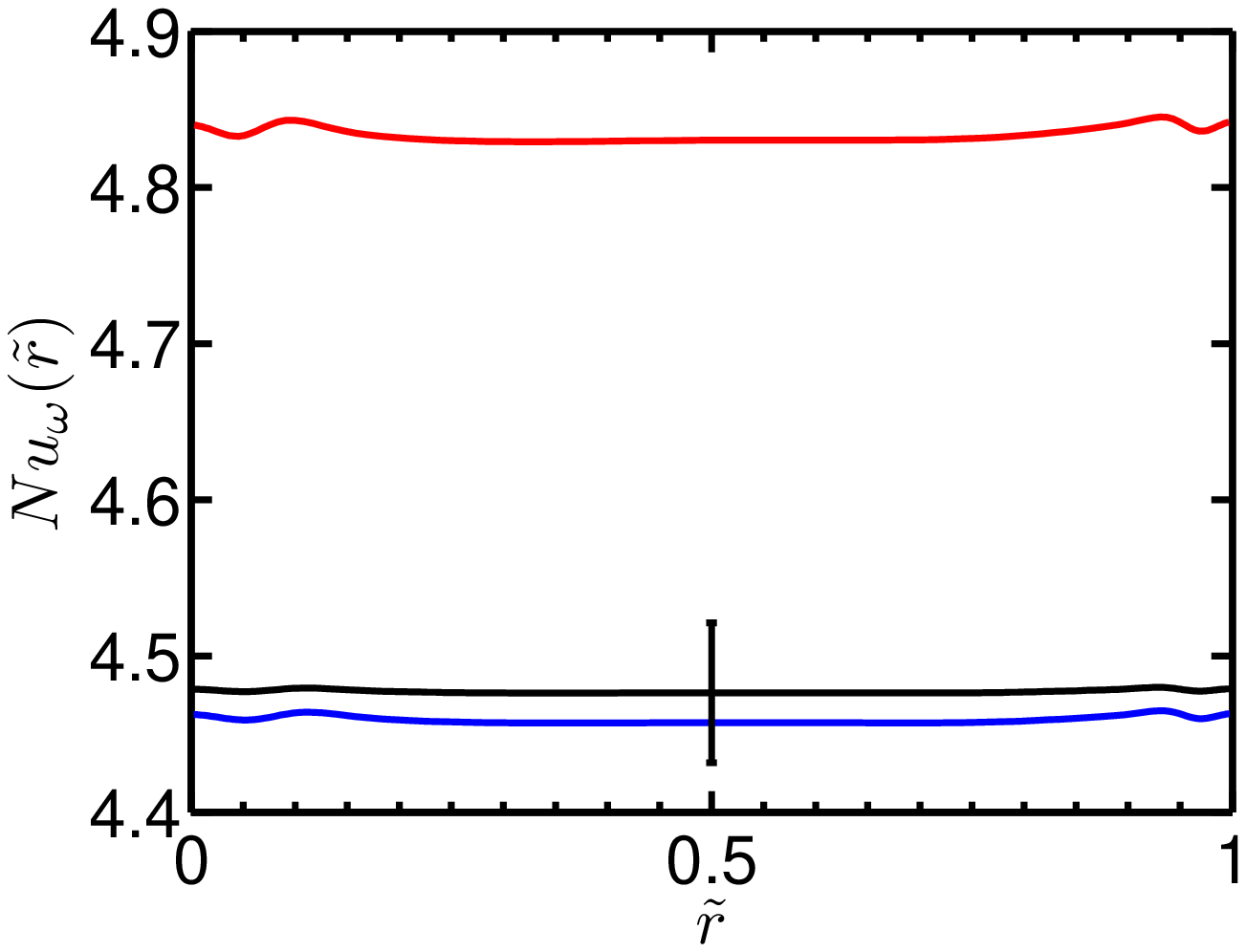}}
  \subfloat{\label{fig:Nurspatialresavg2M}\includegraphics[width=0.49\textwidth,trim = 0mm 5mm 9mm 13mm, clip]{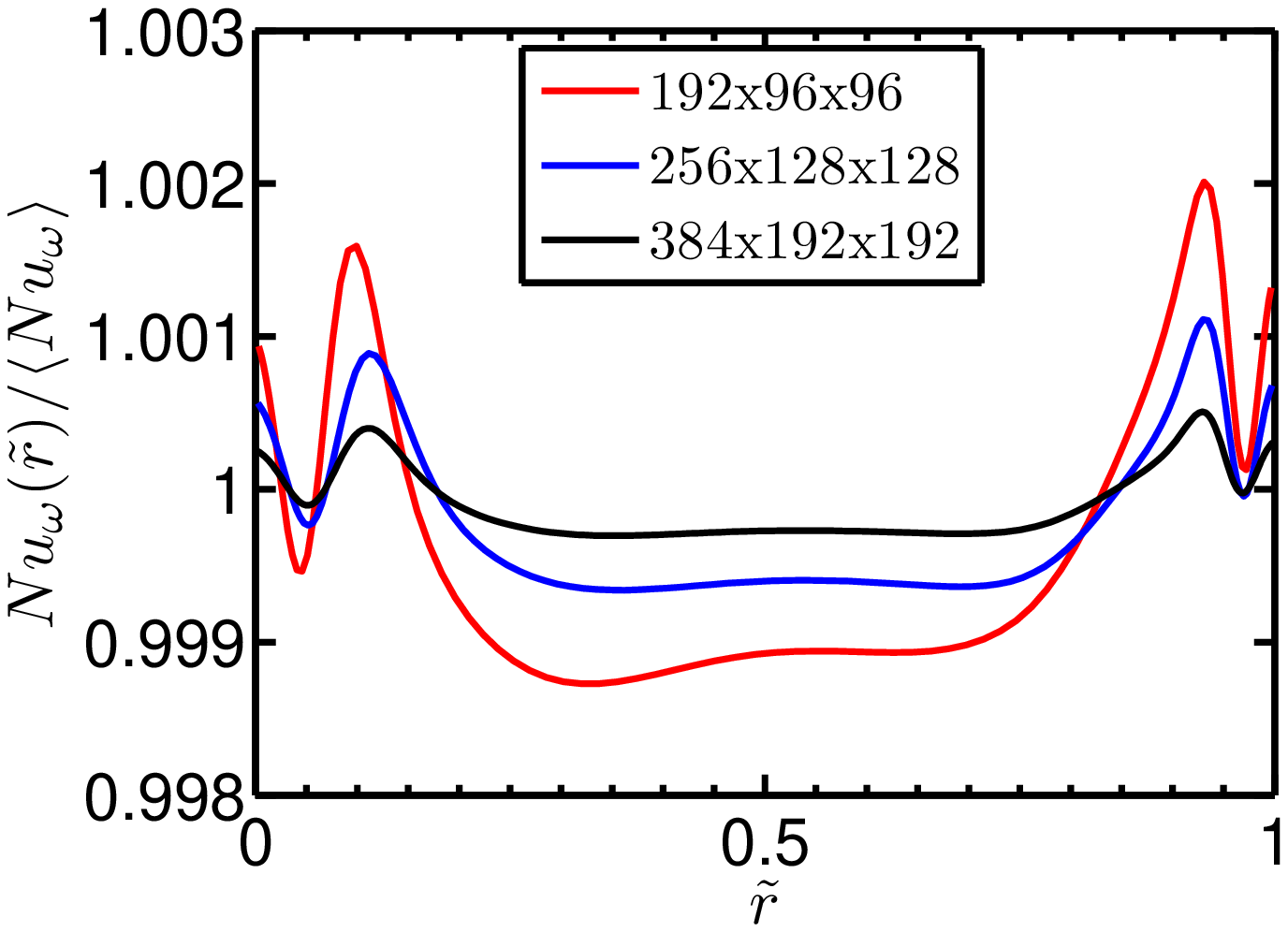}}
  \caption{Radial dependence of $\nom(\hat{r})$ for three different grid resolutions (see legends). The top two figures are for $Ta=2.44\cdot10^5$ and the bottom two are for $Ta=1.91\cdot10^6$. The figures on the left show the absolute values, an error bar indicating a $1\%$ error for reference. The resolved cases lie within this error bar. The figures on the right show the curves normalized by their average value to compare the radial fluctuations of $\nom$.}
  \label{fig:Nurspatialres}
 \end{center} 
\end{figure}

Spatial convergence required more grid points than initially expected as satisfying the torque balance alone is a necessary but not a sufficient condition for grid resolution independence. The top two graphs in figure \ref{fig:Nurspatialres} show a plot of the radial dependence of $\nom(\tilde{r})$ at $Ta=2.44\cdot 10^5$ ($Re_i=400$) for an underresolved case ($100$x$50$x$50$, $\nom=2.70845$), a reasonably resolved case ($200$x$100$x$100$, $\nom=2.76208$) and a extremely well resolved reference case ($300$x$150$x$150$, $\nom=2.77855$). $\nom$ should not be a function of the radius as mentioned previously, but numerically it does show some dependence. For the underresolved case we can see that the torque balance is satisfied very well ($0.06\%$), even if other criteria are not satisfied, E.g. the peak-to-peak variation of $\nom$ is approximately $1\%$ and the relative error in comparison to the reference case is $2.5\%$. The graph also shows that taking the value of $\nom$ at one of the cylinders gives a higher result for the transport current than taking the radial mean.

The bottom two panels in figure \ref{fig:Nurspatialres} show the same plots for $Ta=1.91\cdot 10^6$ ($Re_i=1120$) and the three cases: underresolved ($192$x$96$x$96$, $\nom=4.8354$), reasonably resolved  ($256$x$128$x$128$, $\nom=4.4600$) and extremely well resolved reference ($384$x$192$x$192$, $\nom=4.4776$). For this Taylor number, the underresolved case shows a smaller deviation of $\nom$ from the mean value and the torque difference in comparison to the lower Taylor number case. However, the discrepancy in the mean value of $\nom$ between the underresolved and the reference case is much larger ($7.9\%$). For this $Ta$ the value of $\nom$ at the cylinder walls is larger than the average value of $\nom$, too.

If we look at the $\nom(r)$ profiles at given Taylor numbers $Ta$, they show similar radial dependences, whose magnitudes decrease with increasing resolution. However, the shape of this dependence is different for both Taylor numbers.  The peaks of $\nom(r)$ are located close to the boundaries, indicating that they are probably produced by some boundary layer features and are not a systematic bias of our solver.

According to EGL 2007, dissipation should be equal, irrespective of the way in which it is calculated, directly from its definition or indirectly via the Nusselt number balance cf. eq. \ref{eq:epsiloneck}. \cite{ste10} also mentioned the importance of the corresponding equality in RB flow, especially for low values of $Pr$, as a way to ensure that the flow field is sufficiently resolved and that the gradients are captured adequately. Underresolving a flow in Taylor-Couette will result in a value of $\Delta_\epsilon$ which is too large in magnitude. This can be seen in Table \ref{tbl:resEta0714} for the underresolved simulations at $Ta=2.44\cdot 10^5$ and $Ta=1.91\cdot 10^6$. But as was elaborated at the beginning of this subsection, we should consider the convergence of $\Delta_\epsilon$ towards zero (becoming smaller than any chosen threshold) as a neccesary but not as a necessarily sufficient way to ensure grid convergence.

Besides $\Delta_{\epsilon}$ being small enough, at least also $Nu_{\omega}(r)$ must be converged. 

\subsection{Dependence on initial conditions}
\label{sec:vorstr}

For the lower Taylor numbers $Ta$ the flow was started from rest ($\bu=0$). The Taylor vortices start forming within a couple of revolutions. After enough time, a steady state with three pairs of Taylor vortices was reached. However, the simulations can also be started from non-resting conditions. Depending on these conditions a different number of vortex pairs can arise. This has a strong influence on both the global and the local response of the system. Once the vortices have formed, they are persistent in time during the simulation. Therefore, it is possible to bias a simulation through the initial conditions to have a higher or lower amount of vortex pairs, which results in a different response.

\begin{figure}
 \begin{center}
  \includegraphics[width=0.6\textwidth]{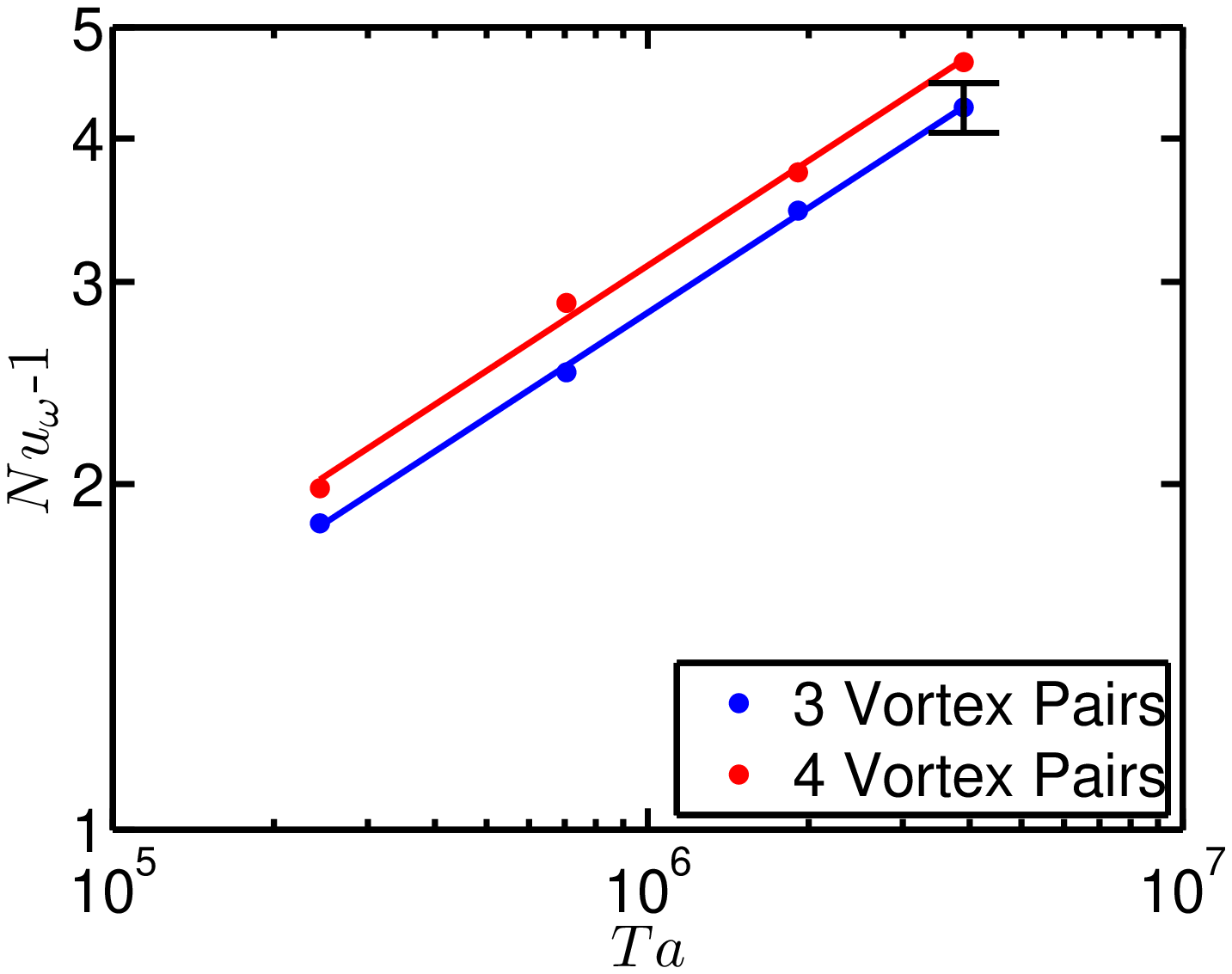}
  \caption{Dependence of $\nom-1$ vs $Ta$ on the number of vortices. Although the values of $\nom$ are different, the scaling behavior with $Ta$ is the same. The error bar indicates a 5\% difference.}
  \label{fig:vortdif}
 \end{center}
\end{figure}

Although the importance of these coherent structures gets smaller and smaller with increasing $Ta$ (Section \ref{sec:cohst}), at lower $Ta$ the number of vortex pairs must be fixed to determine the response. The three vortex pair state has been chosen to be the base state, to keep the aspect ratio of the vortices as close to 1 as possible ($2N_{pairs}/\Gamma=0.96$). The effect of having 3 or 4 vortex pairs on the response for selected $Ta$ is shown in figure \ref{fig:vortdif}. It is important to note that this effect is different from the effect caused by neutral surface stabilization, which occurs when the vortices cannot penetrate the whole flow, and the number of vortices is changed as a result. That will be featured in more detail in Section \ref{sec:nlpushing}. 

\section{Global response}
\label{sec:global}

In this section, the global response of the Taylor-Couette system is shown across the parameter space. 
First the onset of Taylor vortices is analysed. Then the scaling laws are revealed for pure inner cylinder 
rotation. Finally, the effect of the outer cylinder rotation on the scaling laws is investigated and an optimum of $\nom$
as a function of $a$ for given $Ta$ is found, as has been reported for large $Ta$ from experiment, cf. \cite{gil12}.

\subsection{Transport $\nom$ and wind $Re_w$ for pure inner cylinder rotation}

The global response of the system is quantified through $\nom$ and $\rew$. These two quantities measure two different flow responses. $\nom$ quantifies the transport of angular velocity and $\rew$ the ``wind'', i. e. the additional velocity on top of the azimuthal flow. For the purely laminar-azimuthal flow $\nom=1$ by definition, and $\rew=0$ as this laminar flow only has an azimuthal velocity component. 

\begin{figure}
 \begin{center}
  \subfloat{\label{fig:TaNuEta0714a0T}\includegraphics[width=0.49\textwidth,trim = 4mm 0mm 9mm 0mm, clip]{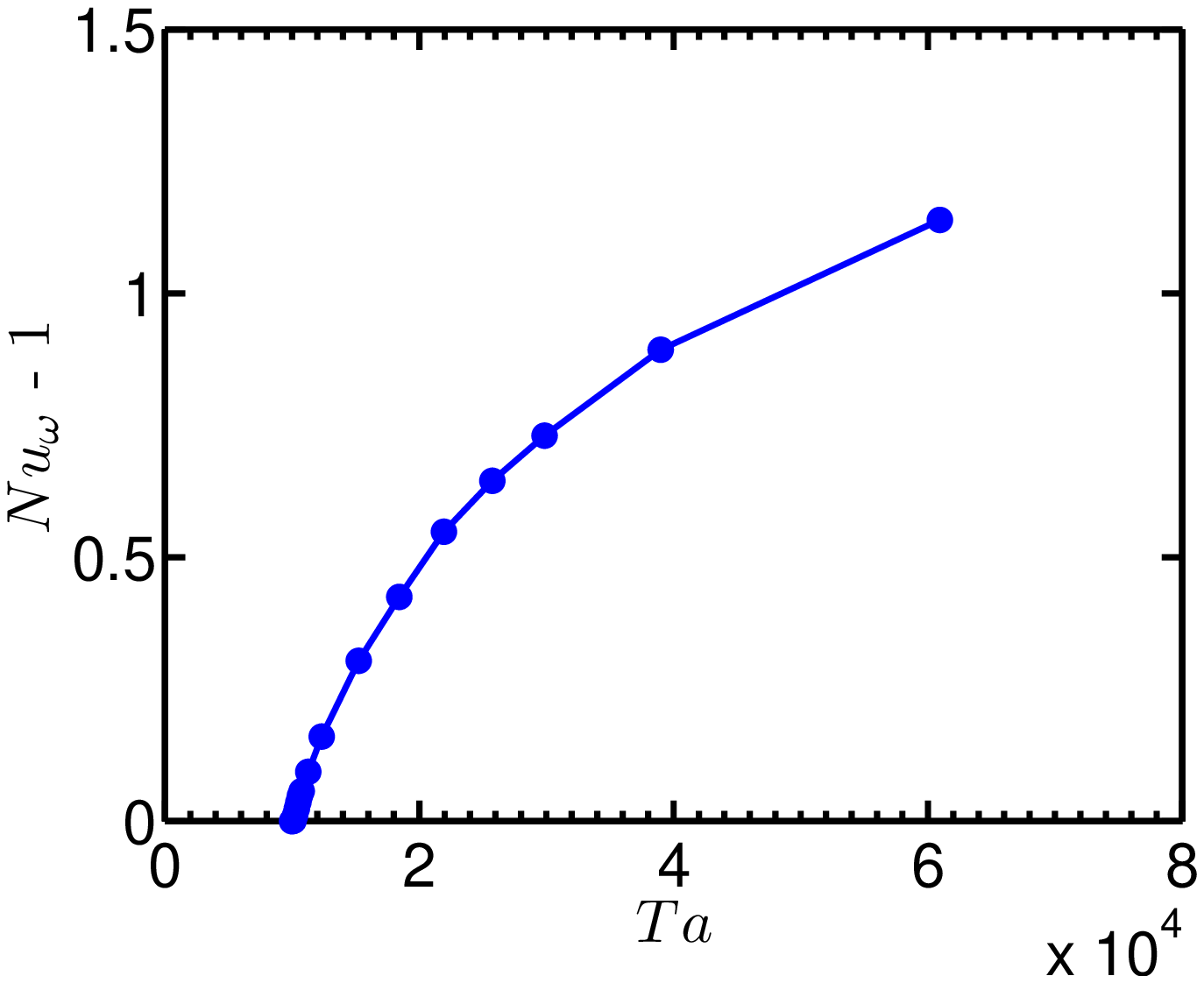}}
  \subfloat{\label{fig:TaRewEta0714a0T}\includegraphics[width=0.49\textwidth,trim = 11mm 3mm 15mm 13mm, clip]{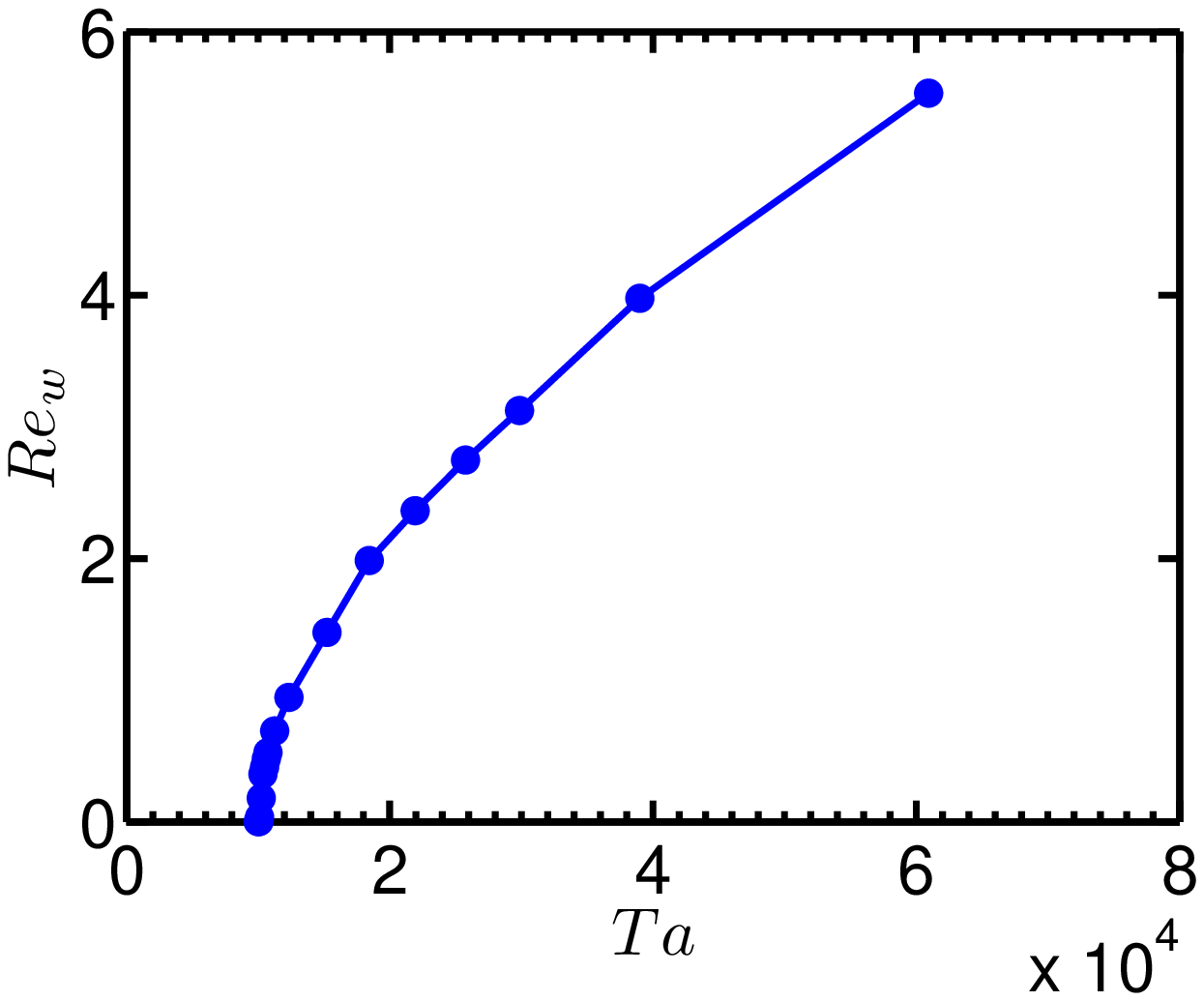}} 
  \caption{System responses $\nom - 1$ (left) and $\rew$ (right) as a function of $Ta$ for pure inner cylinder rotation near the onset of Taylor vortices. Onset in the present DNS occurs at $Ta\approx1020$.}
  \label{fig:TaNuRewEta0714a0Tran}
 \end{center} 
\end{figure}

First of all we analyse how the onset of Taylor vortices is reflected in the global response quantities $\nom$ and $\rew$. $\nom-1$ is is the additional transport of angular velocity on top of the laminar transport and the wind is the fluid motion on top of the purely laminar-azimuthal flow. Figure \ref{fig:TaNuRewEta0714a0Tran} shows the numerically calculated $\nom-1$ and $\rew$ as functions of $Ta$ close to onset of the Taylor-vortex state. The critical Taylor number ($Ta_c$) for the onset of Taylor vortices is calculated to be around $1020$ for our value of $\eta$. This DNS value can be compared with $Ta_c$ as obtained from the analytical approximation of \cite{ess96}, which is $1038$ for the present $\eta$. The agreement is within $1.6\%$. Later on we shall use these analytically calculated onset Taylor numbers.   

\begin{figure}
 \begin{center}
  \subfloat{\label{fig:TaNuEta0714a0}\includegraphics[width=0.45\textwidth]{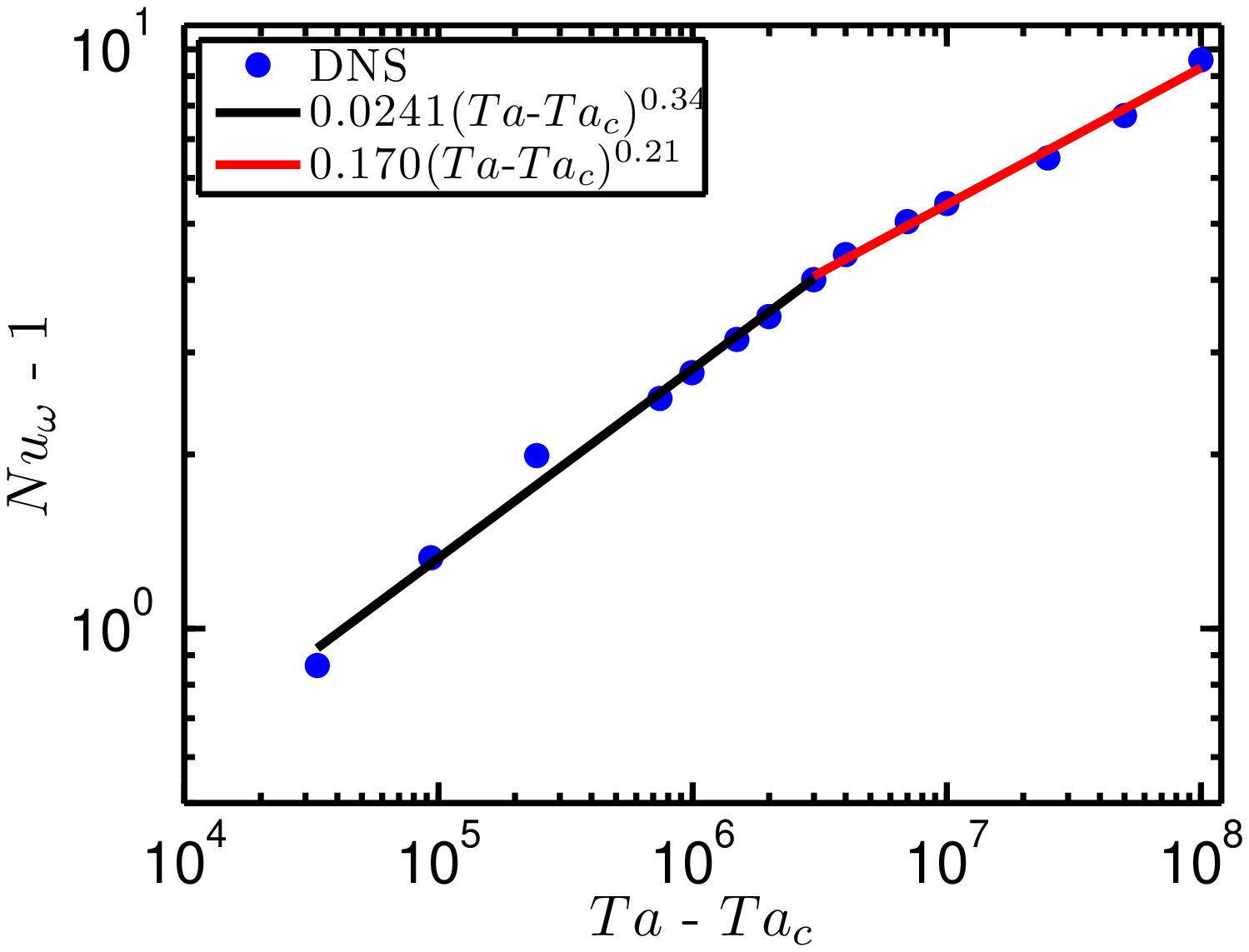}}
  \subfloat{\label{fig:TaRewEta0714a0}\includegraphics[width=0.45\textwidth]{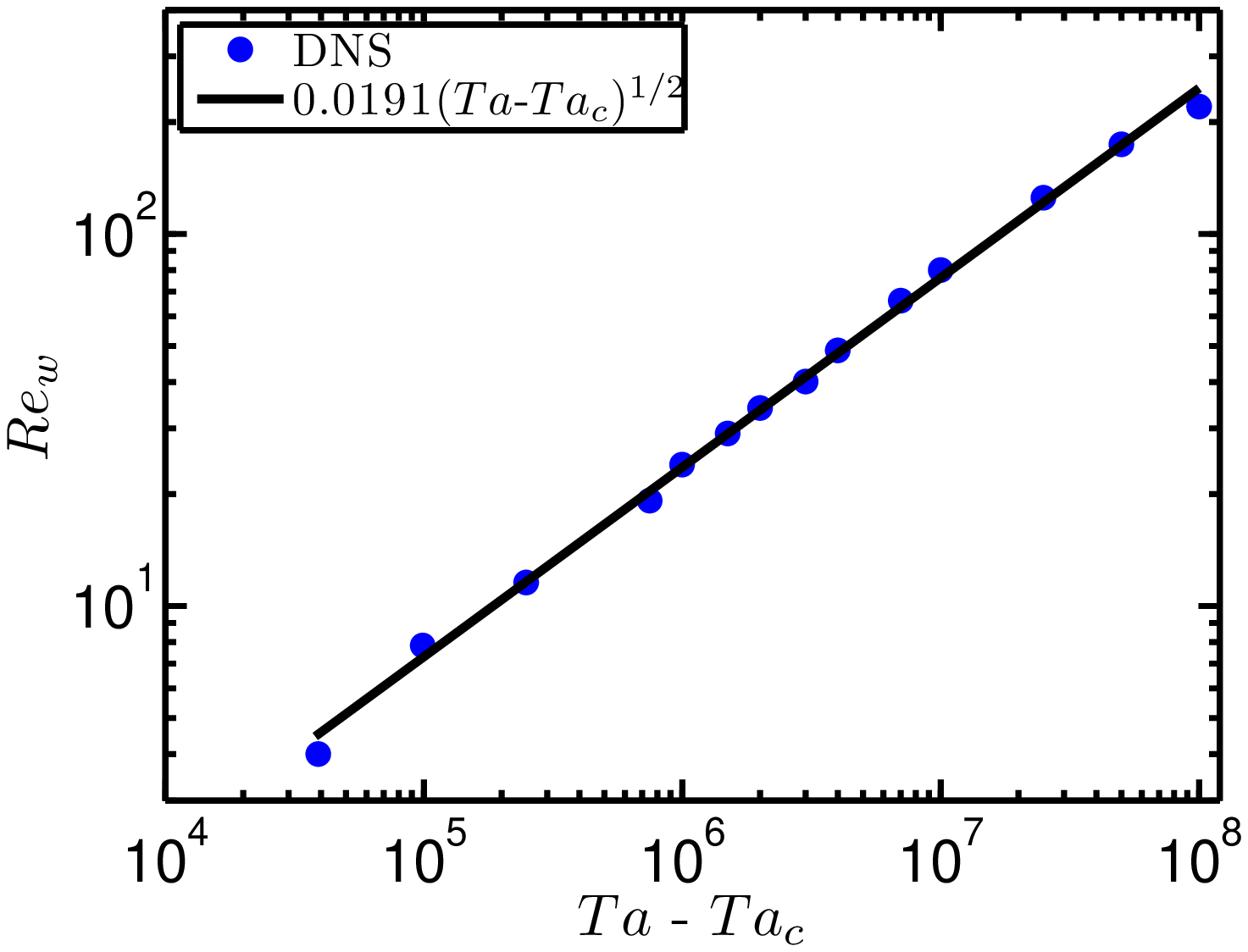}}   \\             
  \subfloat{\label{fig:TaNuEta0714a0C}\includegraphics[width=0.45\textwidth]{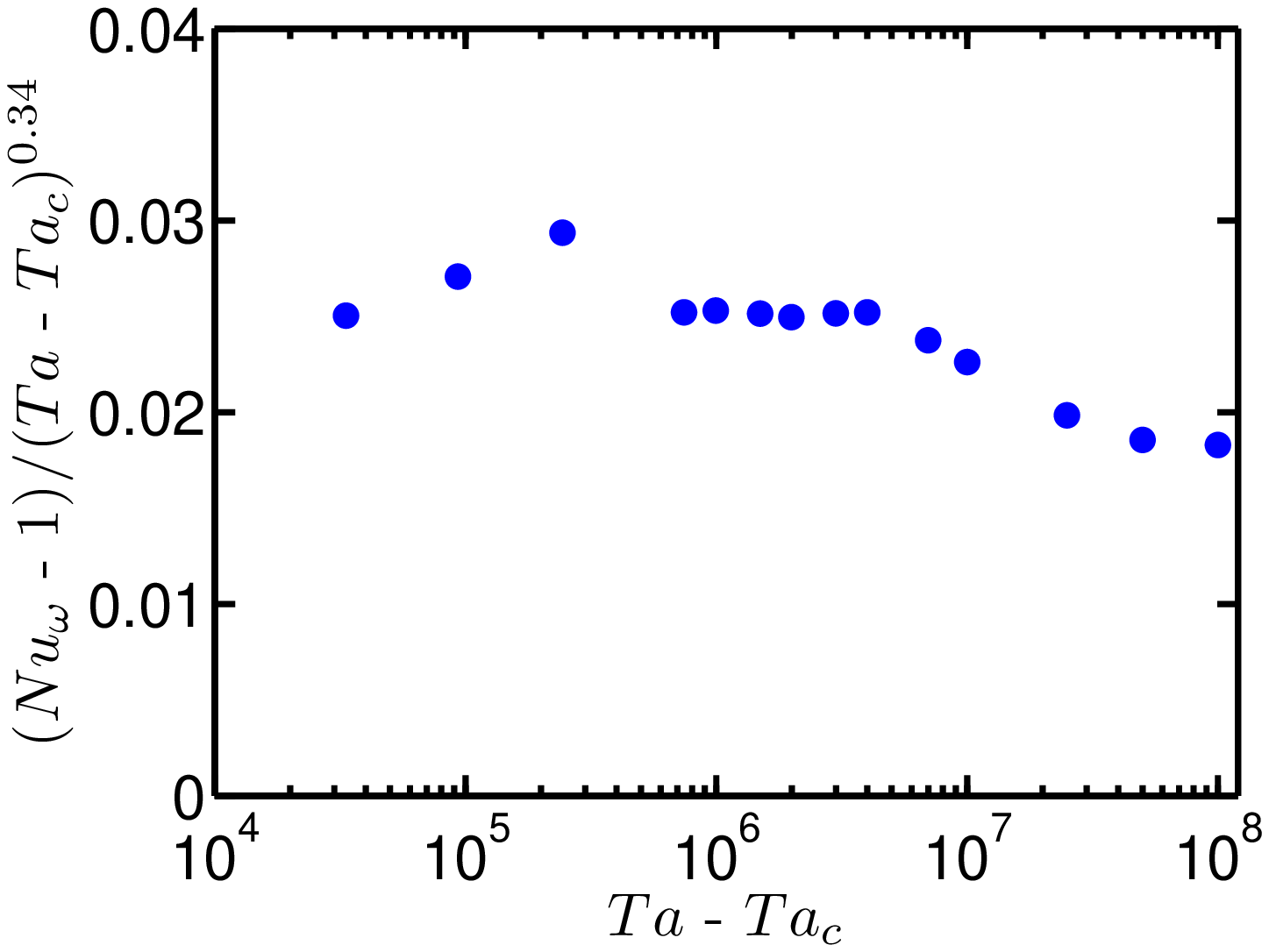}}
  \subfloat{\label{fig:TaRewEta0714a0C}\includegraphics[width=0.45\textwidth]{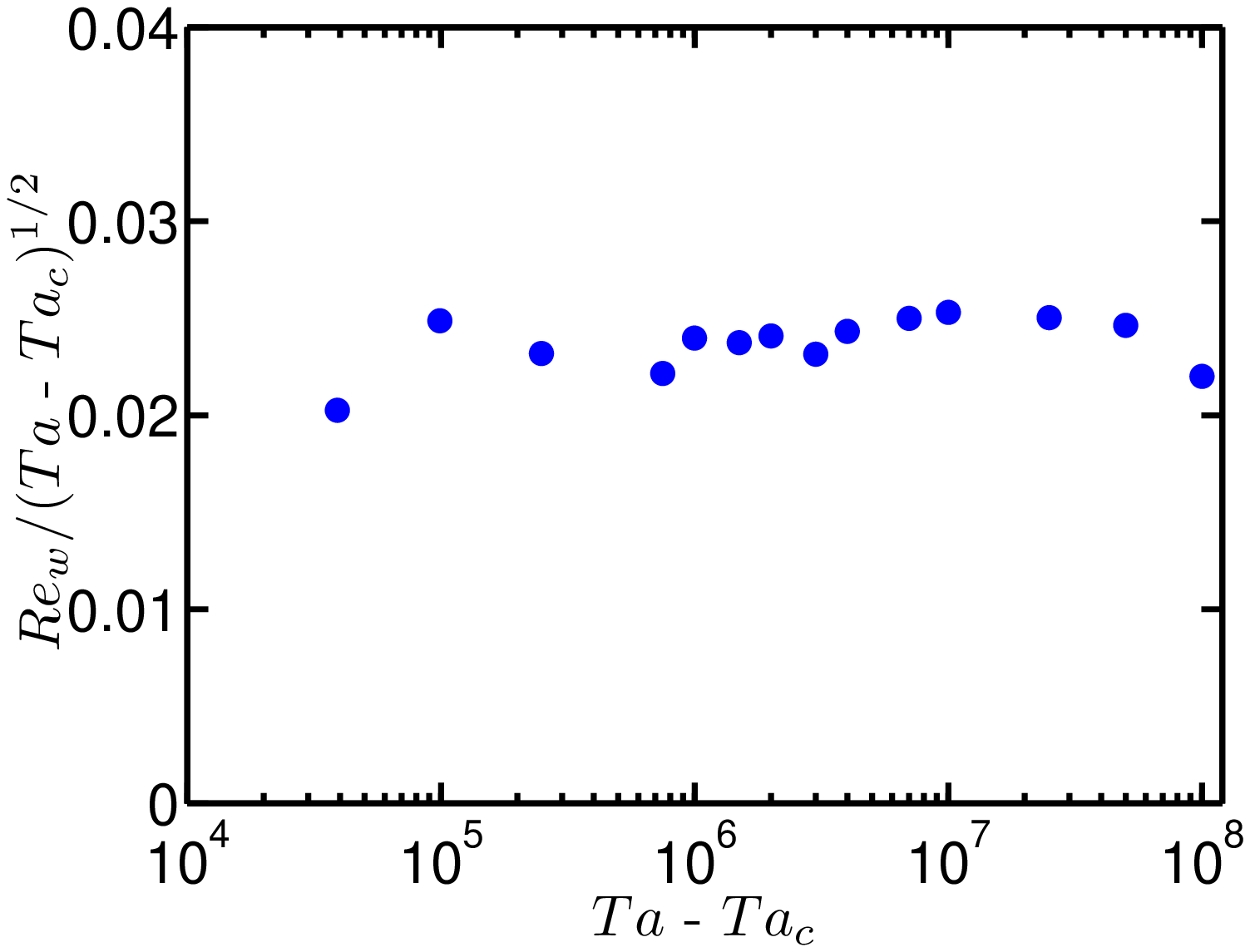}}   \\             
  \subfloat{\label{fig:RewNuEta0714a0}\includegraphics[width=0.45\textwidth]{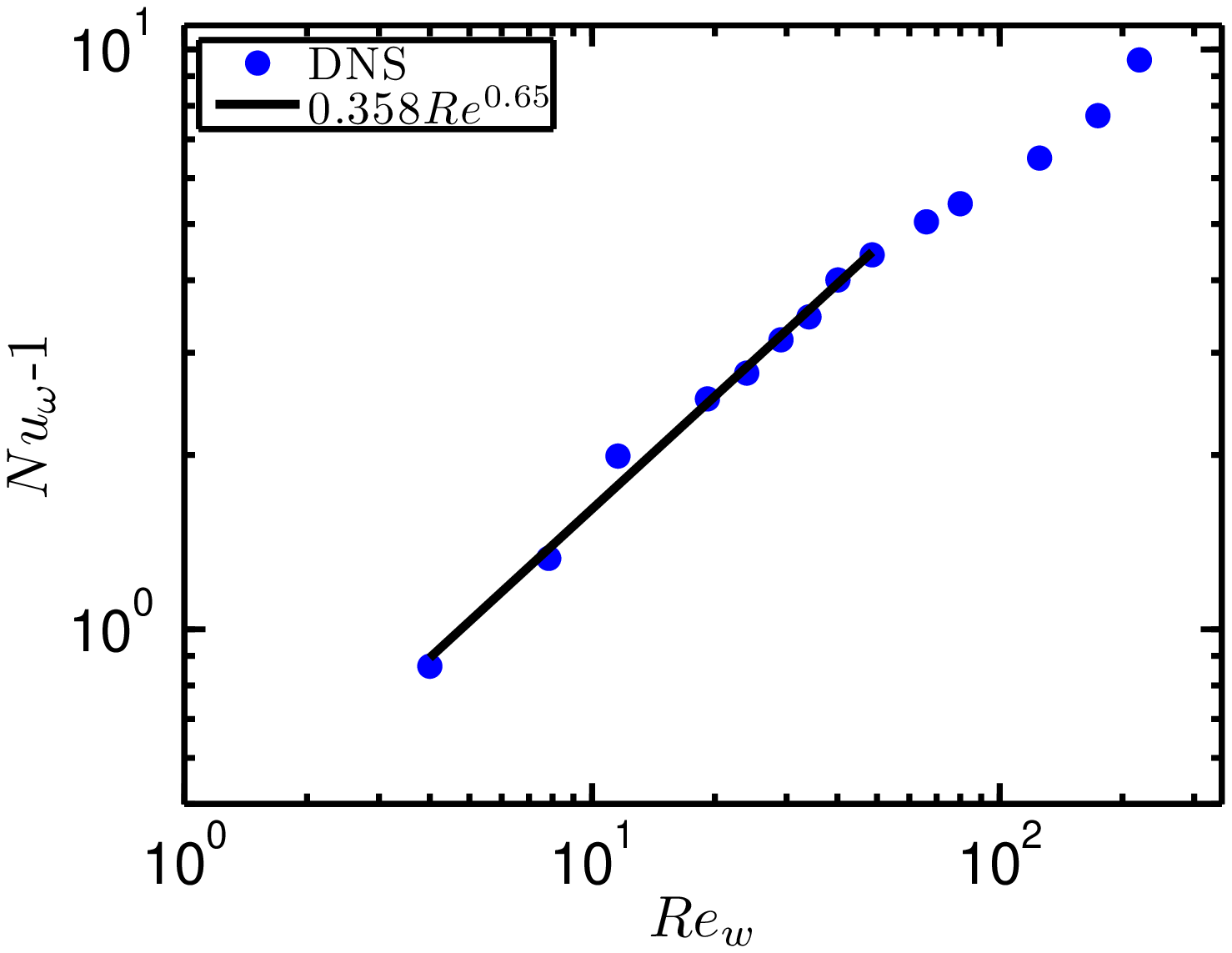}}
  \subfloat{\label{fig:Nuvstime}\includegraphics[width=0.45\textwidth]{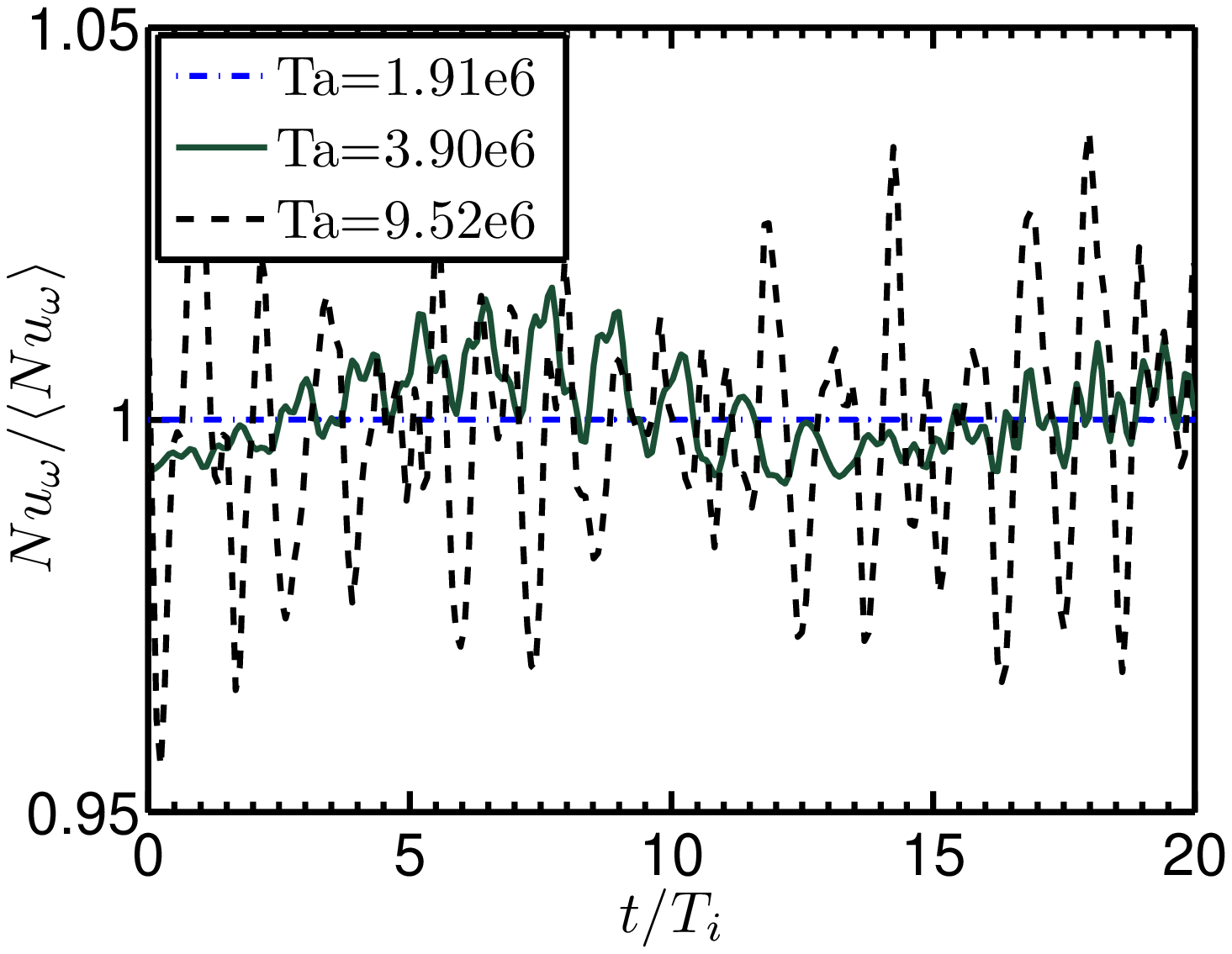}}
  \caption{The top two figures show the system response as a function of $Ta$ with best fit lines for pure inner cylinder rotation, $\nom-1$ on the left and $Re_w$ on the right. The middle plots show the date of the two top plots compensated by the scaling factor to test the quality of the scaling behavior. The bottom left figure presents the  functional relation between the two responses $\nom - 1$ and $\rew$, and the bottom right figure shows the temporal dependence $Nu_\omega(t)$ for three  different $Ta$. The time dependence can be seen to set in between $Ta=1.91\cdot10^6$ and $Ta=3.90\cdot10^6$, just where we see the change in the effective scaling in the two left figures. The analytical approximation of
  \cite{ess96} is used, i.e.\ $Ta_c = 1038$ for the present $\eta$.}
  \label{fig:TaNuRewEta0714a0}
 \end{center} 
\end{figure}

After the Taylor vortices have appeared in the system, they are the dominating feature of the flow for several decades of $Ta$. The top two panels in figure \ref{fig:TaNuRewEta0714a0} show the response of the system with increasing Taylor number in the case of resting outer cylinder and pure inner cylinder rotation. We plot $\nom$ vs. $Ta$-$Ta_c$ rather than vs. $Ta$ as it then shows a better scaling for the points at low $Ta$. 

There seems to be a clear change in the scaling law of $\nom$ versus $Ta$, but not so in the scaling law for the wind $\rew$ as a function of $Ta$. This change occurs between $Ta=1.91\cdot10^6$ and $Ta=3.90\cdot10^6$ and has been seen in other numerical simulations too \citep{cou96}. $Nu_\omega - 1$ scales as $(Ta - Ta_c)^{0.34}$ for $Ta<2 \times 10^6$ and as $Nu_\omega - 1 \sim (Ta - Ta_c)^{0.21}$ for $Ta>2\times 10^6$. 
We attribute the change in scaling to the changes in the coherent flow structures  that affect the angular velocity transport but not the global wind amplitude. As will be discussed later in detail, we expect coherent flow structures to lose importance for increasing $Ta$, see Section \ref{sec:cohst}.  We note already here that although the loss of influence of coherent structures in RB flow (for $Pr=1$) and in TC flow sets in at similar values of $Ra$ and $Ta$ respectively, i.e.\ around $10^7$ \citep{sug07}, there is a large difference in the shear Reynolds numbers $Re_s$ of the boundary layers in these two systems. This will be discussed in the following Section.  

$\rew$ measures the amplitude (strength) of the Taylor vortices, which persist at long time scales. The nondimensional characteristic speed for these vortices remains approximately constant with $Ta$, viz. about $5$-$6\%$ of the inner cylinder velocity $u_i$, throughout the whole Taylor number range considered, and that is why we see a direct scaling law of $\rew \sim U \sim(Ta-Ta_c)^{\frac{1}{2}}$, cf. eq.(\ref{characteristic-u}).

The mutual functional dependence of the two responses $\nom$ and $\rew$ is presented in the bottom left panel of figure \ref{fig:TaNuRewEta0714a0}. As expected from figures \ref{fig:TaNuEta0714a0}-\ref{fig:TaRewEta0714a0}, the relation between $\rew$ and $\nom$ also shows the change in the scaling. We interpret this as follows. Before the change, mainly the Taylor vortices are responsible for the additional transport. Beyond the change, some short time scale fluctuations appear, indicating other structures, which disrupt the flow and finally 
become its dominating features, while the Taylor vortices lose importance. In order to see these time scales, we show the temporal dependence of $Nu_\omega(t)$ in the bottom right panel of figure \ref{fig:TaNuRewEta0714a0}. The Nusselt number shows almost no time dependence for lower Taylor numbers. But it shows two different time scales at $Ta=3.90\cdot10^6$. The short time scale gains much more importance for the highest Taylor numbers, causing fluctuations of about $10\%$. 

%Figure \ref{fig:Nutransitionvstime} shows $Nu(t)$ for two $Ta$ numbers between which the transition to time dependence takes place. For $Ta=2.75\cdot10^6$ the global response is still stable to perturbations and the time dependence dissapears with time. However, for $Ta=3\cdot10^6$, these perturbations do not decay, and a small constant oscillation remains. This oscillation becomes larger and more chaotic with increasing $Ta$ as seen previously in Figure \ref{fig:TaNuRewEta0714a0}.

% \begin{figure}
%  \begin{center}
%   \subfloat{\label{fig:Nutransition1}\includegraphics[width=0.45\textwidth]{EPSfigs/Nu_Nuavg_Eta0714_a0_Tay2p75e6}}
%   \subfloat{\label{fig:Nutransition2}\includegraphics[width=0.45\textwidth]{EPSfigs/Nu_Nuavg_Eta0714_a0_Tay3e6}}
%   \caption{$Nu_\omega (t)$ behaviour with time at $Ta=2.75\cdot10^6$ (left) and $Ta=3\cdot10^6$ (right). Both simulations are initialized with a small perturbation with imposed on the result of a finalized simulation. $\nom (t)$ can be seen to be stable to perturbations at the lower $Ta$. At the higher $Ta$, a small oscillatory behaviour remains. At $Ta = 3.90 \cdot 10^6$ and beyond the time dependence is strongly chaotic, cf. Figure \ref{fig:TaNuRewEta0714a0}, bottom right panel. }
%   \label{fig:Nutransitionvstime}
%  \end{center} 
% \end{figure}

\subsection{The effect of outer cylinder rotation and optimal transport}

In this section the effect of the outer cylinder rotation on the global responses $\nom$ and $\rew$ will be studied. This effect is felt by the flow as a Coriolis force (Equation $\ref{eq:rotatingTC}$), so plots in this section will be done versus $Ro^{-1} \propto - a / |1 + a|$ with $a = - \omega_o / \omega_i$. 
\begin{equation}
Ro^{-1} = \frac{2 d \omega_o}{U} = - 2 \frac{1-\eta}{\eta} \frac{a}{|1+a|} = 2 \frac{1-\eta}{\eta} \frac{\omega_o}{|\omega_i - \omega_o|}~. 
\label{ro-invers}
\end{equation}
\noindent The inverse Rossby number $Ro^{-1}$ runs from $2 \eta / (1 + \eta)$ to $-1$ if $\omega_o$ runs from $\eta^2 \omega_i$, the inviscid Rayleigh-line in the first quadrant of the $(Re_o,Re_i)$-plane, to $-\infty$. It is useful to note that for a given $Ta$ constant $Ro^{-1}$ means constant $a$ and vice versa. Thus $a$ seems the preferable notion as it is more direct; $Ro^{-1}$ will be used only when it provides a clear advantage in visualization or later in the paper when we will trace back the occurrence of the maximum to the Navier-Stokes equation. 

\begin{figure}
 \begin{center}
  \includegraphics[width=0.8\textwidth]{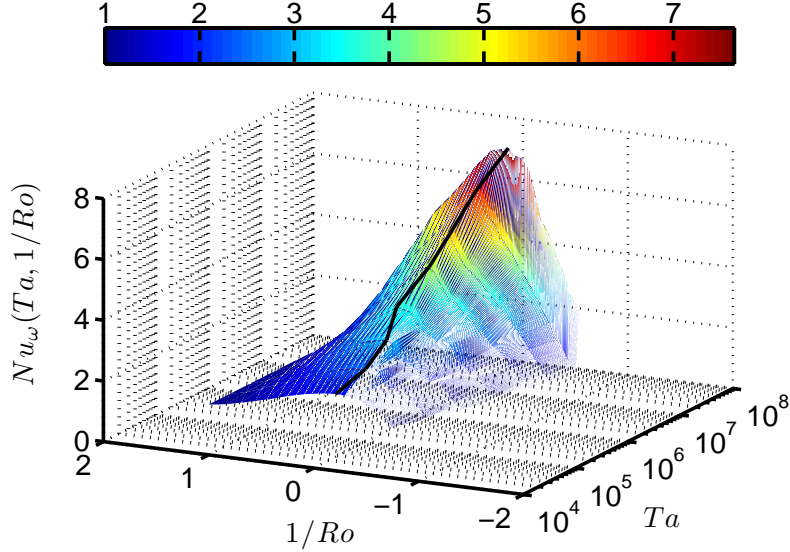}
  \caption{$\nom(Ta,1/Ro)$. The black line shows the position of $Ro^{-1}_{opt}$, the value of $Ro^{-1}$ for which $\nom$ is maximum at a given Taylor number}
  \label{fig:Nu_a_3D}
 \end{center} 
\end{figure}

\begin{figure}
 \begin{center}
  \subfloat{\label{fig:TaNuEta0714ascorr}\includegraphics[width=0.49\textwidth,trim = 0mm 0mm 0mm 0mm, clip]{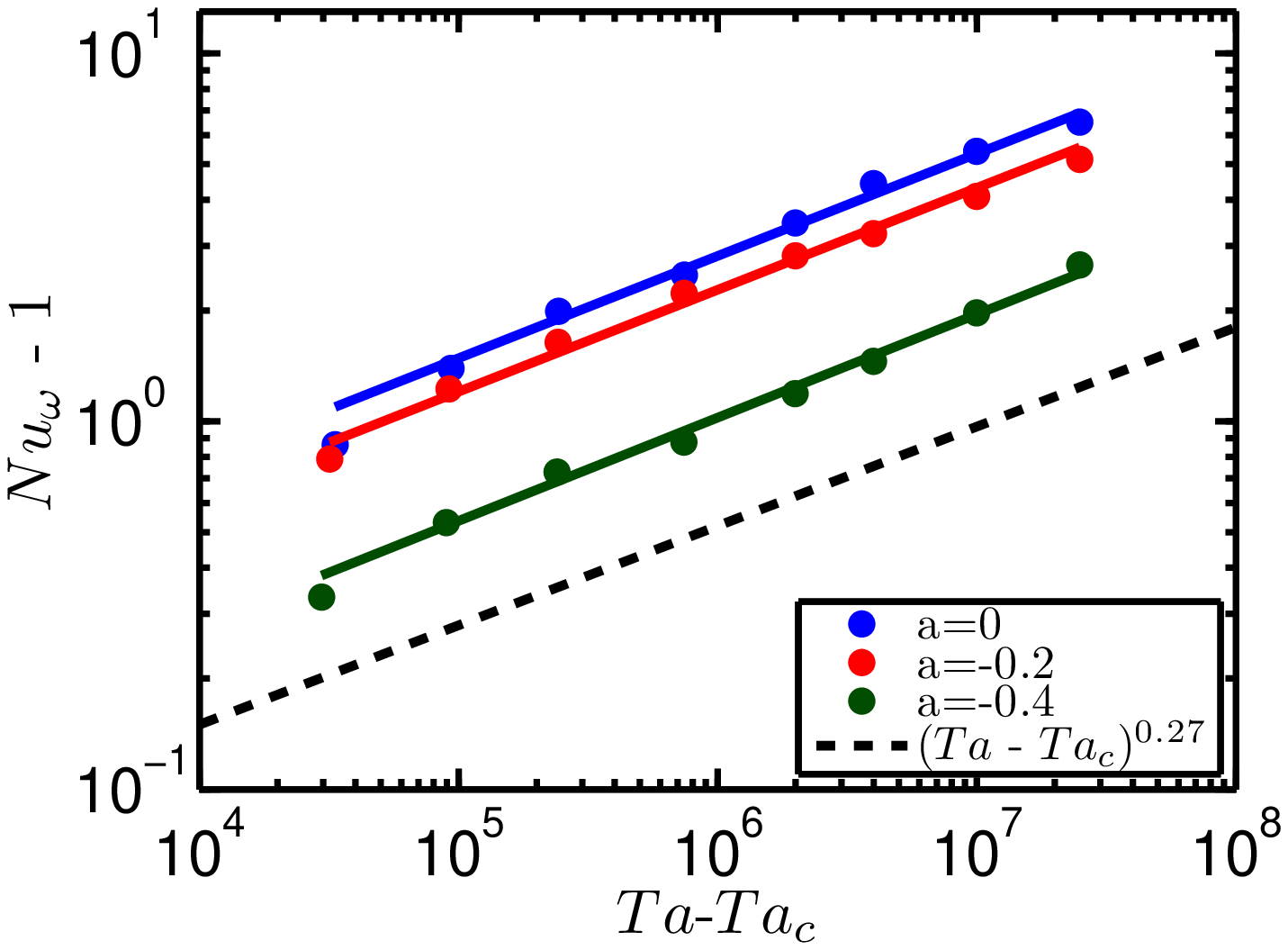}}
  \subfloat{\label{fig:TaNuEta0714ascount}\includegraphics[width=0.49\textwidth,trim = 0mm 0mm 0mm 0mm, clip]{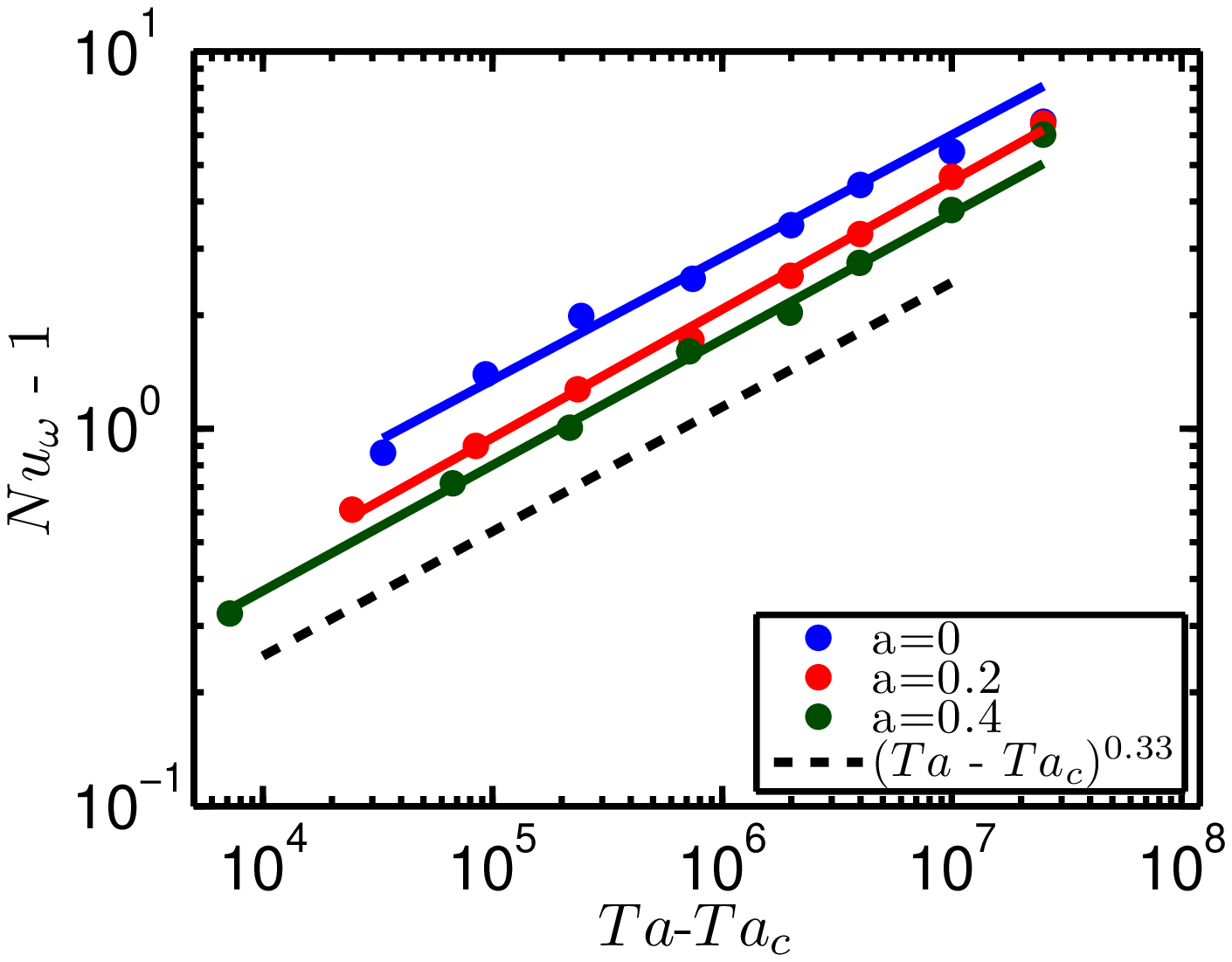}}\\
  \subfloat{\label{fig:TaNuEta0714ascorrcomp}\includegraphics[width=0.49\textwidth,trim = 0mm 0mm 0mm 0mm, clip]{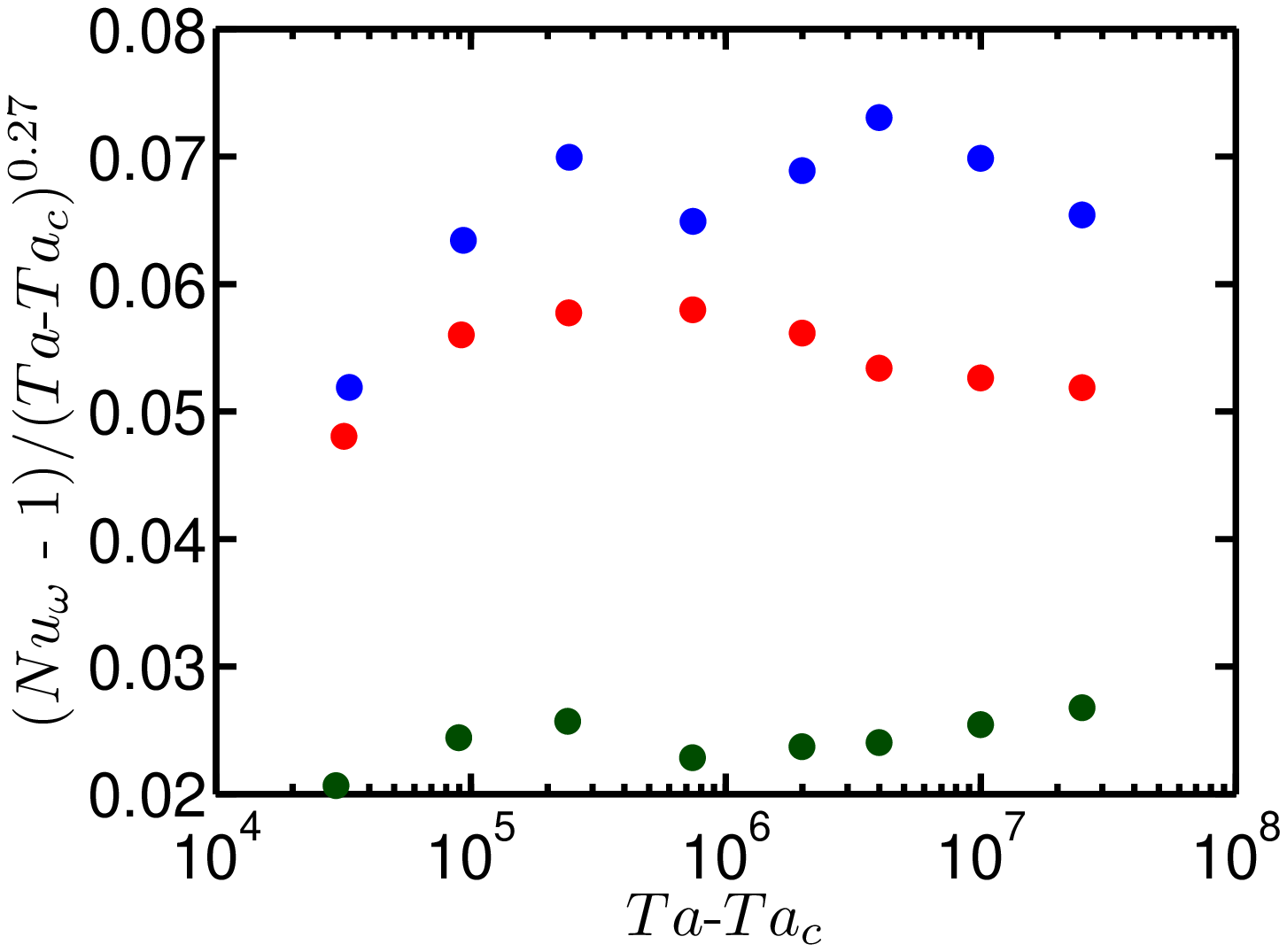}}
  \subfloat{\label{fig:TaNuEta0714ascountcomp}\includegraphics[width=0.49\textwidth,trim = 0mm 0mm 0mm 0mm, clip]{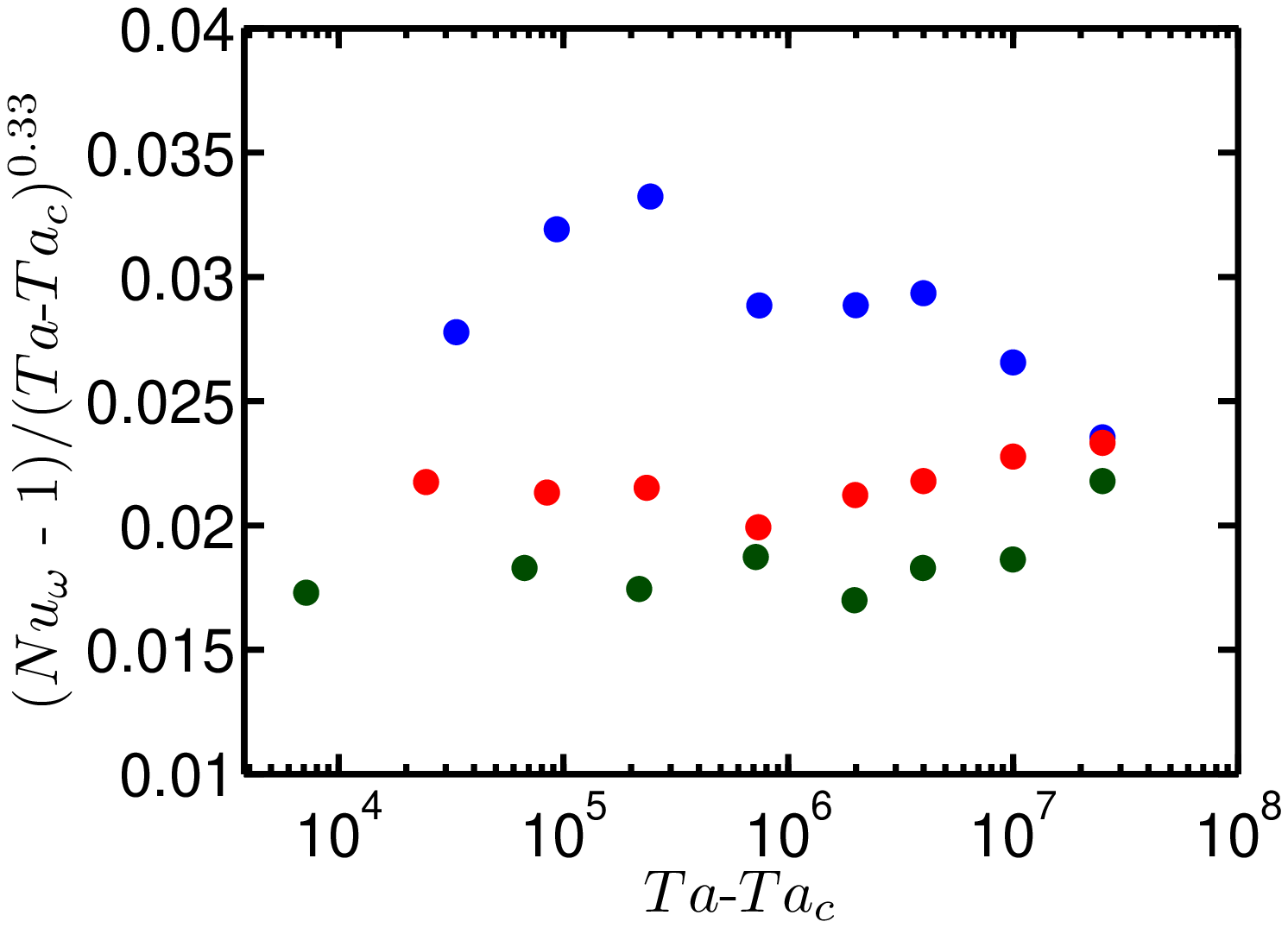}}\\
  \caption{$\nom-1$ versus $Ta-Ta_c$ for three values of co-rotating $a$ (left) and pure inner cylinder and two values of counter-rotating $a$ (right). Compensated plots are shown below. Numerical uncertainties are less than 1\%. $Ta_c$ depends on a and is determined respectively from the analytical approximation \citep{ess96}; for a=0 it is $Ta_c = 1038$. }
 \label{fig:TaNuEta0714as}
 \end{center} 
\end{figure}

Figure \ref{fig:Nu_a_3D} shows the complete results for $\nom$ as a function of $Ro^{-1}$ and $Ta$. In order to better quantify the results from figure \ref{fig:Nu_a_3D}, cross sectional cuts are taken. By taking cross sections of constant $a$, scaling laws can also be discovered for non-zero values of $a$, i.e.\ for co- and counter-rotation $\omega_o \neq 0$. This is shown in figure \ref{fig:TaNuEta0714as} for five different values of $a$, two under co-rotation $\omega_o > 0$, two under counter-rotation $\omega_o < 0$, and as reference 
case $\omega_o=0$. 

For counter-rotating cylinders and Taylor numbers prior to  the change in scaling happens, a universal scaling of approximately $\nom - 1 \sim (Ta - Ta_c)^{0.34}$ is seen. However, the change in scaling and its exponent happens earlier in $Ta$ for $a=0$, while the scaling prevails to larger $Ta$ for the other $a$ (0.2 and 0.4). The scaling is different for co-rotating cylinders, and is approximately $\nom - 1 \sim (Ta - Ta_c)^{0.27}$. $Ta_c = 1038$ has been subtracted as done previously so that the scaling is not lost for the first points. 

The time independence of $\nom$ is broken for much smaller $Ta$, if the outer cylinder is rotating. For both co- and counter-rotating cylinders time dependence can be noted to set in at $Ta$ as low as $10^5$. Also, the scaling of $\nom$ with $Ta$ is maintained throughout a much larger range of the Taylor number. Therefore, the breakdown of time independence cannot be associated anymore to the change in scaling, as one could conclude when only considering pure inner cylinder rotation, where the loss of time-independence and change in scaling happened at about the same $Ta$.

\begin{figure}
 \begin{center}
  \includegraphics[width=0.8\textwidth]{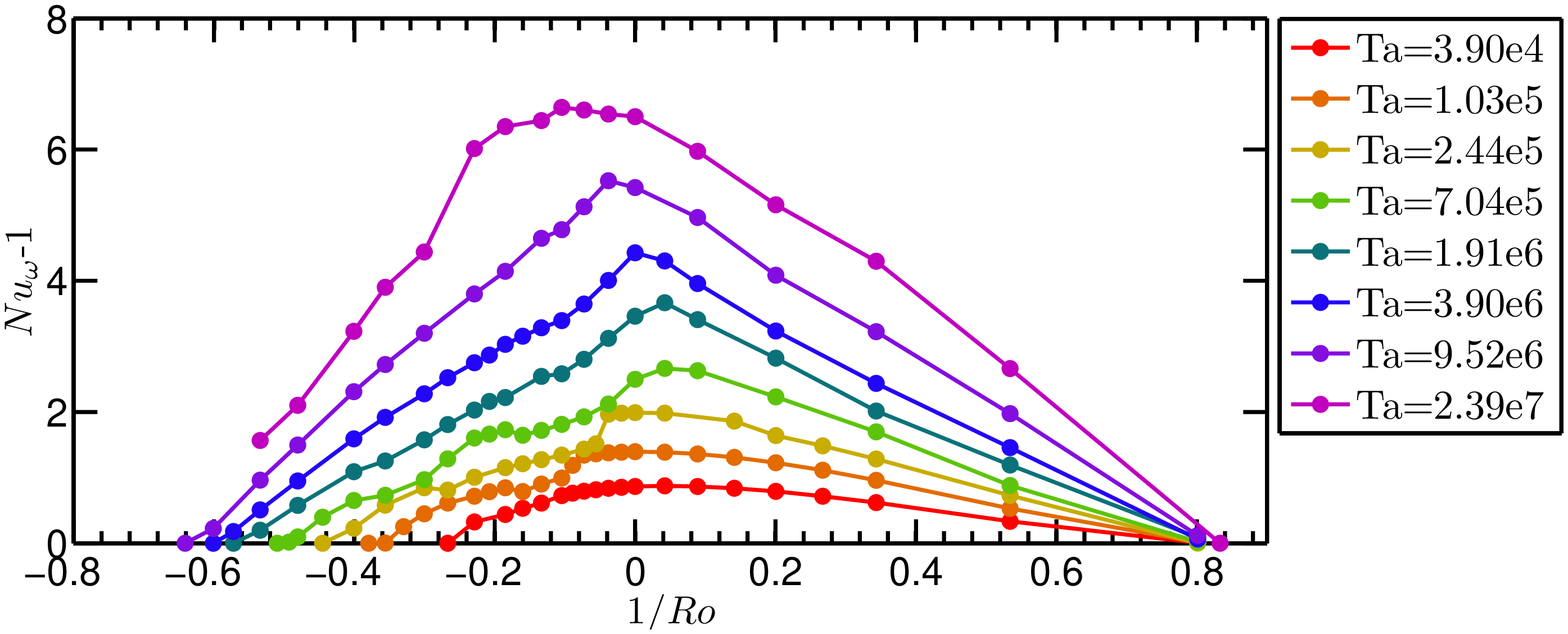}
  \caption{$\nom-1$ as a function of $Ro^{-1}$ across the available $Ta$ range. The advantage of plotting versus $\usro$ can be seen as some of the graphs show piecewise linear behaviour and the value of $\usro$ at the inviscid Rayleigh stability limit $a = - \eta^2$ (denoted as $\usro_{Ra}$) is $\usro_{Ra}=0.833$, appears for all $Ta$. }
  \label{fig:NunnaEta0714}
 \end{center} 
\end{figure}

Cross sections of constant $Ta$ are shown in figure \ref{fig:NunnaEta0714}. $\nom-1=0$ indicates points for which the flow is purely laminar-azimuthal. For co-rotating cylinders, the maximum value of $Ro^{-1}$ reaches the inviscid Rayleigh stability line $Ro^{-1}_{Ra}=0.833$ for even the lowest values of $Ta$. On the other hand, the minimum $Ro^{-1}$ which destabilizes the laminar state can be seen to decrease (become more negative) with increasing $Ta$. 

In order to better visualize the results it is useful to define a normalized Nusselt number as $\nomh=(\nom(\usro)-1)/(\nom(\usro=0)-1)$, which allows easier visualization of the dependence of $\nom$ on $\usro$ across the $Ta$ range of interest. The numerator of $\nomh$ goes to zero, if $Ro^{-1}$ becomes too large or too small, i. e. reaches the stability lines (where $\nom$ becomes 1, since the flow is laminar-azimuthal in the stable ranges), while the denominator is always larger than zero, as long as $Ta > Ta_c$.  

\begin{figure}
 \begin{center}
  \includegraphics[width=0.8\textwidth]{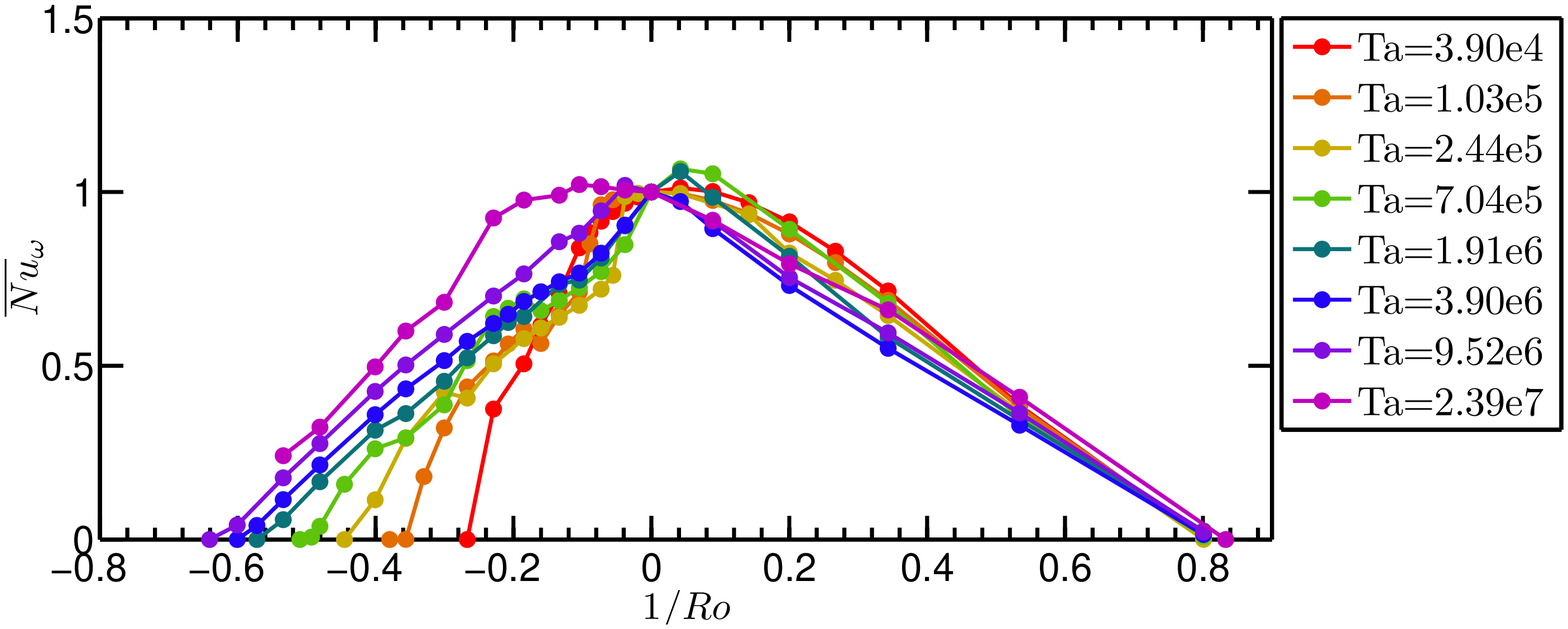}
  \caption{$\nomh$ versus $\usro$. The shift of $\usro_{opt}$ with increasing $Ta$ can be appreciated here.}
  \label{fig:RoNomh}
 \end{center} 
\end{figure}

\begin{figure}
 \begin{center}
  \subfloat{\label{fig:ReoReioptPhasespace}\includegraphics[width=0.49\textwidth]{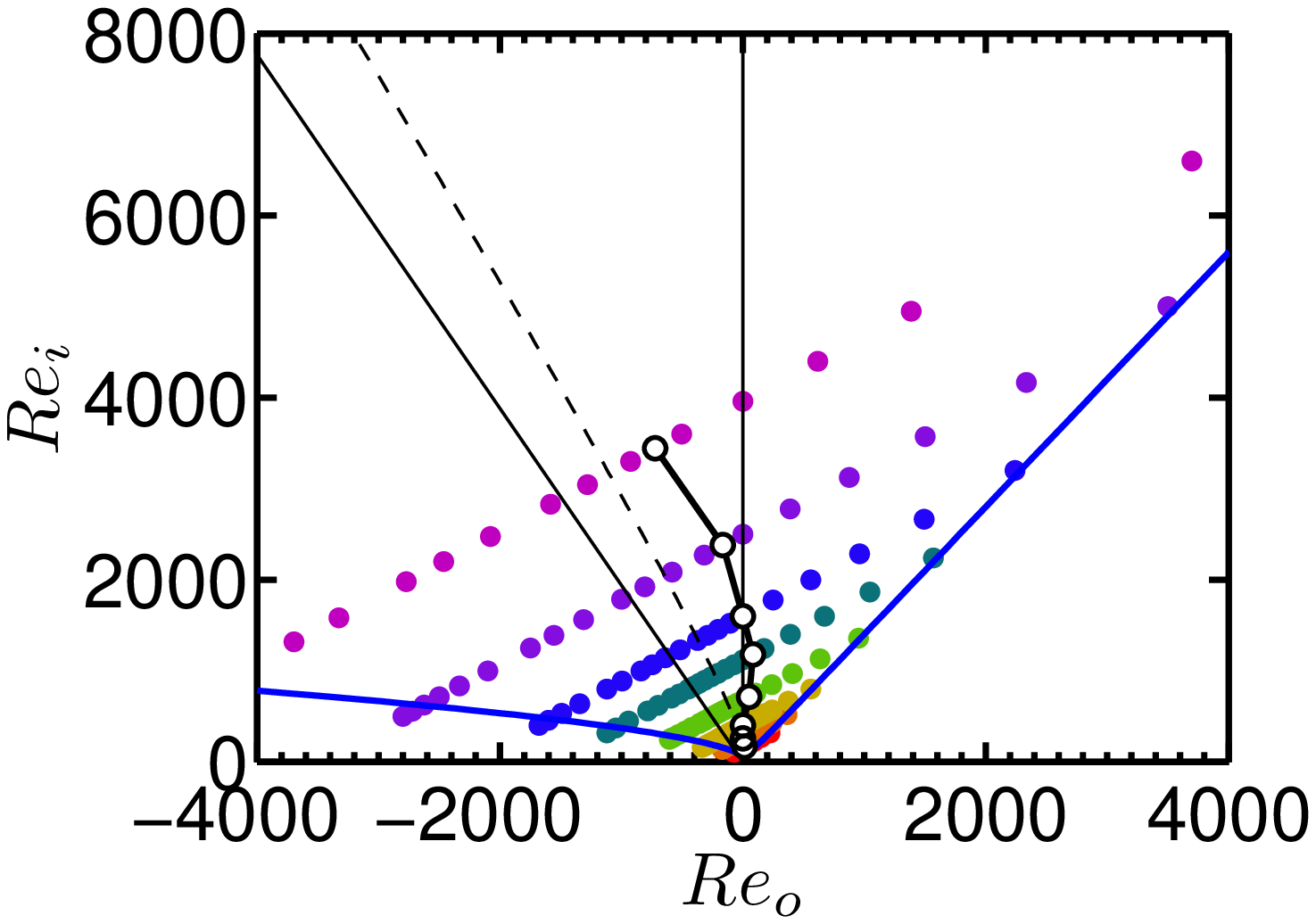}}
  \subfloat{\label{fig:TaRooptPhasespace}\includegraphics[width=0.49\textwidth]{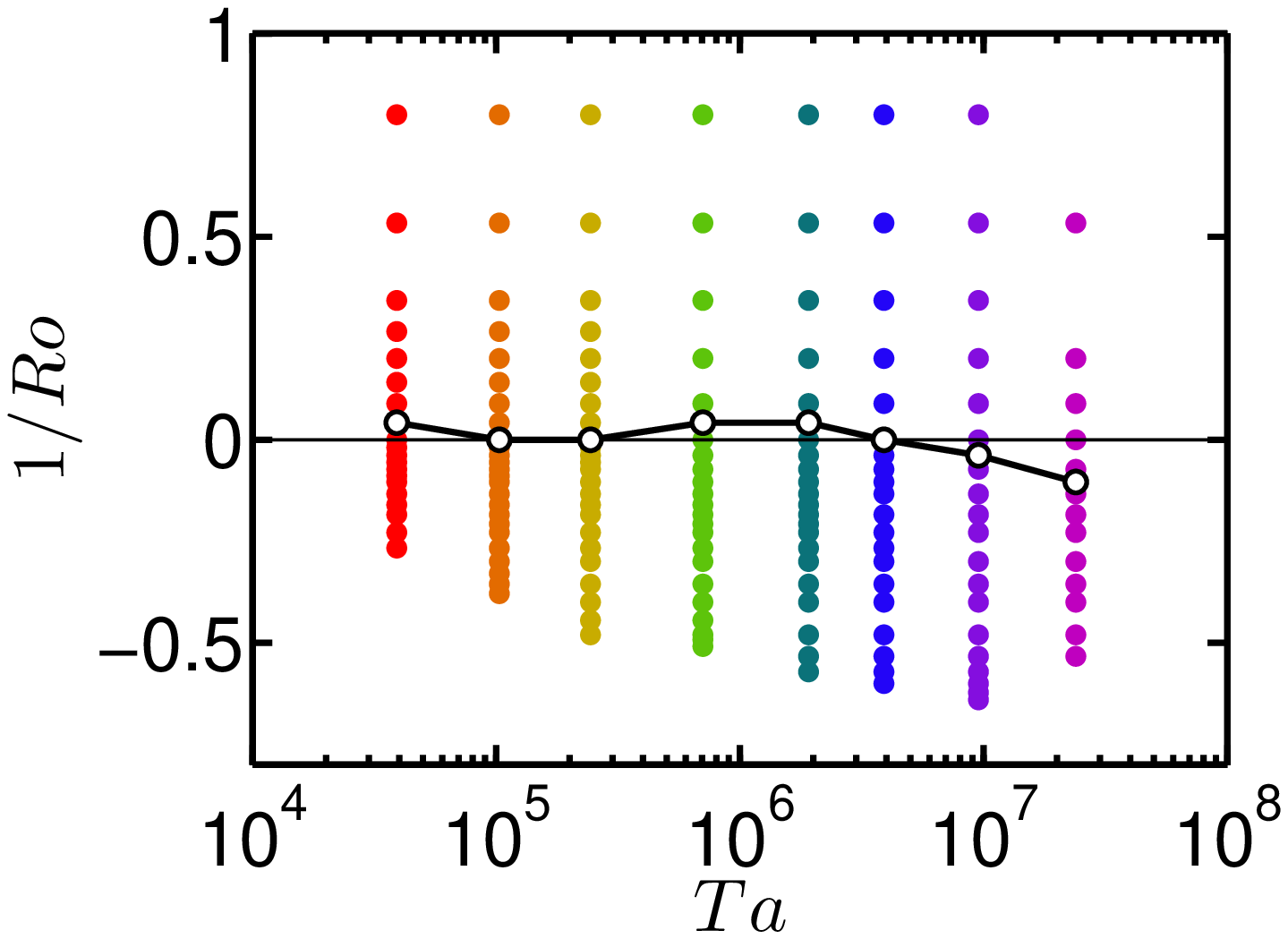}}\\
  \caption{Location of the  optimal transport (hollow black dots connected by thick black line) in the $(Re_o,Re_i)$ phase space of figure \ref{fig:PhaseSpace}a, and in the $(Ta,1/Ro)$ phase space of figure \ref{fig:PhaseSpace}c. On the left, the solid blue line represents the onset of instability according to \cite{ess96} and the thin dashed line is the line of equal distance to the left and right branch of the
  Esser-Grossmann instability line. The solid line is the bisector of the Rayleigh instability line ($a=- \eta^2$) and the line of pure outer cylinder rotation ($a= \infty$).
   }
  \label{fig:minima-in-phase-dia}
 \end{center}
\end{figure}

%\begin{figure}
%include here the new figure 
%\caption{The $Ta$ dependence of the optimal transport, black dots, in the $Re_o,Re_i$ phase space. For small and moderate $Ta$ the influence of the %coherent structures is visible, leading to an optimum even for slight co-rotation. For increasing Ta the curve of optimum positions bends over, %following the bisector between the (viscid) stability borders as discussed in \citep{gil12}. It may approach this bisector or stay slightly upwards %shifted, which can be decided only for larger $Ta$.
%\end{figure}

If $Ta$ is large enough the shape of the graph resembles two straight lines from the maximum value of $\nom$ to $\usro_{Ra}$ and $\usro_{min}$. These straight lines have already been seen when plotting $\nom$ versus a slightly different version of $\usro$ in \cite{pao11} and are the reason we chose to plot $\nom$ versus $\usro$ in this section. 

For smaller $Ta$ an optimum transport for co-rotation can be seen, i. e., for positive values of $\usro$. This holds   for $Ta$ less than the discussed change of the scaling behavior of $\nom$ versus $Ta$. For the two lowest values of $Ta$ the deviation of $\usro$ beyond 0 may still be within numerical uncertainties. However, $\usro_{opt}$ is definitely positive
for $Ta=7.5\cdot10^5$ and $Ta=2\cdot10^6$ and seems to fit with the piecewise linear shape of the graphs. At around $Ta=4\cdot10^6$, which is beyond
 the change in scaling, the maximum $\nomh$ begins to drift to negative $Ro^{-1}$, i. e., towards counter-rotation. This can be seen in figure \ref{fig:RoNomh}. 
 In fig.\ \ref{fig:minima-in-phase-dia} we plotted the positions of the optimum transport in the $(Re_o,Re_i)$ phase space. 
Clearly, the curve does not have equal distance to the two instability branches of the Esser-Grossmann approximation, as was speculated in \cite{gil12}. 
Another feature of the drift of $\usro_{opt}$ is the following: While the curve of $\nomh$ has a prominent peak at $\usro_{opt}$ for values of $Ta$ of around $10^6-10^7$, this turns into a plateau for the highest value of $Ta$ and $\usro_{opt}$ becomes hard to identify. For the highest value of $Ta$, the lower border $\usro_{min}$ already is beyond our parameter range of negative $\usro$.

Experiments \citep{gil11,pao11} have found an optimum transport $a_{opt}\approx0.33-0.35$, corresponding to $\usro_{opt}\approx-0.20$ for Taylor numbers of the order of $10^{12}$. Thus the position of the maximum shifts for higher Taylor numbers. 

Figures \ref{fig:NunnaEta0714} and \ref{fig:RoNomh} show some anomalous jumps in the graph around $\usro\approx-0.2$ which corresponds to $a\approx 0.3$.  These are
 caused by different vortical states as mentioned in section \ref{sec:vorstr}. These may be present even if the simulations are started from the same initial conditions for different values of $\usro$ and $Ta$. If the number of vortices is higher, the vortices become stronger, and the value of $\rew$, which measures their strength, also becomes higher. Since $\nom$ is monotonously related to $\rew$, it also increases. We will further analyse this multi-vortex state in Section \ref{sec:nlpushing}.

\section{Characterization of the flow state}
\label{sec:cohst}

In this section we will analyse two characteristic Taylor number ranges in TC flow. The first, lower one, is the range in which the
 importance of coherent flow structures is lost, since these have become too small in size. In section \ref{sec:global} we have  observed a change in the scaling law for the angular velocity flux from $\nom\sim (Ta-Ta_c)^{0.34}$ to $\nom\sim (Ta-Ta_c)^{0.21}$. Although the Taylor number for this change coincides with the onset of time dependence for pure inner cylinder rotation, when adding outer cylinder rotation the onset of the time dependence is much earlier, and a transition in the scaling laws cannot even be seen.

Therefore, the onset of time dependence cannot be linked satisfactorily to the change in the $\nom$-scaling. Another way of explaining this change is by estimating
 when the spatially coherent flow structures loose influence because their size becomes too small. We do this by defining an average, global coherence length in terms of the Kolmogorov length scale \citep{sug07} resulting from the volume averaged dissipation rate:

\begin{equation}
 \ell_c=10 \eta_K = 10 (\nu^3/\epsilon_u)^{1/4}=10 d (\sigma^{-2} Ta (\nom-1) + \hat{\epsilon}_{u,0})^{-1/4}.
 \label{eq:coh-length}
\end{equation}

\noindent where eq. (\ref{eq:epsiloneck}) has been used for the second equality. We compare 
the global coherence length $\ell_c$ with the gap width $d$  or, equivalently with the extension of the remnants of the Taylor vortices, whose length can also be estimated as $d$, since they tend to have a square aspect ratio.  

\begin{figure}
 \begin{center}
  \subfloat[]{\label{fig:etaKTa}\includegraphics[width=0.47\textwidth]{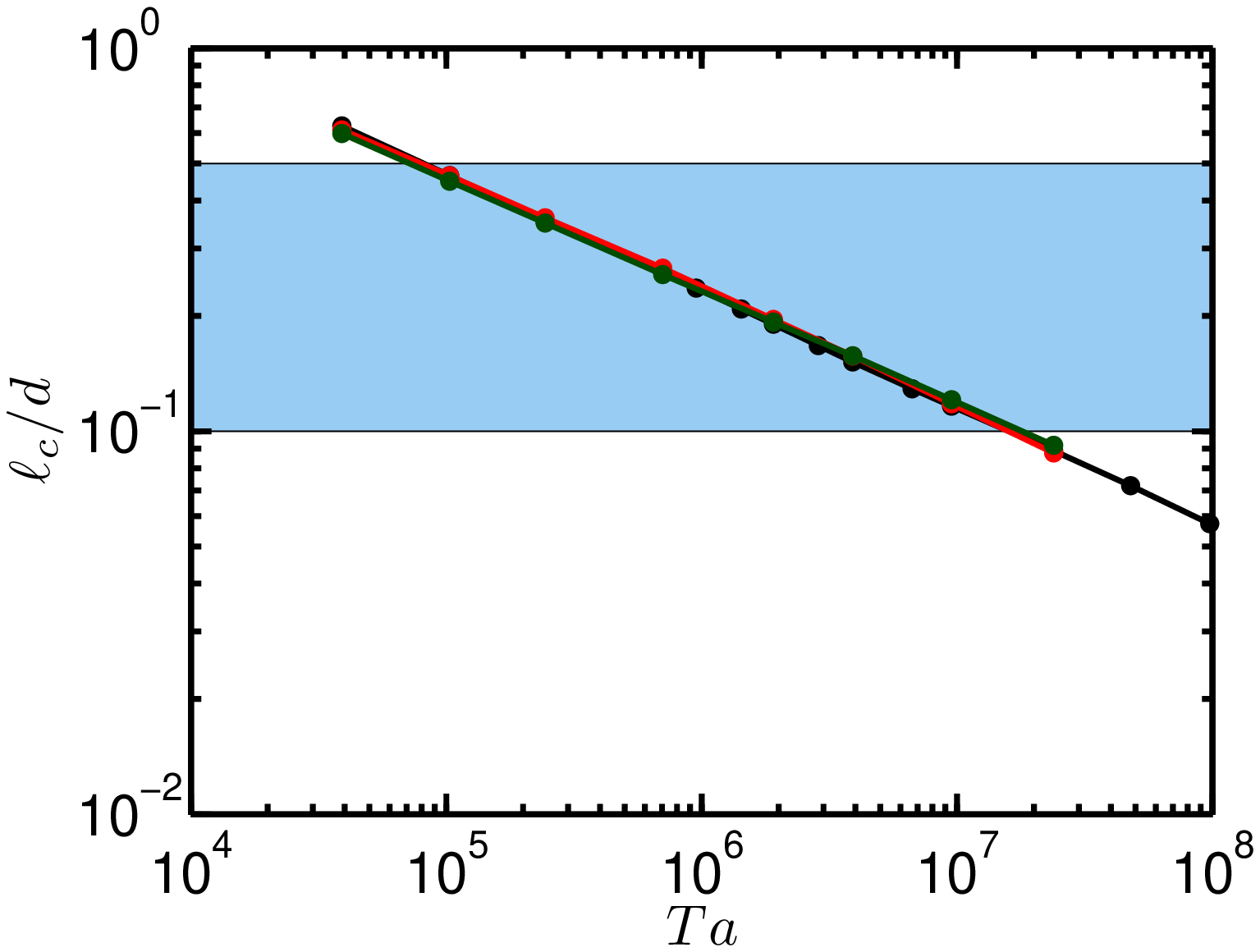}}
  \subfloat[]{\label{fig:ResTa}\includegraphics[width=0.47\textwidth]{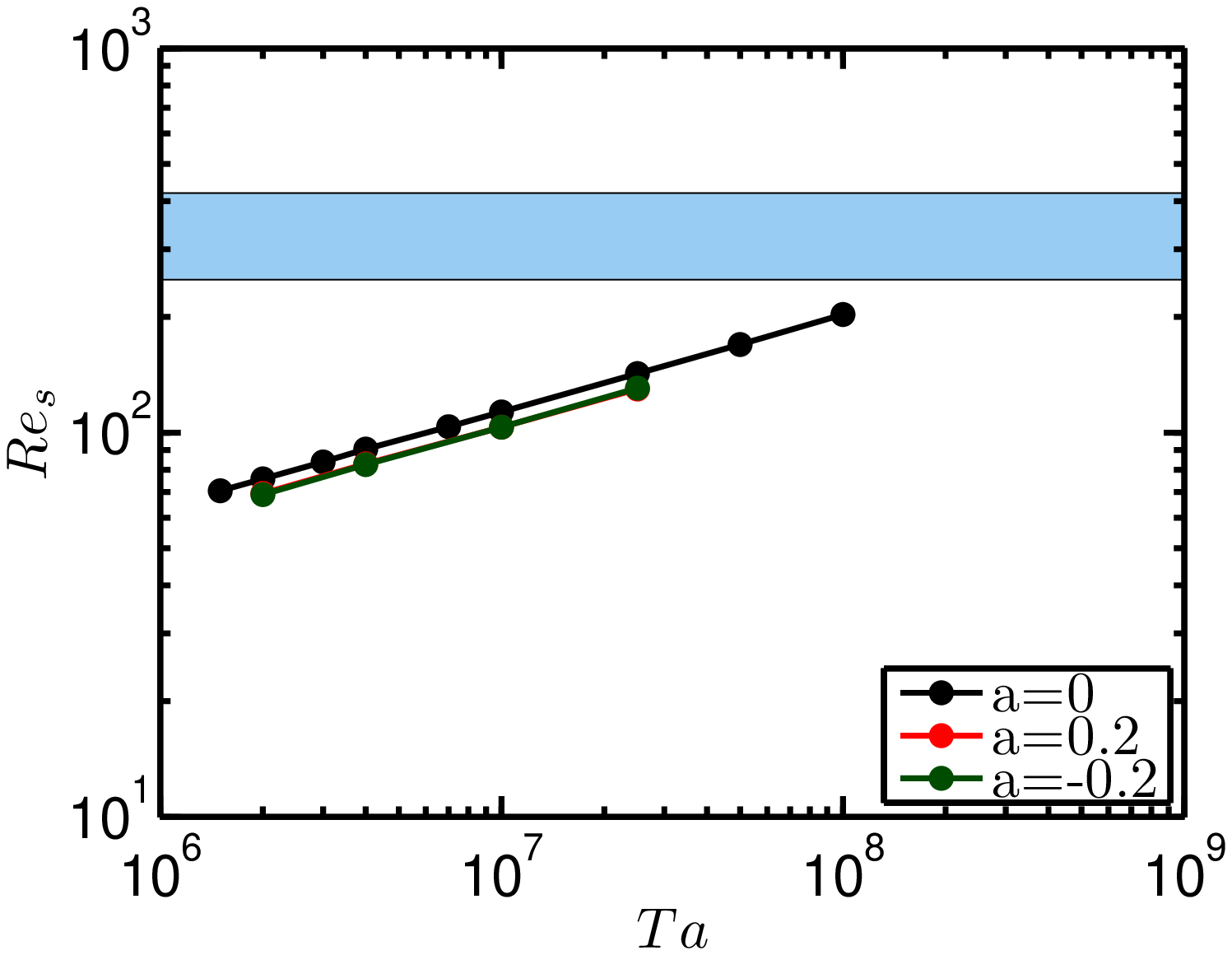}}
  \caption{The graph on the left shows $\ell_c/d$ versus $Ta$ for $\eta=5/7$ and three values of $a$ as shown in the inset. The blue region indicates the range of decreasing importance of the coherence structures. -- The graph on the right shows the shear Reynolds number $Re_s$ of the boundary layers versus $Ta$. In the blue shaded region we expect a shear instability of the boundary layers, in which the Prandtl laminar boundary layers become turbulent. The TC system then is in the so-called `ultimate' regime, cf.\ \cite{gro11}.}
  \label{fig:etaKResTa}
 \end{center} 
\end{figure}

Figure \ref{fig:etaKResTa} shows the variation of $\ell_c/d$ with increasing Taylor number. The loss of importance of coherent structures happens in the range where $\ell_c/d$ is between $0.1-0.5$, corresponding to $Ta \approx 10^6-10^7$. It is just within this $Ta$ range, where the change in the $\nom$ scaling law occurs. The graph is consistent with  that change taking place at approximately the same $Ta$ for different values of $a$, which is what is seen in figure \ref{fig:TaNuEta0714as}. This transition is further analysed in section \ref{sec:localcohlen}.

A second characteristic Taylor number is connected with the shear instability of the boundary layers (BL). Here the laminar Prandtl type BLs become turbulent. 
Beyond that $Ta$ the flow is fully turbulent throughout and this state 
is known as the ultimate state, cf. \cite{gro11}. This happens if the BL shear Reynolds number becomes $Re_s > 250 - 420$ and $Re_s$ is calculated 
from the shear velocity $U_s$ as  \citep{gil12}:
\begin{equation}
Re_s = \displaystyle\frac{U_s \delta}{\nu} = {a_{PB}\sqrt{\Rey_i-\rew}}~.
\end{equation}
\noindent The empirical constant $a_{PB}$ is taken as $2.3$ as in \cite{gil12} for Prandtl-Blasius type boundary layer in TC flow. This value is obtained by a fit to experimental data, detailed in \cite{gil12}.

For RB flow this transition is expected at $Ra \approx 10^{14}$ \citep{gro01,ahl09,gro11}, 
while figure \ref{fig:ResTa} shows that the transition in TC is expected for $10^8 \lesssim Ta \lesssim 10^9$. Recently, experiments have confirmed the ultimate scaling both for $\nom$ and $\rew$. \cite{hui12} have shown that in TC flow $Nu_\omega \sim Ta^{0.38}$ and $\rew \sim Ta^{0.50}$ when $Ta \gtrsim 10^9$. A confirmation of the analogy between RB and TC is obtained by the high $Ra$ number experiments by \cite{he11} as they measured that $Nu \sim Ra^{0.38}$ and $\rew \sim Ra^{0.50}$ for $Ra \gtrsim 5\times 10^{14}$. These measured scaling exponents agree exactly with the predictions by \cite{gro11}. In contrast to the experiments of \cite{gil11,gil12}, in our present numerical simulations the
ultimate state is not yet achieved, as clearly seen from fig.\ \ref{fig:ResTa}. 

\section{Local results}

This section contains the analysis of local results. For convenience we skip in this section the "hat" on the dimensionless flow field variables, but still understanding them as being dimensionless. The angular velocity profiles are shown and the ratio of the BL thickness is calculated and compared
with the theory of EGL (2007).
 The angular velocity 
profiles reflect the interplay of bulk and boundary layers and that of the mean flow and added perturbations.
 The importance of convective versus diffusive transport is quantified through the bulk slope of the angular velocity profile, and again we will
find a maximum as function of $a$, which we will connect with the maximum in the angular velocity transport $Nu_\omega$.

\subsection{Local coherence length and vortex characterization}
\label{sec:localcohlen}

\begin{figure}
 \begin{center}
  \subfloat[]{\label{fig:cohlength2dTa2e6}\includegraphics[width=0.4\textwidth,trim = 35mm 0mm 35mm 0mm, clip]{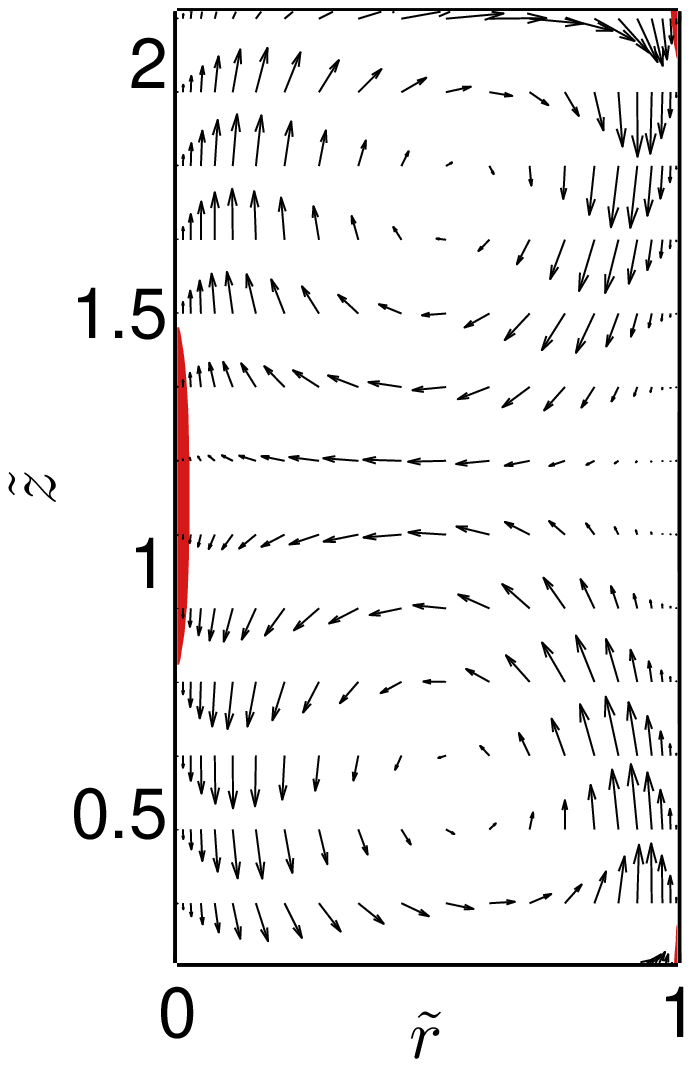}}
  \subfloat[]{\label{fig:cohlength2dTa1e7}\includegraphics[width=0.4\textwidth,trim = 35mm 0mm 35mm 0mm, clip]{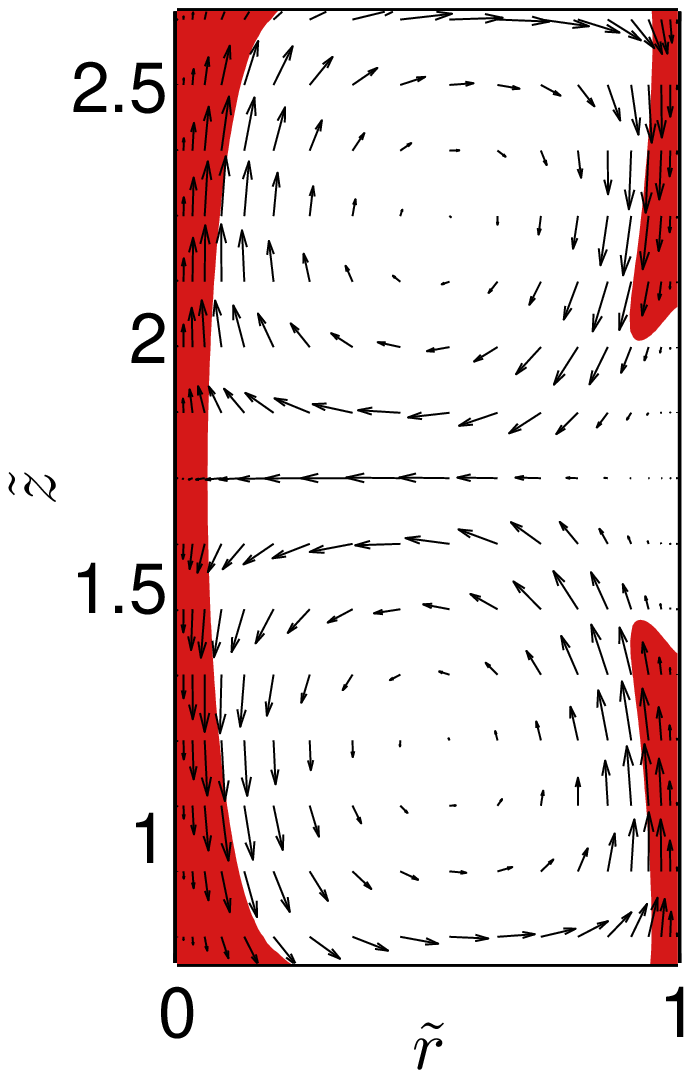}}
  \caption{The two plots show the areas in red where $\ell_c/d<0.1$ for pure inner cylinder rotation and Taylor numbers of (a) $Ta=1.91\cdot10^6$ and (b) $Ta=9.52\cdot10^6$.  In these areas the breakup of the coherent structures is likely to occur. The arrows indicate $\bu$ in magnitude and direction. }
  \label{fig:cohlength2d}
 \end{center} 
\end{figure}

Figure \ref{fig:cohlength2d} shows the \emph{local} coherence length calculated from the local dissipation in analogy to eq. (\ref{eq:coh-length}). This figure adds details on where we expect the Taylor vortices to break down. At low Taylor number, the local coherence length is smaller than 0.1 only very near to the wall, where the highest local dissipation takes place. With increasing Taylor number, the highest local dissipation still is near the wall, but the dissipation rate is large enough in the whole domain to break up the dominance of the coherent structures, even if they do not fully disappear but become small enough. 

From figures \ref{fig:etaKResTa}-\ref{fig:cohlength2d} we expect coherent structures to break up at Taylor numbers in the range of $Ta \approx 10^6-10^7$. This may at first sight contradict the earlier finding that the scaling of the wind remains constant across the whole Taylor range studied (cf. section \ref{sec:global}), especially as the characteristic wind velocity is defined from a time-averaged field. One might expect that the perturbations destroy the large scale structures and as a consequence completely modify the wind. In order to analyse this transition in more detail, we investigate the instantaneous velocity fields before and after the breakdown of coherence. For this, vortices will be characterized employing the so-called $\lambda_2$-criterion \citep{jeo95}. 

\begin{figure}
 \begin{center}
  \subfloat{\label{fig:lambda23dTay750K}\includegraphics[width=0.4\textwidth,trim = 40mm 0mm 25mm 0mm, clip]{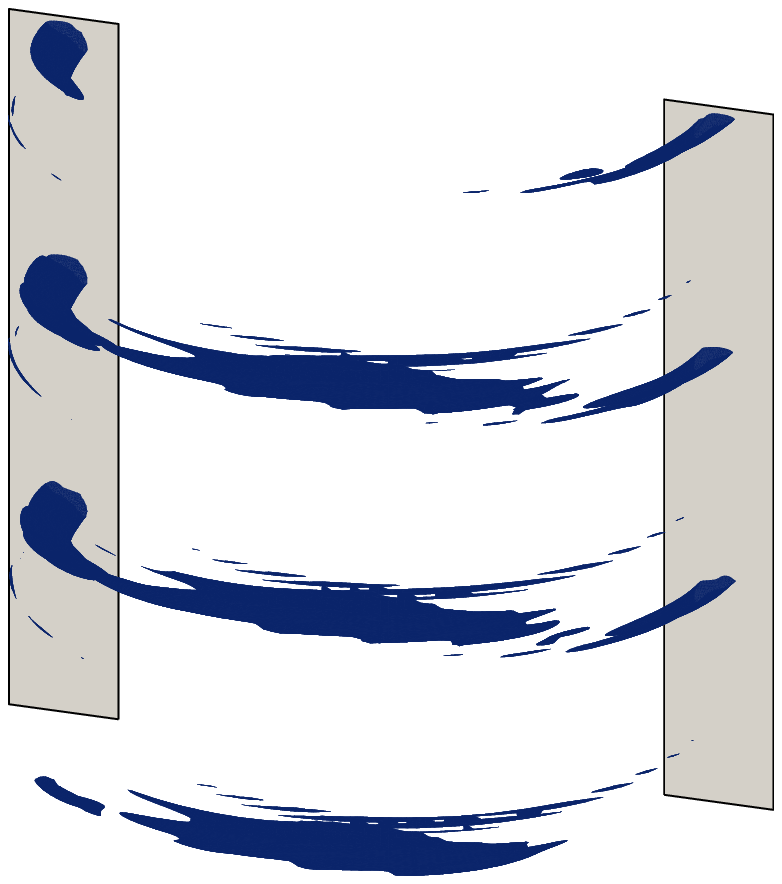}}
  \subfloat{\label{fig:lambda23dTay25M}\includegraphics[width=0.4\textwidth,trim = 30mm 0mm 35mm 0mm, clip]{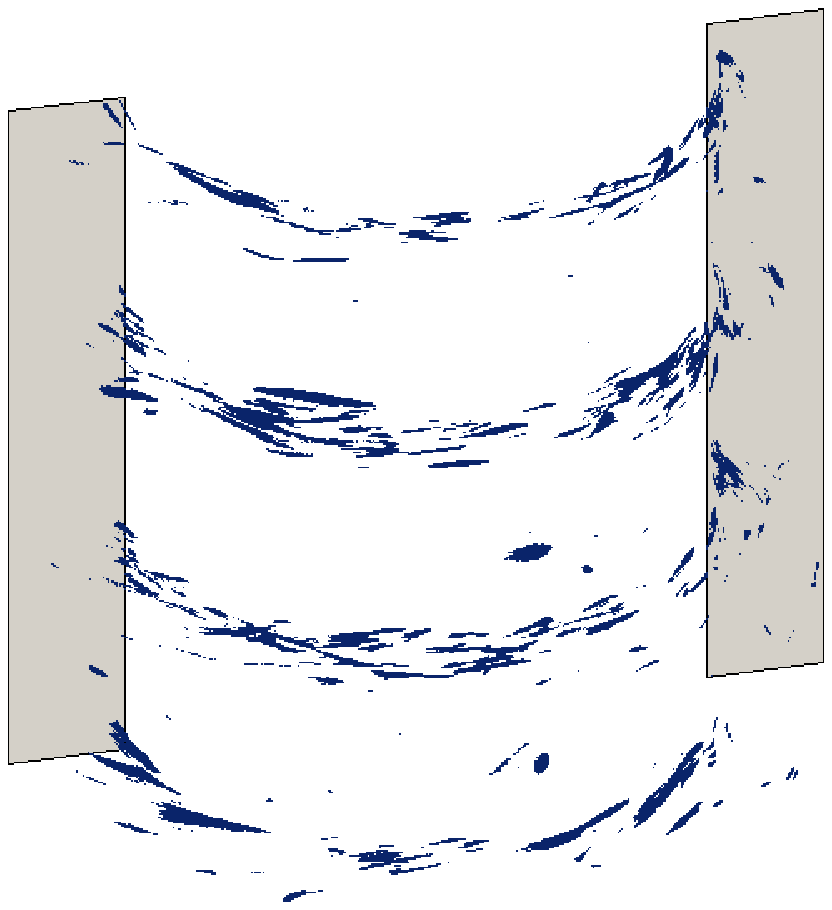}} \\
  \subfloat{\label{fig:lambda2Tay750K}\includegraphics[width=0.4\textwidth,trim = 35mm 0mm 35mm 0mm, clip]{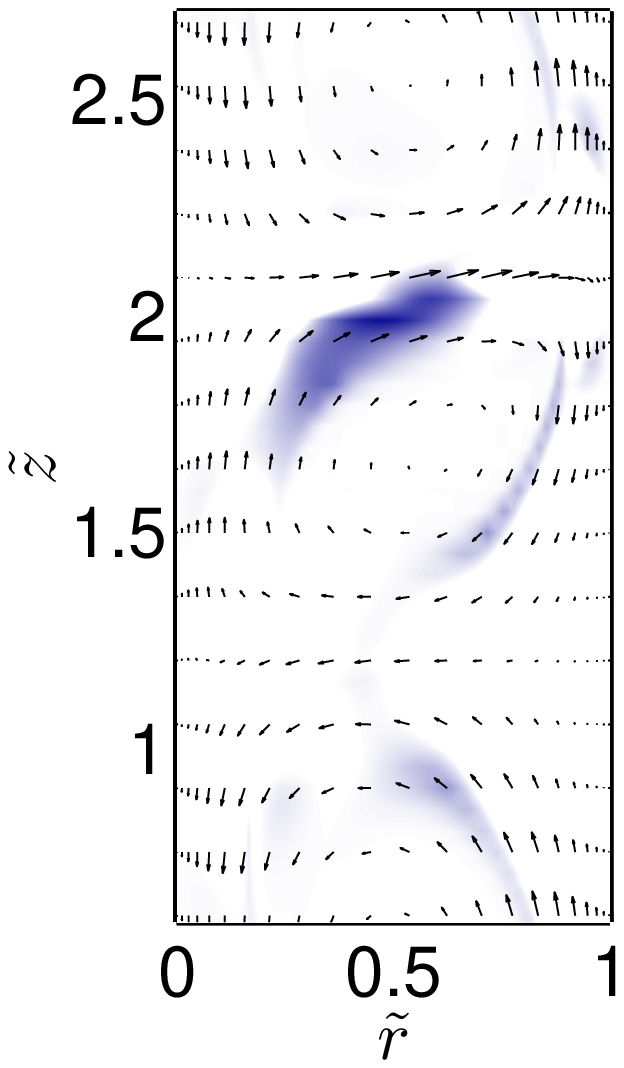}}
  \subfloat{\label{fig:lambda2Tay25M}\includegraphics[width=0.4\textwidth,trim = 35mm 0mm 35mm 0mm, clip]{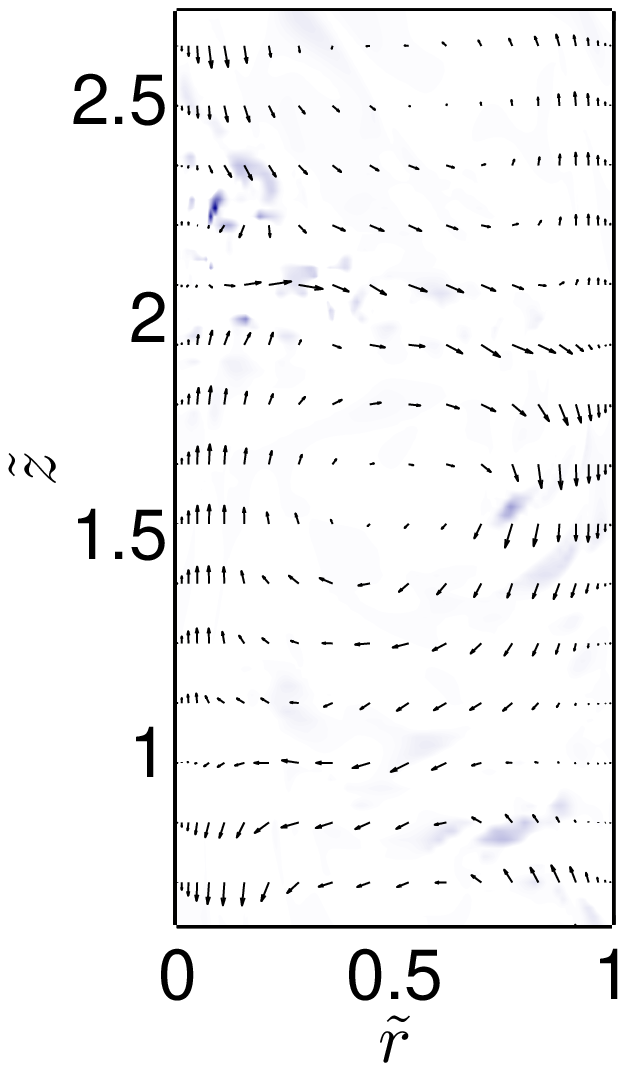}}
  \caption{The two top plots show isosurfaces of $\lambda_2$ for pure inner cylinder rotation and Taylor numbers of (a) $Ta=7.04\cdot10^5$ and (b) $Ta=2.39\cdot10^7$. The bottom two plots show contour plots of $\lambda_2$ truncated for $\lambda_2>0$.  The darkest blue represents the minimum value of $\lambda_2$ in each plot and white represents $\lambda_2\ge0$. The arrows indicate \textbf{u} in magnitude and direction. The topography of the negative $\lambda_2$ areas changes from large coherent regions in the centre of the gap -indicating Taylor vortices- to small regions near the inner cylinder -indicating hairpin vortices. The underlying wind, however, remains unchanged.}
  \label{fig:lambda2}
 \end{center} 
\end{figure}

The top two panels of figure \ref{fig:lambda2} shows full 3D isosurfaces of $\lambda_2$ for two Taylor numbers, on the left for $Ta=7.04\cdot10^5$ before the transition, and on the right for $Ta=2.39\cdot10^7$, after the transition. The bottom two panels of figure \ref{fig:lambda2} show an azimuthal-cut contour plot of $\lambda_2$ for two Taylor numbers. The instantaneous ``wind'' is superimposed. It is important to note that for the left panel time dependence has not yet set in, so the instantaneous and mean velocity fields are indistinguishable. In this panel we can see that the lowest values of $\lambda_2$ are located in the centre of the gap, coinciding with an area of positive $u_r$ wind and almost no $u_z$ wind. Structures of negative $\lambda_2$ occupy a significant portion of the space between the cylinders.

On the right panels, we can see a different picture. The structures of negative $\lambda_2$ are now much smaller, and no longer occupy a significant region of the domain, unlike in the left panel. These stuctures are also in a different place- clustered near the inner cylinder, from where they seem to originate. The instantaneous wind is superimposed on the contour plot. A similar structure as the one in the left panel is seen, indicating that even though the coherent structures are no longer dominant, the underlying wind behaves in a similar manner. Indeed, once the velocity field is averaged in time, the large scale Taylor vortices are recovered. This result is consistent with the findings reported by \cite{don07}.

\subsection{Angular velocity profiles}

\begin{figure}
 \begin{center}
   \includegraphics[width=0.99\textwidth]{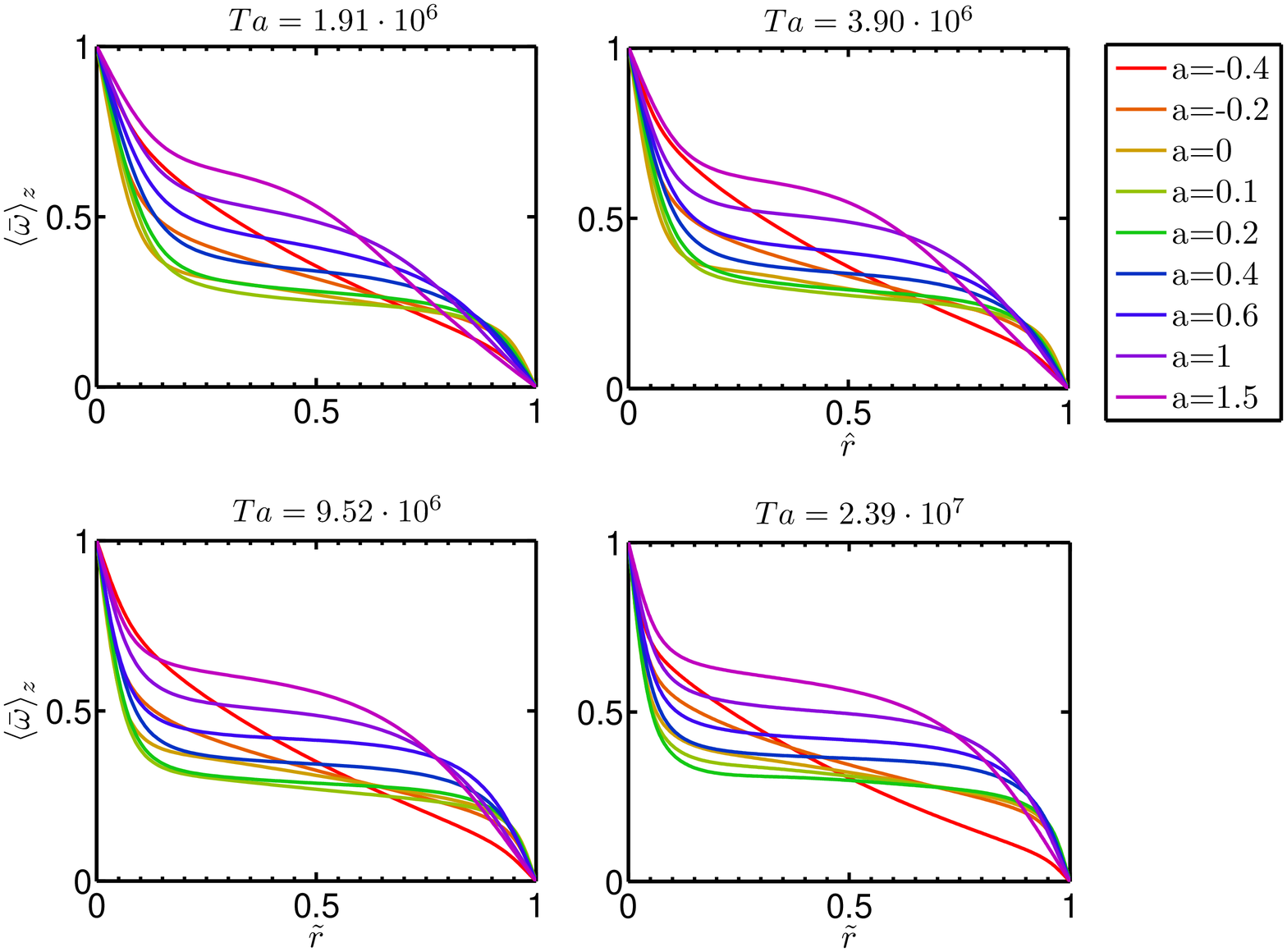}
   \caption{The $t, \theta$ and also $z$-averaged angular velocity $\langle\bar{\omega}\rangle_z$ versus $\tilde{r}$ for four Taylor numbers. The boundary layers of the $\omega$ profiles become thinner around slight counter-rotation than they are for higher values of $a$ as well as for co-rotation as an indication of increased transport. For strong co-rotation $a=-0.4$ as well as high $a$, i. e., strong counter-rotation, at low $Ta$ there is not yet a flat bulk zone since the flow is not yet turbulent enough.}
  \label{fig:romegaprofEta0714}
 \end{center} 
\end{figure}

\begin{figure}
 \begin{center}
  \includegraphics[width=0.99\textwidth]{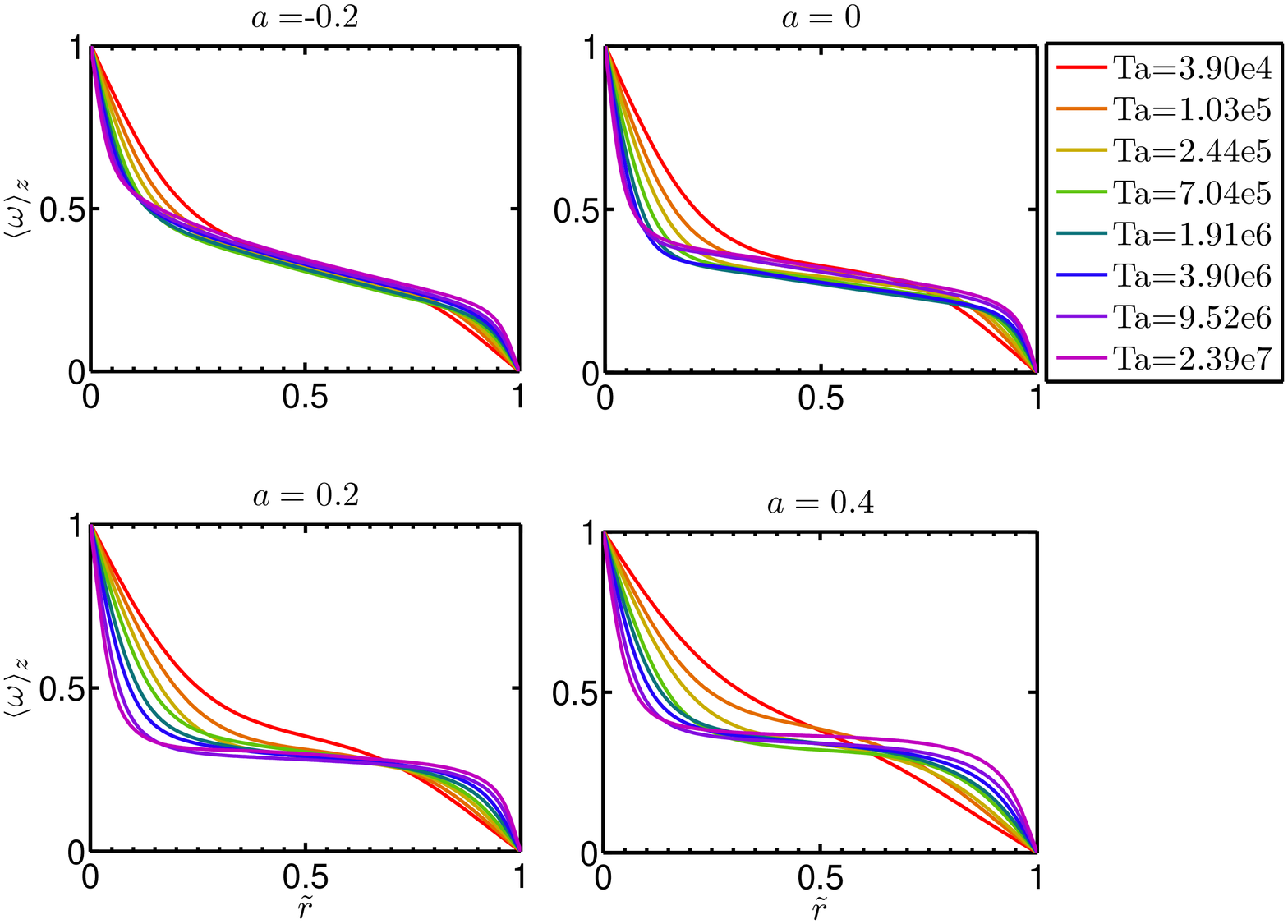}
  \caption{$\langle\bar{\omega}\rangle_z$ versus $\tilde{r}$ for four values of $a$ and increasing $Ta$. The boundary layers for the $\omega$ transport become thinner with increasing $Ta$ indicating increased angular velocity transport. The boundary layers get steeper with increasing $Ta$ and the bulk region becomes more extended. For low values of $Ta$ the bulk region is rather small. The $\langle\bar{\omega}\rangle_z$ profile in the center remains approximately unchanged with $Ta$.}
  \label{fig:romegaprofEta0714a0}
 \end{center} 
\end{figure}

The angular velocity $\omega$ is the transported quantity in Taylor-Couette flow. Analysing the dependence of the $\omega(r)$ profiles on the driving parameters $Ta$ and $a$ seems useful to understand how transport takes place in the flow. $\omega(r)$ profiles are shown in figures \ref{fig:romegaprofEta0714} and \ref{fig:romegaprofEta0714a0}. Beyond the breakdown of the laminar, purely azimuthal flow, three distinct regions in the gap appear. These are the inner and outer 
boundary layers (BL), in which  the transport mechanism is dominantly diffusive, and a flatter bulk zone, in which 
the transport mechanism is dominantly convective, see  figure \ref{fig:blobliex} for a sketch, in which 
we approximate the profile of the mean azimuthal velocity $\langle \bar{u}_\theta\rangle_z$ by three straight lines, 
one for each boundary layer and one for the bulk. For the boundary layers, we calculate the slope of the lines by fitting (least mean square) a line through the first three computational grid points. For the bulk, we first force the line to pass through the inflection point of the profile (the nearest grid point). Then, its slope is taken from a least mean square fit using two grid points on either side of this inflection point. The respective boundary layer line will cross with the bulk line at a point which then defines the thickness of that boundary layer. 

In the next two subsections we will discuss the BL and bulk regimes separately.

\subsection{Angular velocity profiles and resulting boundary layer thicknesses}
\label{sec:blobli}
\begin{figure}
 \begin{center}
  \includegraphics[width=0.9\textwidth]{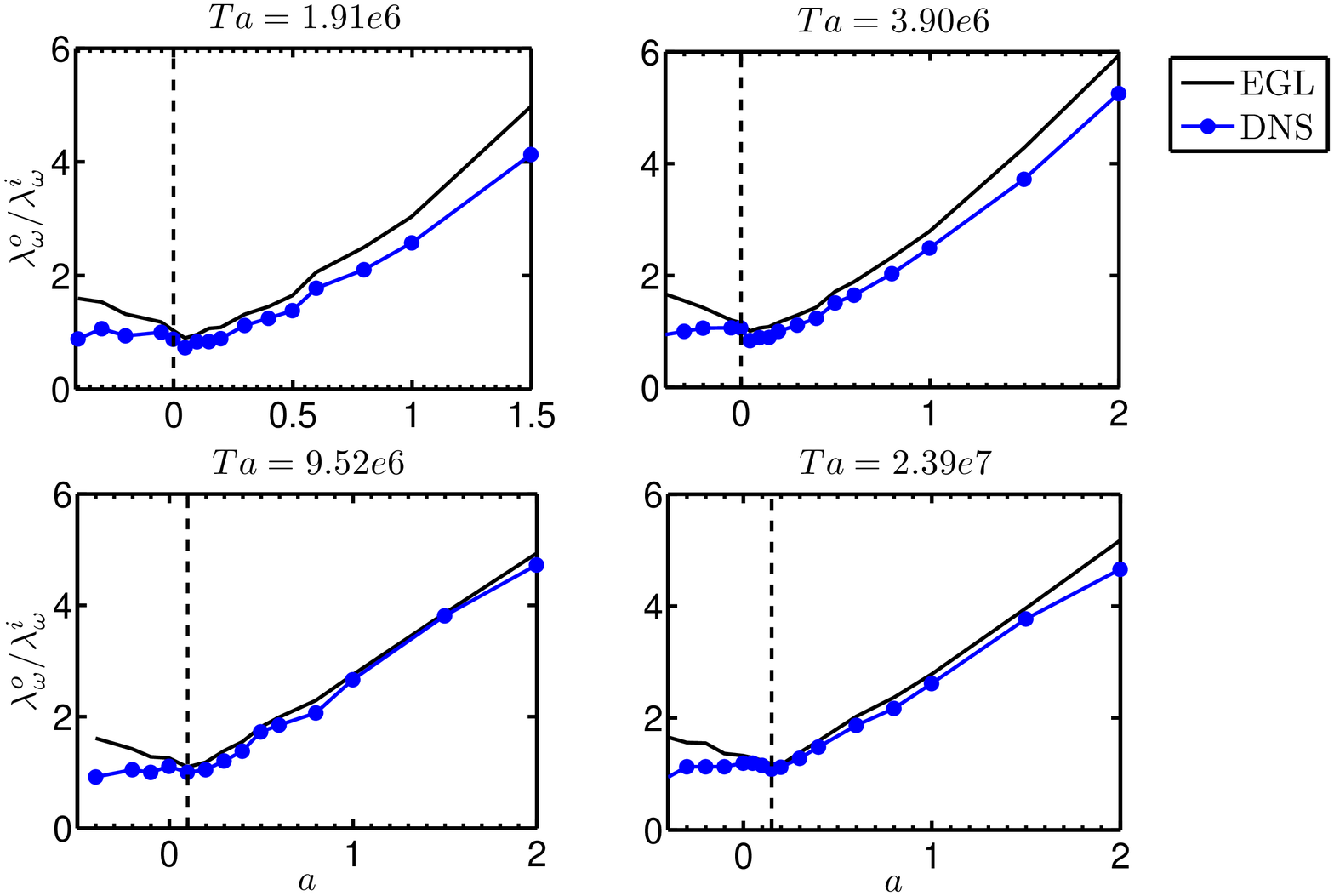}
  \caption{$\lambda^o_\omega/\lambda^i_\omega$ versus $a$ for four values of $Ta$. The agreement between theory and simulation is better for counter-rotation and with increasing $Ta$, but does not match for co-rotation. A dashed line indicates the optimal rotation ratio $a_{opt}$.}
  \label{fig:blobliEta0714}
 \end{center} 
\end{figure}

 With increasing $Ta$, in order to accommodate the increasing angular velocity transport, 
  the boundary layers become thinner and the $\omega$-slopes steeper. 
 What is striking is the strong asymmetry between the inner BL and the outer BL, which is much thicker. 
Figure \ref{fig:blobliEta0714} shows the ratio between the outer and the inner boundary layer thicknesses versus rotation ratio $a$ for two Taylor numbers. 
 This asymmetry is a consequence of  the exact relation $\partial_r \langle \omega \rangle |_o = \eta^3 \partial_r \langle \omega \rangle |_i$, 
   obtained from the $r$-independence of $J^{\omega}$, cf.\ eq.\ (\ref{eq:j-definition} ): 
 The slope at the inner cylinder is a factor of $\eta^{-3}$ larger than  at the outer one and thus 
the outer boundary layer is much more extended than the inner one.
 Since for the present $Ta$-range the shear Reynolds number $Re_s$ is  still below 
the  threshold value range for the transition to turbulence in the boundary layers 
(see  section \ref{sec:cohst}),
we can compare the numerically obtained boundary thickness ratio with that one obtained by EGL (2007), which had been derived in the spirit
of the  Prandtl-Blasius (i.e.\, laminar-type) boundary layer theory, namely
\begin{equation}
\label{eq:thicknessratio}
  \displaystyle\frac{\lambda^o_\omega}{\lambda^i_\omega}=\eta^{-3}\displaystyle\frac{|\omega_o-\omega_{bulk}|}{|\omega_i-\omega_{bulk}|}.
\end{equation}
Here the value of $\omega_{bulk}$ is a characteristic bulk angular velocity chosen to be the value of $\omega$ at the inflection point of  the $\omega$-profile (see figure \ref{fig:domegablobliex}). It is calculated from the numerical simulations.
The result for the BL thickness ratio $\lambda_\omega^o/\lambda_\omega^i$ is shown in figure  \ref{fig:blobliEta0714}. 
 The agreement with the numerically obtained ratio 
is satisfactory for the counter-rotating $a$-cases, getting even better with increasing $Ta$. This is because the estimate is based on a flat profile in the bulk, and indeed the profile becomes flatter with increasing $Ta$. For co-rotation, formula (\ref{eq:thicknessratio}) apparently fails. This had to be expected, because the  approximation of the profile of $\langle \bar{u}_\theta\rangle_z$ by three straight lines,
which  was assumed in the derivation of (\ref{eq:thicknessratio}), 
 is then no longer appropriate.

\subsection{Angular velocity profiles in the bulk}
We now come back to the mean profiles in the bulk. As can be clearly seen from comparing figures\ 
\ref{fig:romegaprofEta0714}   and \ref{fig:romegaprofEta0714a0}, 
both the mean angular velocity and its  slope are  controlled by $a$ (or $\usro$) rather than by $Ta$. This behavior can be understood from
 equation (\ref{eq:rotatingTC}):
The outer cylinder rotation is reflected in that  equation
 as a Coriolis force. This force is present in the whole domain, while $Ta$ controls the strength of the viscous term, 
which is dominant  in the boundary layer.
Therefore  the 
 profile is controlled by the Coriolis force, i.e.\, $\usro$ or $a$, and not by $Ta$.

\begin{figure}
 \begin{center}
  \subfloat{\label{fig:domegadrexample}\includegraphics[width=0.49\textwidth,trim = 10mm 7mm 9mm 5mm, clip]{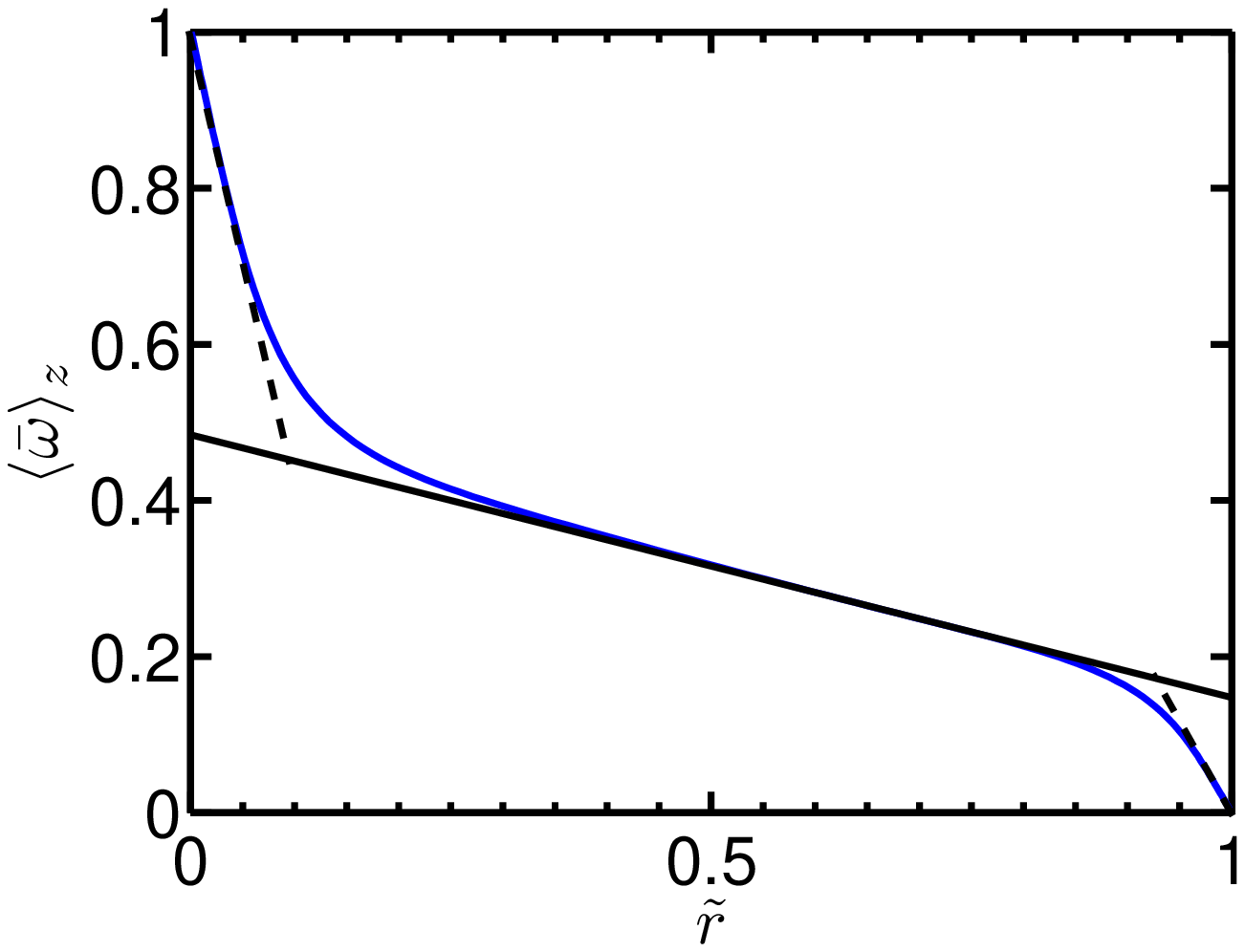}}
  \subfloat{\label{fig:blobliex}\includegraphics[width=0.49\textwidth,trim = 10mm 7mm 9mm 5mm, clip]{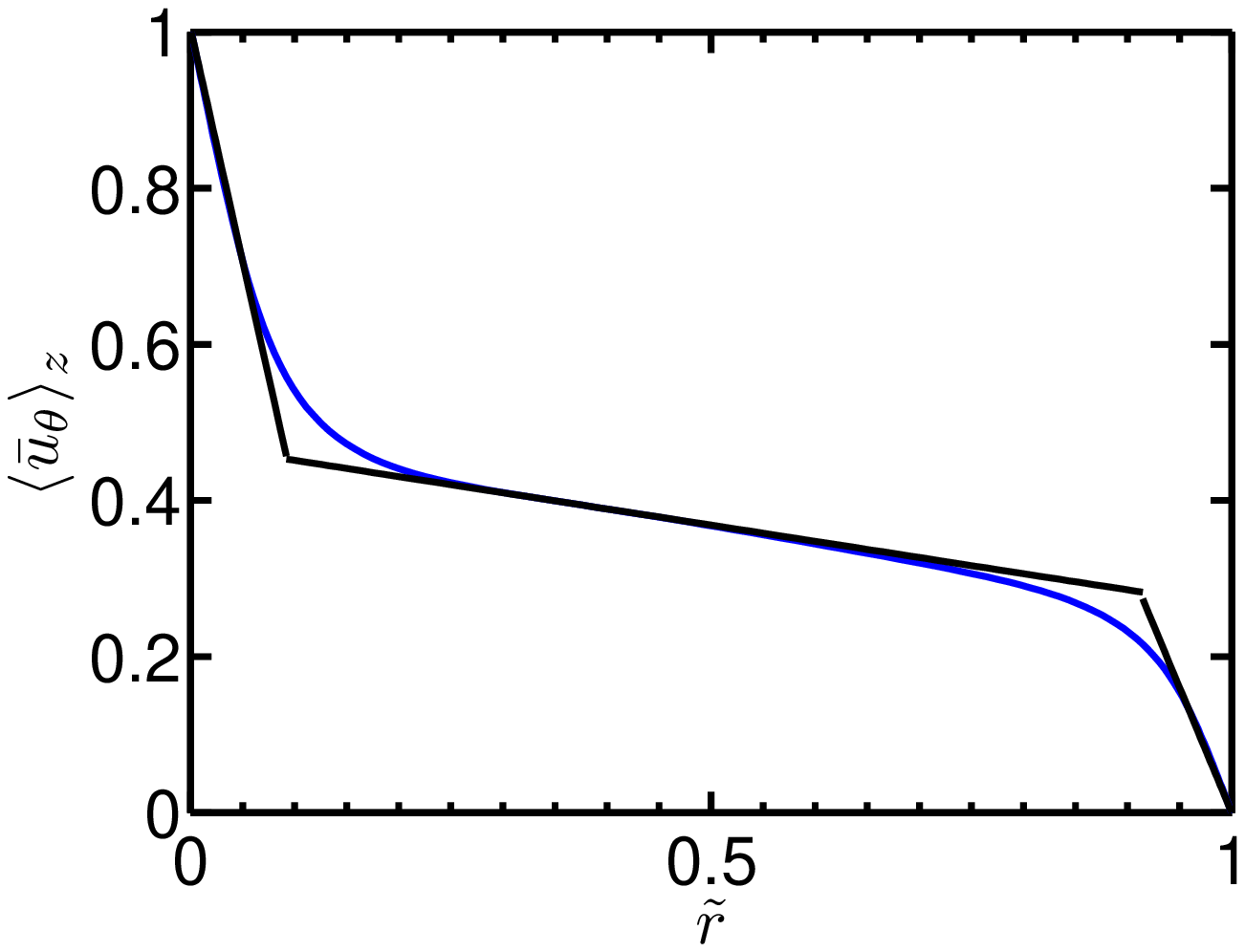}}\\
  \caption{Example of the two fitting procedures for $Ta$=$1.91\cdot10^6$ and $a$=$-0.2$. The left panel shows a linear fit to the bulk part of the  angular velocity $\langle \bar{\omega}\rangle_z$. The right panel shows in addition the fit to $\langle \bar{u}_\theta\rangle_z$ for its boundary parts, which will be used in Section \ref{sec:blobli}. Both bulk fits are done at the inflection point, but for  different variables ($\bar{\omega}$ or $\bar{u}_\theta$), which gives a slight difference.}
  \label{fig:domegablobliex}
 \end{center} 
\end{figure}

\begin{figure}
 \begin{center}
  \subfloat{\label{fig:adwdrEta0714Tas}\includegraphics[width=0.49\textwidth]{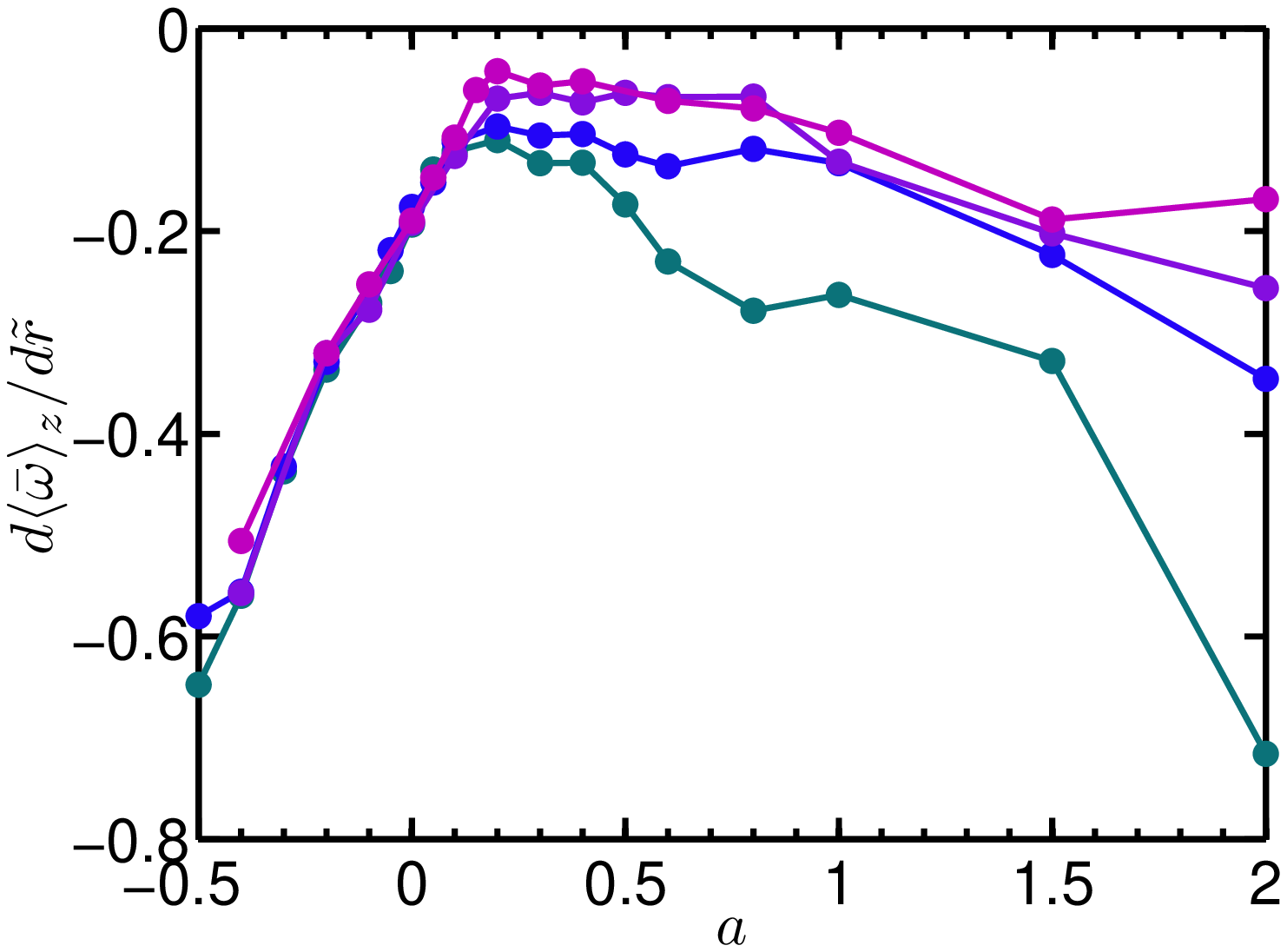}}
  \subfloat{\label{fig:RodwdrEta0714Tas}\includegraphics[width=0.49\textwidth]{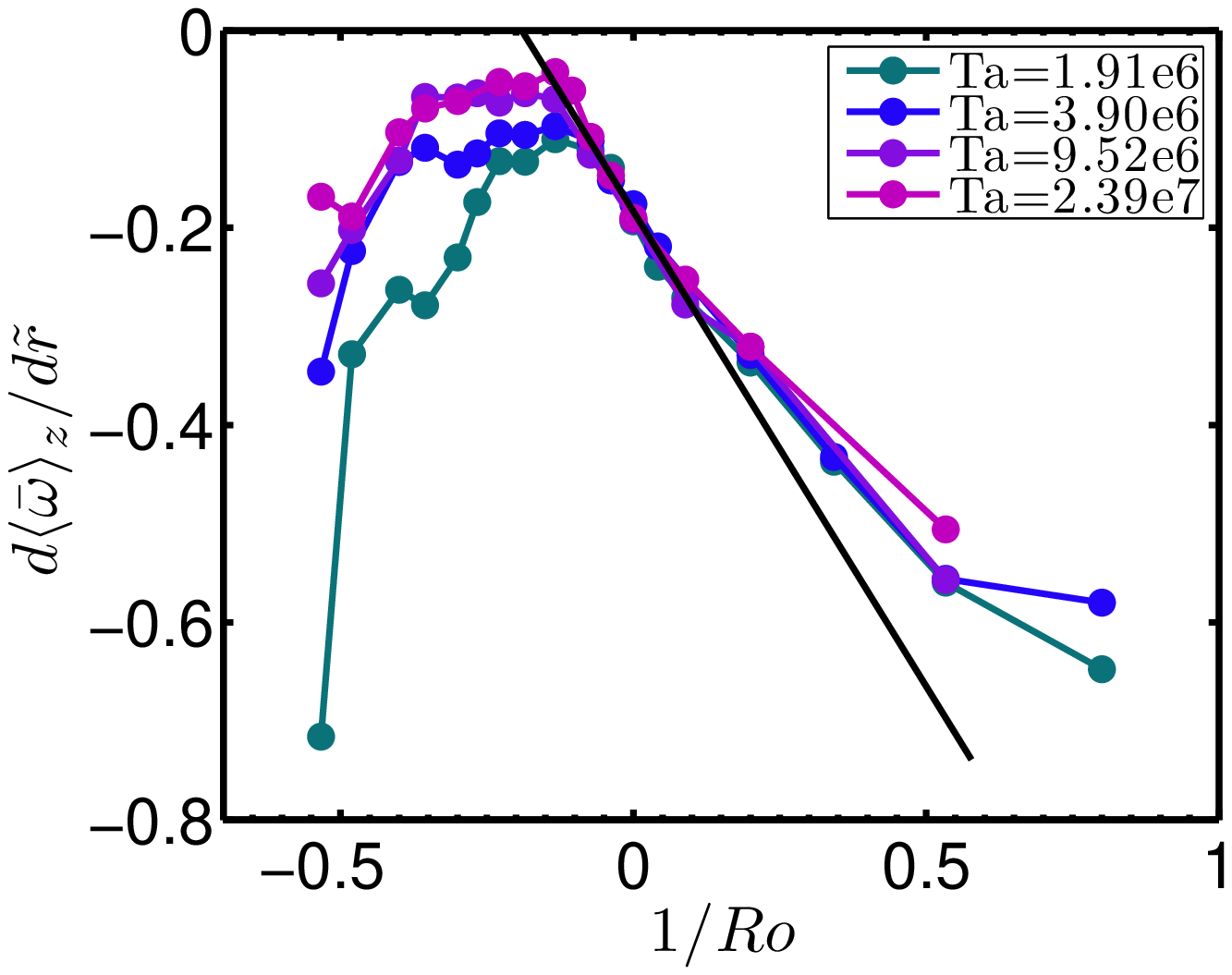}}
  \caption{ The average slope of the angular velocity $d\langle\bar{\omega}\rangle_z/d\tilde{r}$ versus $a$ (left) and $\usro$ (right) in the bulk. For co-rotation and only slight counter-rotation, there is a nearly linear relationship linking $\usro$ with $d\langle\bar{\omega}\rangle_z/d\tilde{r}$. A black straight line has been added in this region to artificially extend this relationship up to $d\langle\bar{\omega}\rangle_z/d\tilde{r}=0$. There is also a plateau of $d\langle\bar{\omega}\rangle_z/d\tilde{r}$ at counter-rotation. Here the $r$-slopes are smaller, i. e., the profiles flatter, for increasing $Ta$, reflecting an increased convective transport.  
}
  \label{fig:aRodwdrEta0714Tas}
 \end{center} 
\end{figure}

To further quantify  this, the gradient of $\langle \bar{\omega}\rangle_z$ is calculated. This is done by fitting a straight line to $\langle \bar{\omega}\rangle_z(\tilde{r})$ at the point of the profile's inflection, numerically using the grid points around it. An example of how this is done can be seen in figure \ref{fig:domegadrexample}. The results for the profile slopes in the bulk as functions of $Ta$ and $a$ or $\usro$ are shown in figure \ref{fig:aRodwdrEta0714Tas}.

The graphs collapse on each  other for co-rotation ($a<0$), which is what we expected from figure \ref{fig:romegaprofEta0714a0} and our previous analysis. An almost linear relationship between $\usro$ and the bulk slope of $\omega(r)$ is found. 
If pure inner cylinder rotation ($a=0$) is approached, the graphs for the various $Ta$ start to differentiate and reach a plateau.  
The absolute value of the slope decreases with increasing $Ta$. This is due to the increasing 
 importance of convection at higher $Ta$. Note however that also this counter-rotating case, for large enough $Ta$ the center slopes again
lose their $Ta$ dependence, i.e.\, are again mainly controlled by $Ro^{-1}$ and thus the Coriolis term.

We now come back to the corotating regime $a<0$ and want to connect the numerically found approximately 
linear relationship between the slope of $\omega(r)$ in the bulk and $Ro^{-1}$ with the dynamical equation (\ref{eq:rotatingTC}), which for the $\theta$-component of
the velocity can be rewritten as
\begin{equation}
 \label{eq:nsazimuthal}
 \partial_t u_\theta + u_r \partial_r u_\theta + \displaystyle\frac{u_\theta}{r} \partial_\theta u_\theta 
+ \displaystyle\frac{u_r u_\theta}{r}  + u_z \partial_z u_\theta =  \usro u_r -\partial_\theta p ~.
\end{equation}
The linear relationship can be obtained if we assume that the Coriolis force term $\usro u_r$ and convective term $u_r \partial_r u_\theta + u_r u_\theta/r$ balance each other, i.e.\,
we assume that the axial, azimuthal, and temporal dependences are small in eq. (\ref{eq:nsazimuthal}), which then boils down to
$u_r ( \partial_r u_\theta + u_\theta /r) \sim u_r Ro^{-1}$. Next, we use the fact 
 the radial velocity component $u_r$ -- the wind - in its non-dimensional form is constant along the present $Ta$-range (cf. Section \ref{sec:global}, seen also in experiment of \cite{hui12}). Therefore, and as $u_\theta$ hardly depends on $Ro^{-1}$, 
 an increased Coriolis force can only  be balanced by a larger slope $\partial_r u_\theta$.
 The only alternative is that the wind $u_r$ vanishes altogether, $u_r=0$,  and the flow state returns to the 
  purely azimuthal, laminar case.

To further substantiate this argument, we now 
decompose the flow field into a $(t,\theta,z)$-averaged mean azimuthal flow component $U_\theta$, depending on the radial position $r$ only, plus fluctuations $u^\prime$, as well as a decomposition of the pressure into a mean pressure $P$ plus the pressure fluctuations $p^\prime$. Inserting these Reynolds type
decompositions into the radial and azimuthal momentum equations -- in which besides $U_\theta$ only its $r$-derivative survives and ignoring viscosity for now, we arrive at the following equations:
\begin{equation}
\label{eq:inviscid-1}
 \partial_t u^\prime_r + u^\prime_r \partial_r u^\prime_r + \displaystyle\frac{u^\prime_\theta}{r} \partial_\theta u^\prime_r - \displaystyle\frac{U^2_\theta}{r} 
- \displaystyle\frac{U_\theta u^\prime_\theta}{r}
- \displaystyle\frac{u^{\prime2}_\theta}{r} + u^\prime_z\partial_z u^\prime_r = -\partial_r (P+p^\prime) - \usro (U_\theta + u^\prime_\theta),
\end{equation}

\begin{equation}
\label{eq:inviscid-2}
 \partial_t u^\prime_\theta + u^\prime_r \partial_r U_\theta + u^\prime_r \partial_r u^\prime_\theta 
+ \displaystyle\frac{u^\prime_\theta}{r} \partial_\theta u^\prime_\theta + \displaystyle\frac{u^\prime_r U_\theta}{r} 
+ \displaystyle\frac{u^\prime_r u^\prime_\theta}{r}  + u^\prime_z \partial_z u^\prime_\theta =  \usro u^\prime_r -\partial_\theta p^\prime.
\end{equation}

It is important to note that $U_r$ and $U_z$ are both equal to zero, so $u_r=u^\prime_r$ and $u_z=u^\prime_z$. 
As long as $\usro$ is larger than $\usro_{opt}$, we assume that 
already the mean  flow contributions  alone balance in eqs.\ (\ref{eq:inviscid-1}) and  (\ref{eq:inviscid-2}),
\begin{equation}
 -\displaystyle\frac{U_\theta^2}{r} \sim -\partial_r P +  \usro U_\theta~, 
\end{equation}
and
\begin{equation}
\label{eq:first-U-balance}
 u^\prime_r \partial_r U_\theta = u^\prime_r \left (r \displaystyle\frac{d\langle\bar{\omega}\rangle_z}{dr}  + \langle\bar{\omega}\rangle_z \right ) \sim -u^\prime_r \left (\usro+\displaystyle\frac{U_\theta}{r}\right ). 
\end{equation}
 As  in the bulk $r\approx r_a $ is almost constant,  the linear relationship
 $\partial_r \omega \propto const - Ro^{-1}$ 
  between $\usro$ and bulk slope $d\langle\bar{\omega}\rangle_z/d\tilde{r}$ is obtained.

 Figure \ref{fig:RodwdrEta0714Tas}, displaying this linear relationship,
 can be used to obtain a quantitative estimate for optimal transport for large $Ta$. We can see two distinct features in the slope $d\langle\bar{\omega}\rangle_z/d\tilde{r}$ versus $\usro$ curve. There is a plateau, where the value of the slope is linked to $Ta$ (and therefore to the viscous term in the equation of motion), and there is a line to the right of the plateau where the value of $d\langle\bar{\omega}\rangle_z/d\tilde{r}$  is independent of $Ta$ and thus linked only to the Coriolis force. From the previous discussion and from the experimental evidence of \cite{gil12} 
 we know that optimal transport is linked to the flattest $\omega$-profile. We can interpret the shift of $\usro_{opt}$ with $Ta$ as that value of $\usro$ where the plateau meets the co-rotation linear relationship, i.e.\ the flattest possible $\omega$-profile that does not break the large scale balance discussed before. If $\usro$ becomes more negative, the profile would have to become flatter to keep on satisfying the large scale balance. As this does not happen, the transport decreases for more negative $\usro$.

With increasing $Ta$, the value of $d\langle\bar{\omega}\rangle_z/d\tilde{r}$ at the plateau increases, and the curves cross at a smaller value of $\usro$, which corresponds 
to  a shifted maximum. Eventually, the plateau value of $d\langle\bar{\omega}\rangle_z/d\tilde{r}$ will tend to zero as seen in the experiments of  \cite{gil12}, and the co-rotation line will cross the plateau at the x-axis. We can  extend the straight line to get an estimate for when this happens
 and obtain $\usro_{opt}(Ta\to\infty)=-0.20$, corresponding to $a_{opt}\approx 0.34$, an estimate consistent with the experimental values $a_{opt} \approx 0.33 \pm 0.05$ of \cite{gil12} and $a_{opt} \approx 0.35$ of \cite{pao11}.
  
If $\usro$ is too negative, $\usro<\usro_{opt}$,  the large scale balance of equation (\ref{eq:first-U-balance})
 can no longer be satisfied. This can be seen as the Coriolis force now has values which would require a smaller (or even a negative) value of $d\bar{\omega}/d\tilde{r}$ for the balance to hold. Since this is not possible to accommodate, bursts will  originate from the outer cylinder towards the inner cylinder, because the flow tries to accommodate a large Coriolis force. These bursts increase in importance until they end up stabilizing parts of the flow, or even the whole flow which will drastically reduces the transport. Therefore, a maximum transport is reached just when the Coriolis force balances the large scale convective term. If it is further increased, stabilized regions start appearing. This is linked to the appearance of a neutral surface, which is analyzed in the next section. 

The large scale balance cannot be satisfied either if $\usro$ is too positive. The Coriolis force then requires a value of $d\bar{\omega}/d\tilde{r}>1$ to be balanced through the convective acceleration forces due to the average flow, and this cannot be accomodated for. Unlike the previous case, the flow cannot be separated into stable and unstable regions. Instead, this can be linked to the complete dissapareance of the radial and axial components of the flow (the so-called wind). This causes $\nom$ to drop to the purely azimuthal value, as seen in figure \ref{fig:NunnaEta0714}.

\subsection{Neutral surface}
\label{sec:nlpushing}

In this subsection, we will take a break from using the rotating reference frame, and return 
to the inertial reference frame to analyze the neutral surface which  is defined 
as that surface where, in the inertial reference frame, $\omega$ is zero. This surface only exists 
for non-negative values of $a$ and coincides with the outer cylinder in the case of pure inner 
cylinder rotation.  In an inertial reference frame, it marks the  division between the Rayleigh (inviscid)
stable and unstable regions. This means that this surface separates two regions, an unstable inner 
region and a stable outer region. In the stable region, perturbations to the azimuthal flow
(both large  scale wind as in Taylor vortices and small  scale perturbations 
such as plumes) cannot grow. Therefore we expect this surface to play a significant role for the behaviour 
of the flow. It was already shown to be important in controlling optimal transport by  \cite{gil12}.

\begin{figure}
 \begin{center}
  \subfloat{\label{fig:omegaEta0714Tay100Ka02}\includegraphics[height=0.6\textwidth,trim = 70mm 2mm 55mm 3mm, clip]{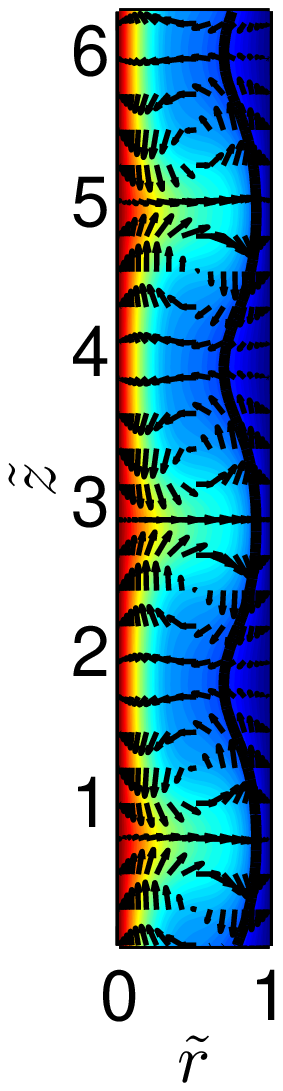}}
  \subfloat{\label{fig:omegaEta0714Tay100Ka04}\includegraphics[height=0.6\textwidth,trim = 72mm 2mm 57mm 3mm, clip]{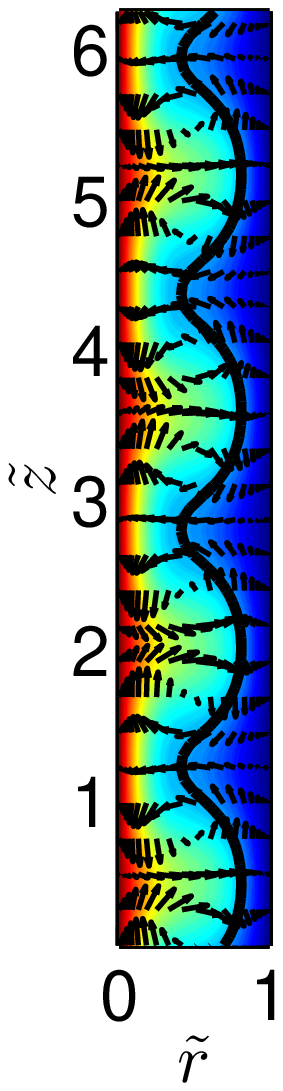}}                
  \subfloat{\label{fig:omegaEta0714Tay100Ka06}\includegraphics[height=0.6\textwidth,trim = 63mm 2mm 50mm 3mm, clip]{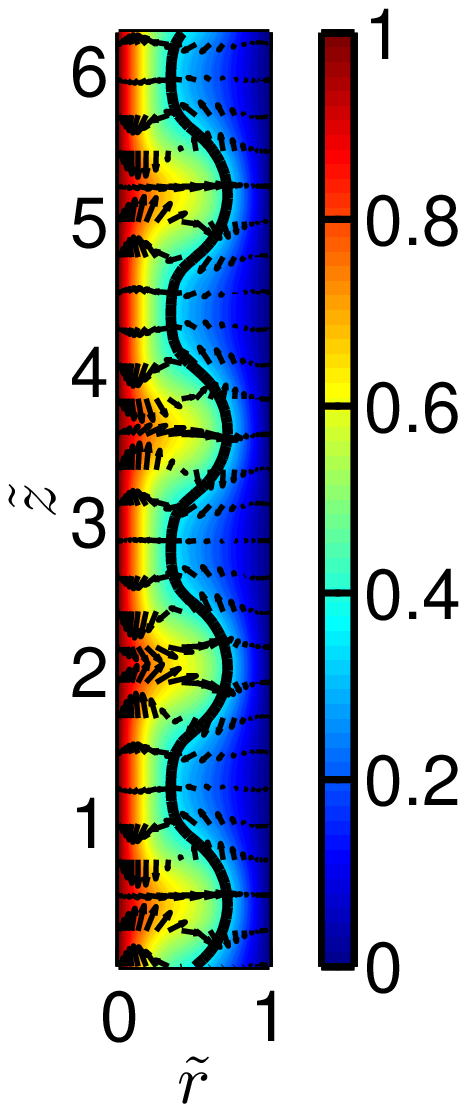}}                                
  \caption{Contour plots of the dimensionless angular velocity $\bar{\omega}(\tilde{r},\tilde{z})$ with indicated neutral surface (black line) for $Ta=1.03\cdot10^5$. Left: $a =0.2$ and three vortex pairs. Middle: $a = 0.4$ and four vortex pairs with a non-square aspect ratio and a highly distorted neutral surface. Right: $a$ $=$ $0.6$, four vortex pairs which do not penetrate the whole gap. See also figure \ref{fig:nlpushing} for a zoom-in.}
  \label{fig:omegaEta0714Tay100K}
 \end{center} 
\end{figure}

\begin{figure}
 \begin{center}
  \subfloat{\label{fig:omegaEta0714Tay250Ka02}\includegraphics[height=0.67\textwidth,trim = 70mm 2mm 55mm 3mm, clip]{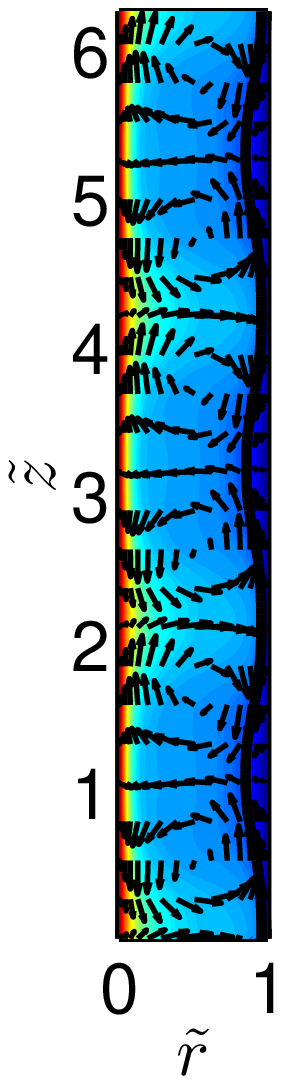}}
  \subfloat{\label{fig:omegaEta0714Tay250Ka04}\includegraphics[height=0.67\textwidth,trim = 72mm 2mm 57mm 3mm, clip]{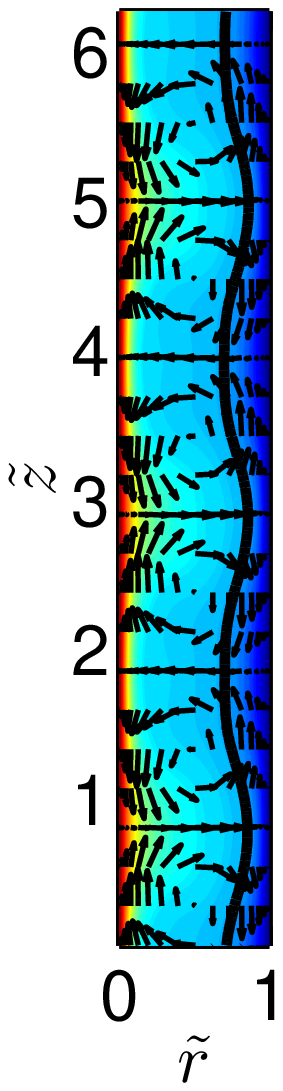}}                
  \subfloat{\label{fig:omegaEta0714Tay250Ka06}\includegraphics[height=0.67\textwidth,trim = 63mm 2mm 50mm 3mm, clip]{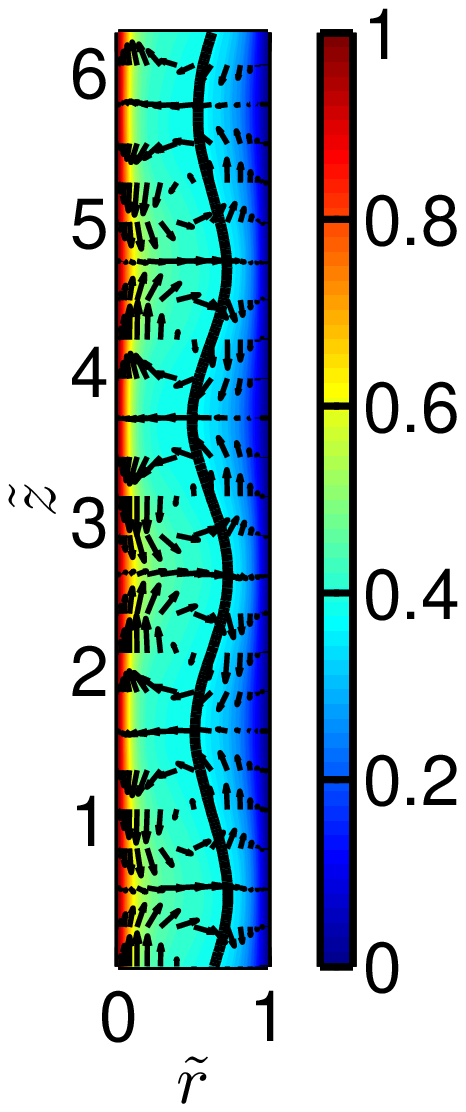}}                                
  \caption{Contour plots of $\bar{\omega}(\tilde{r},\tilde{z})$ with neutral surface indicated for $Ta=3.90\cdot10^6$ and $a$-values (left to right) of $0.2$, $0.4$, and $0.6$. All of them have three vortex pairs.}
  \label{fig:omegaEta0714Tay250K}
 \end{center} 
\end{figure}

In general, the position $r_N$ of the neutral surface depends on the height $z$. 
figure \ref{fig:omegaEta0714Tay100K} shows contour plots of the angular velocity with $r_N(z)$ indicated. 
 Indeed, $r_N(z) $ shows a strong axial dependence,  showing heights with positive or negative $u_r$, 
at which the neutral surface is pushed more outside or inside, respectively. This strong axial dependence of $r_N$ is a measure 
for the vortex strength and becomes weaker, when the vortices lose importance at very high $Ta$.
By comparing figures \ref{fig:omegaEta0714Tay100K} and \ref{fig:omegaEta0714Tay250K} the effect of the Taylor number on the position of the neutral surface can be seen. Its distortion happens at larger $a$ for increasing $Ta$, as expected.

\begin{figure}
 \begin{center}
  \subfloat{\label{fig:nlpushinga02}\includegraphics[height=0.4\textwidth,trim = 35mm 0mm 35mm 3mm, clip]{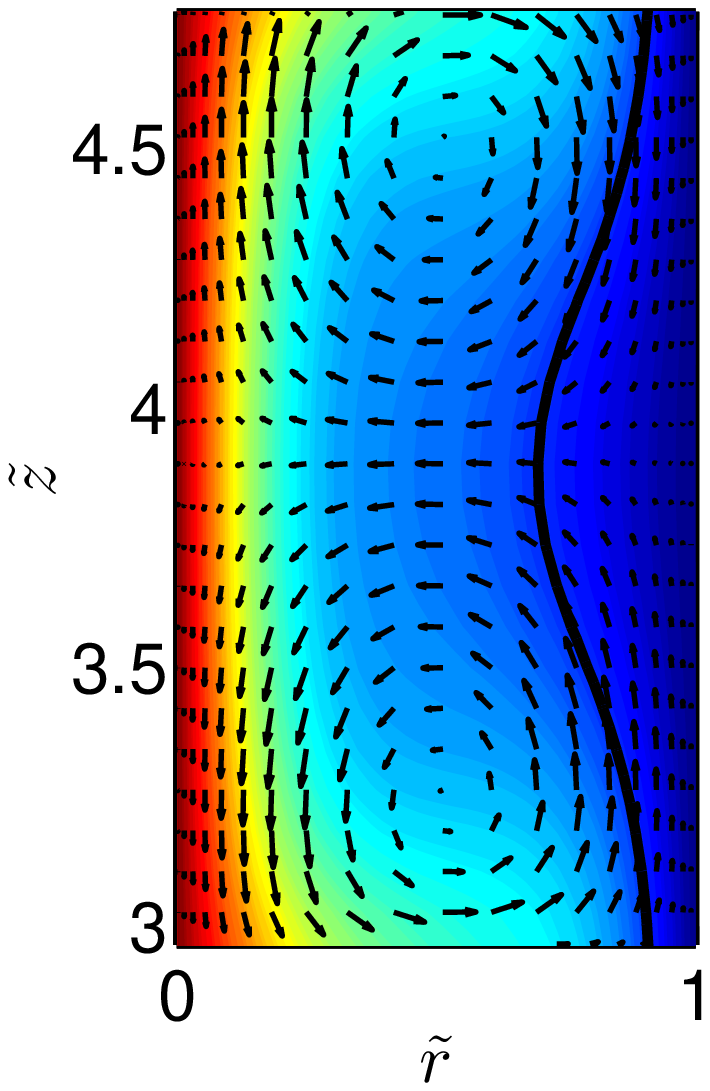}}
  \subfloat{\label{fig:nlpushinga04}\includegraphics[height=0.4\textwidth,trim = 35mm 0mm 35mm 3mm, clip]{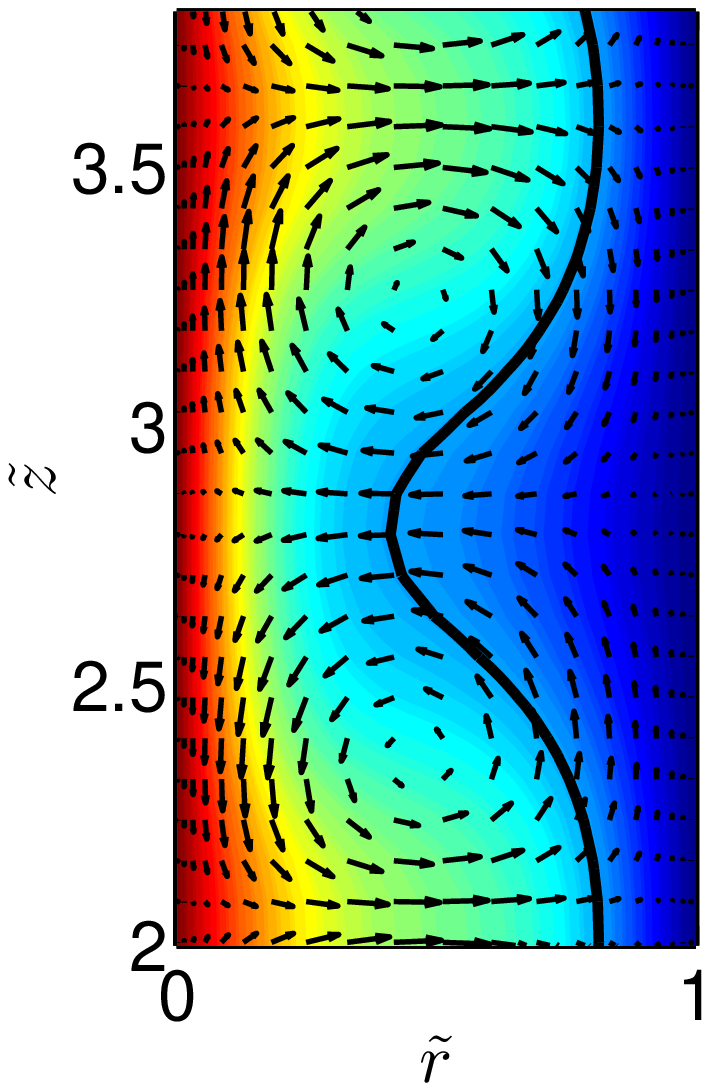}}
  \subfloat{\label{fig:nlpushinga06}\includegraphics[height=0.4\textwidth,trim = 30mm 0mm 30mm 3mm, clip]{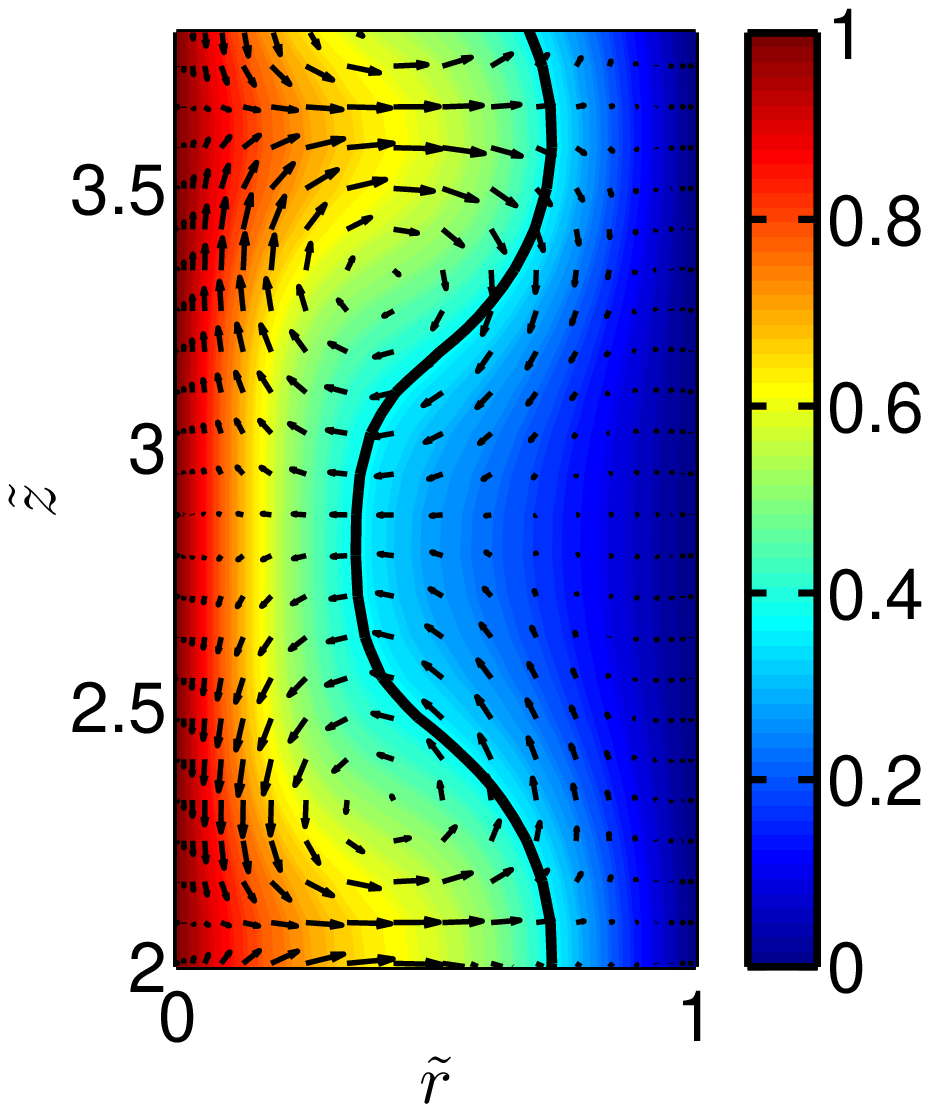}}
  \caption{Zoomed-in contour plots of $\bar{\omega}(\tilde{r},\tilde{z})$ for $Ta=1.03\cdot10^5$ with the neutral surface indicated as a black line. Left: $a = 0.2$, normal state with three vortices. Middle: $a = 0.4$, the distorted vortices are strong enough to fully penetrate the gap. Right: $a = 0.6$, the distorted vortices cannot penetrate the whole gap due to the stabilizing effects beyond the neutral line.
Moreover, with increasing $a$ the distance between the vortex centers  shrinks. 
}
  \label{fig:nlpushing}
 \end{center} 
\end{figure}

\begin{figure}
 \begin{center}
  \subfloat{\label{fig:arnEta0714Tas}\includegraphics[height=0.37\textwidth,trim = 8mm 0mm 18mm 0mm, clip]{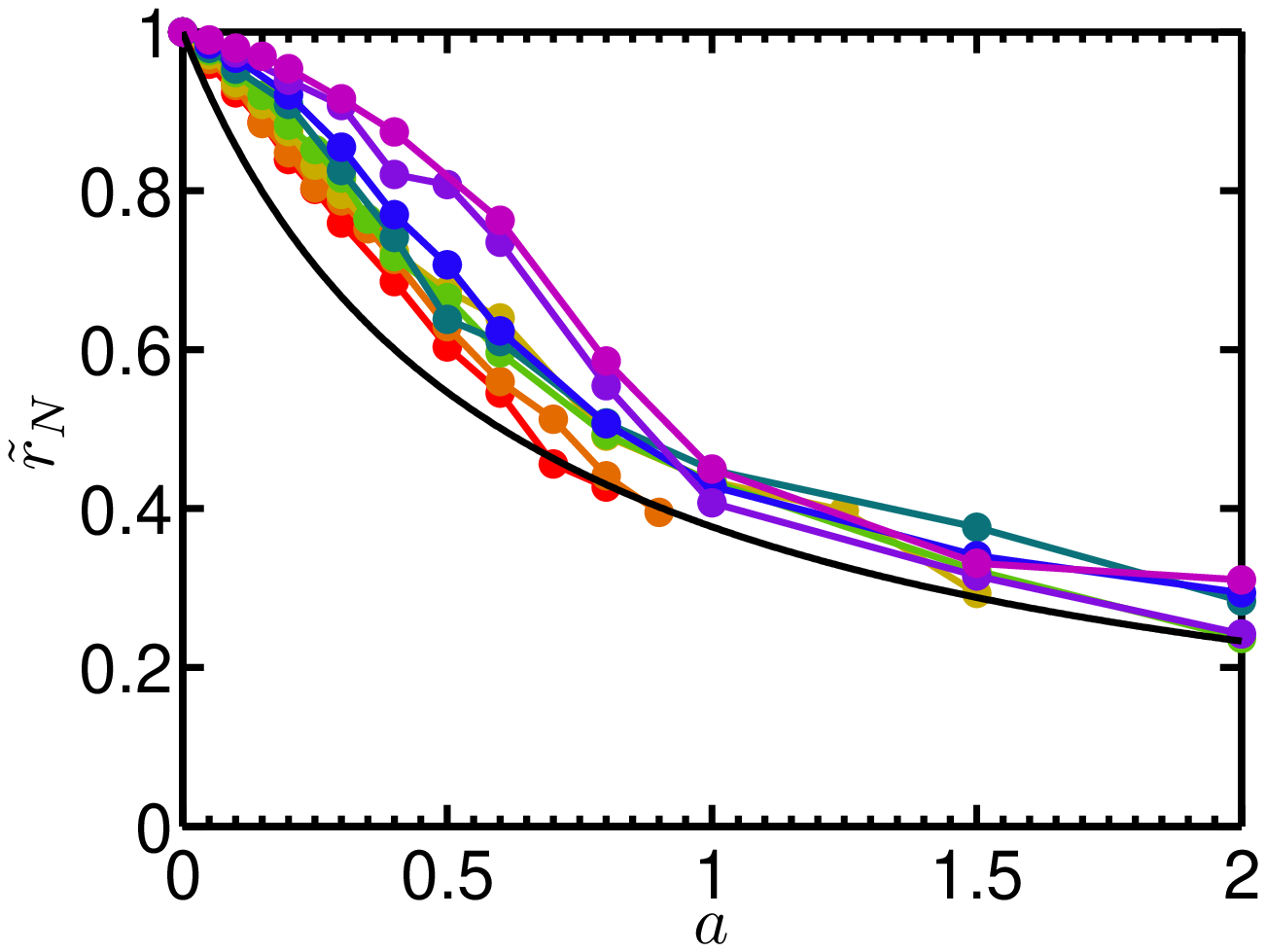}}
  \subfloat{\label{fig:arnrlamEta0714Tas}\includegraphics[height=0.37\textwidth,trim = 0mm 0mm 3mm 0mm, clip]{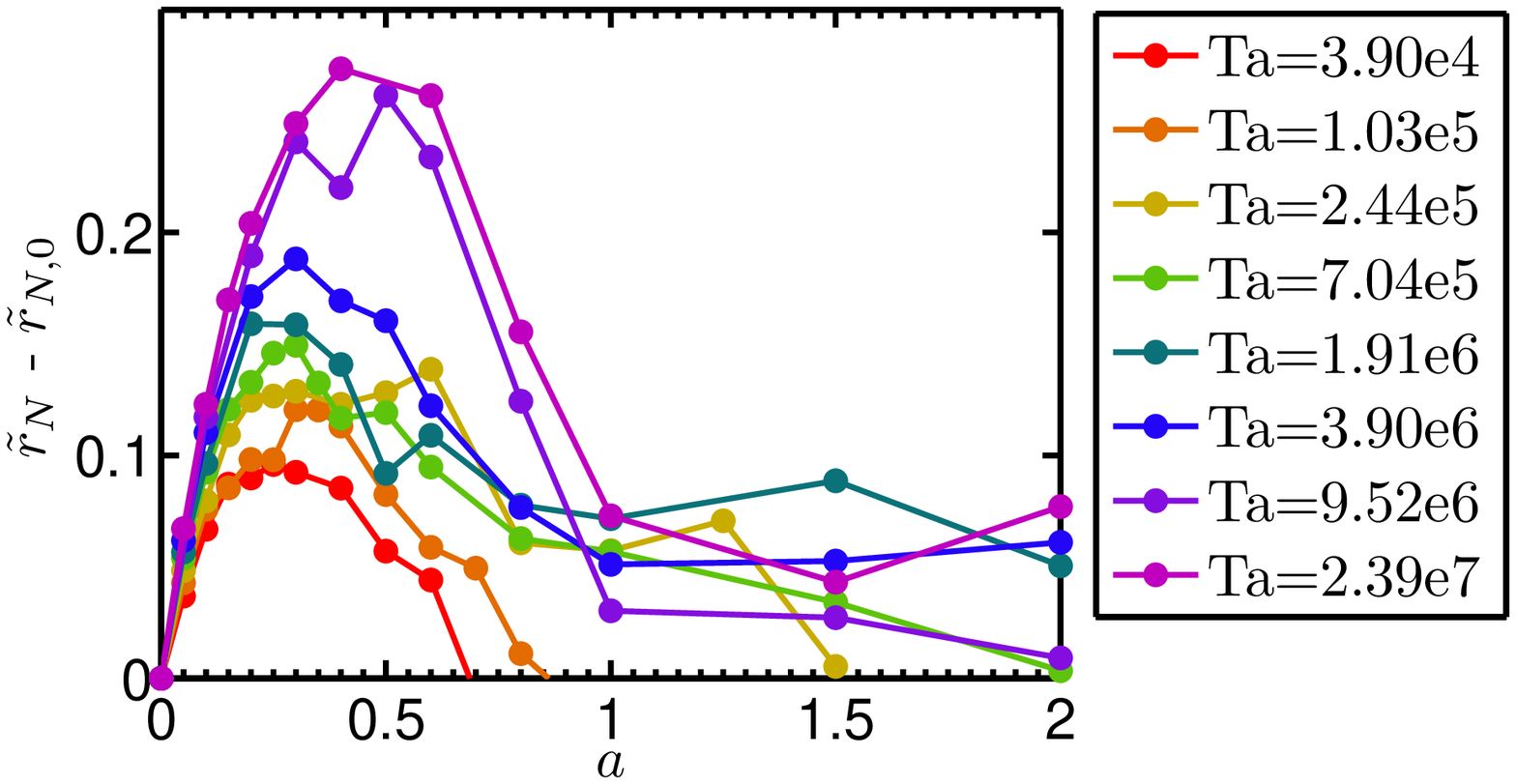}}
  \caption{Left panel, the neutral surface location $\tilde{r}_N$ of the time, azimuthally, and axially averaged angular velocity versus rotation ratio $a$ for various $Ta$. The neutral radius $\tilde{r}_N$ is moving inwards with increasing counter-rotation (increasing $a$), but is pushed back to the towards the outer cylinder for increasing $Ta$ at given $a$. The black line shows the position of neutral radius $\tilde{r}_{N,0}$ in laminar, purely azimuthal flow. Right panel, $\tilde{r}_N-\tilde{r}_{N,0}$ is shown, quantifying 
 the pushing of the neutral line towards the outer cylinder with increasing $Ta$. }
   \label{fig:arnEta0714Tasfull}
 \end{center} 
\end{figure}

If $a$ is large enough, the vortices are no longer able to penetrate the whole gap. There the neutral surface separates 
the Rayleigh stable and unstable regions. The vortices are mainly located in the unstable range, but enter partially also into  
the Rayleigh stable region. Being restricted to part of the gap, they also shrink in horizontal direction. 
Because the vortices try to remain as square-like as possible, their height (wave length) also shrinks, allowing new vortices 
to appear in the available given height. This is visible in the right panel of figure \ref{fig:nlpushing}. 
These vortices are also associated with a stronger wind. If the value of $a$ is not very large, 
they thus will fill up again the distance between the two cylinders but with a distorted aspect ratio. 
A zoom-in of this effect can be seen in the middle panel of figure \ref{fig:nlpushing}. This causes both 
the rise in $\nom$ for positive $a$ seen in figure \ref{fig:RoNomh} around $a\approx0.3$ at low $Ta$ 
and the crossovers seen in figure \ref{fig:arnEta0714Tas}.

Next, in addition to the temporal and azimuthal average, we also average $r_N(z)$ in axial direction and call this average 
$\bar r_N$. Figure \ref{fig:arnEta0714Tas} shows how $\bar{r}_N  $ varies with $a$ and $Ta$. The position of the neutral surface for the laminar, purely azimuthal flow is plotted for comparison.
For slight counter-rotation and fixed $a$, the mean neutral surface is increasingly pushed towards 
the outer cylinder with increasing $Ta$ due to enhanced turbulence. On the other hand, 
 with increasing $a$ the Coriolis force pushes the neutral
surface more and more towards the inner cylinder. Once the neutral surface reaches the laminar and purely
azimuthal flow value, the flow is stabilized.

The curves $\tilde{r}_N(a)$ for different $Ta$ can also  cross. At a constant
rotation ratio, some of the lower $Ta$ have a neutral surface which is further away from the inner cylinder than for some of the higher $Ta$. 
This is due to changes in the number and strengths of vortex pairs in the flow, 
which happen earlier for lower $Ta$. With a further increase of $a$ the 
trend reverses again, since respectively smaller values of $a$ stabilize the flow at already lower $Ta$. 
In an inertial reference frame, this simply means that there is no longer
a radial velocity which can push the neutral surface outwards, so $\bar{r}_N$ falls back
to the laminar, purely azimuthal flow value.

\section{Conclusions}

An extensive direct numerical simulation (DNS) exploration of the parameter space of a Taylor-Couette (TC) system at Taylor numbers in the range of $10^4<Ta<10^8$ was presented. First  the code was validated versus existing numerical and 
 experimental data. After this, the transition from the laminar but still purely azimuthal regime to the Taylor vortex state was analyzed. The regime where these vortices dominate the flow was studied in detail, revealing scaling laws between the Taylor number $Ta$, the angular velocity flux $\nom-1$, and the wind Reynolds number $\rew$. These scaling laws ceased to be valid when the Taylor number was increased beyond $Ta\approx 3\cdot10^6$. At this driving strength 
 the coherence structures become so small that they lose importance and are no longer the dominating feature of the flow.

Then the effect of the outer cylinder rotation on these scaling laws was analyzed. If both cylinders are co-rotating, the scaling laws were (slightly) modified, but for counter-rotating cylinders no significant differences could be seen. After the shrinking of the coherent structure and loss of their importance the value for 
optimal transport $a_{opt}$ shifted towards counter-rotation. This drift is expected to continue at higher Taylor numbers and will be the course of future DNS investigations.

Next, the behavior of local flow variables was studied. 
 Analyzing the profiles $\omega(r)$ sheds light on the two transport mechanisms, convective and diffusive, cf.\ the two contributions in (\ref{eq:ch1_J_w}). The optimal transport of $\omega$ could be linked to a balance between the Coriolis force and the inertial terms in the equations of motion. This balance is best achieved when the bulk profile is flattest and is broken with increasing counter-rotation. This leads to the appearance of a neutral line and of ``stabilizing'' bursts.

The outer boundary layer of the $\omega$-profile is much thicker than the inner boundary layer. The quality of the 
approximation of the $\omega$-profiles by three straight lines was found to improve with increasing $Ta$, as the (bulk-)turbulence becomes stronger. But although the bulk is turbulent, the boundary layers are still of Prandtl-Blasius type. TC flow only reaches the ultimate state, if also the boundary layers undergo a shear instability and become turbulent, too. The present analysis showed that this transition is expected to happen in the $Ta$ range between $10^8$ and $ 10^9$, which is just outside the range of
 the present DNS. It will be analyzed in future work.

Our ambition is to further extend the $Ta$ number range in our DNS of TC in order to allow a one-to-one comparison between experiments and simulations in the ultimate regime of TC turbulence and to explore the physics of this ultimate regime, in particular to understand the transition to this regime, and the bulk-boundary layer interaction in that regime. This ultimate regime in TC flow has recently been observed and analysed in the experiments by \cite{hui12} and \cite{gil12}, as well as in Rayleigh-B\'enard (RB) experiments of \cite{he11}. As the mechanical driving in TC is more efficient than heating in RB convection, it is easier to reach the ultimate regime in TC experiments than in RB experiments. Therefore also numerically we expect to reach the ultimate regime earlier in TC flow than in RB flow.

Acknowledgements: We would like thank G. Ahlers, H. Brauckmann, B. Eckhardt, D. P. M. van Gils, S. G. Huisman, E. van der Poel, M. Quadrio, and C. Sun for various stimulating discussions during these years. The large-scale simulations in this paper were possible due to the support and computer facilities of the Consorzio interuniversitario per le Applicazioni di Supercalcolo Per Universita e Ricerca (CASPUR). We would like to thank FOM, COST from the EU and ERC for financial support through an Advanced Grant.

\bibliographystyle{jfm}

\bibliography{literatur}

\end{document}